\documentclass{jfm}

\usepackage{graphicx}
\usepackage{newtxtext}
\usepackage{newtxmath}
\usepackage{bm}
\usepackage{natbib}
\usepackage{color,xcolor}
\usepackage{hyperref}
\usepackage{siunitx}
\usepackage{booktabs}
\usepackage{float}
\usepackage[percent]{overpic}
\usepackage{xcolor}
\usepackage{tikz}  
\hypersetup{
    colorlinks = true,
    urlcolor   = blue,
    citecolor  = black,
}

\newcommand{\RomanNumeralCaps}[1]

\definecolor{mygreen}{RGB}{0, 104, 55}

\AtBeginDocument{
\captionsetup{
justification=justified,
width=\linewidth,
singlelinecheck=false
}
}
\makeatletter
\def\footerflagdefns#1{}
\makeatother

\usepackage{overpic}
\newcommand{\s}{§}

\title{Modelling velocity gradient invariants in turbulence with Lundgren spiral vortex ensemble}

\author{Zhi Zheng\aff{1}, Weiyu Shen\aff{1,2}, Jin-Han Xie\aff{1}, Yue Yang\aff{1,3}~\corresp{\email{yyg@pku.edu.cn}}}

\affiliation{
\aff{1}State Key Laboratory for Turbulence and Complex Systems, School of Mechanics and Engineering Science, Peking University, Beijing 100871, PR China
\aff{2}Max Planck Institute for Solar System Research, Göttingen 37077, Germany
\aff{3}HEDPS-CAPT, Peking University, Beijing 100871, PR China
}

\begin{document}
\maketitle
\thispagestyle{plain}
\pagestyle{plain}

\begin{abstract}
We investigate the joint probability density function $P(Q,R)$ of invariants $Q$ and $R$ of the velocity gradient tensor in the synthetic turbulent field formed from the Lundgren spiral vortex ensemble, which is referred to as the woven Lundgren turbulence (WLT).  
The WLT inherently captures non-Gaussian vorticity correlations and multiscale interactions, which are essential for reproducing $P(Q,R)$ observed in homogeneous isotropic turbulence. By analyzing the contributions of stretching, winding and dissipation processes in WLT, we demonstrate that the $Q$-$R$ diagram can be divided into three distinct physical regions: a head region dominated by stretching, a body region dominated by spiral vortex arms, and a tail region dominated by dissipation. We systematically vary three key parameters in WLT, the background strain rate, the cutoff age of spiral vortices and the viscous decay factor, and monitor the evolution of $P(Q,R)$ in different topological regions. It is revealed that increasing the stretching rate broadens the head region, larger cutoff ages enhance the inclination of the body region, and reducing viscous factors extends the tail region, thereby mimicking the effects of strain self-amplification and vortex stretching across scales. Based on WLT statistics, we establish and validate an explicit model of $P(Q,R)$ consisting of the stretching, spiral and dissipation contributions. Thus, the WLT provides a physically interpretable framework for understanding the relation between local geometry and global statistics of turbulence. 
\end{abstract}


\section{Introduction}\label{sec:introduction}
Velocity gradients in turbulent flows are central to both practical applications and turbulence theory \citep{Sree1997,Wallace2001}. From an applied perspective, the evolution of sub-Kolmogorov-scale objects embedded in turbulence is closely tied to the velocity gradients they encounter along Lagrangian or inertial particle trajectories. These gradients influence, for example, the deformation and fragmentation of bubbles and droplets \citep{Biferale2010,Maniero2012}, the extension of polymer chains \citep{Balk2000,Bagh2012}, and the rotation rates of small motile organisms \citep{Pedley1992,Chev2013}. From a theoretical perspective, characterizing velocity gradients and velocity increments in homogeneous isotropic turbulence (HIT) is essential for explaining intermittency and multifractality \citep{K62,Meneveau1991,Buaria2026}.

Velocity gradients also provide a rich characterization of rotational and fluid deformation properties at small scales of turbulence~\citep{Chong1990,Meneveau2011}. Deformation rates dictate the dissipation of turbulent kinetic energy, whereas rotation rates define the small-scale coherent vortices~\citep{Jimenez1993,Hosokawa1997,Jimenez1998}. 
The joint probability density function (PDF) $P(Q,R)$ of the second and third invariants $Q$ and $R$ of the velocity gradient tensor has proven to be an effective diagnostic for identifying coherent structures and quantifying local flow topologies~\citep{Bershad1993,Suman2010,Lawson2015}. The $Q$-$R$ phase plane offers a universal framework for classifying flow patterns across various scales and Reynolds numbers~\citep{Martin1998b,Ooi1999}.  Extensive experimental and numerical studies have focused on understanding $P(Q,R)$ in turbulent flows~\citep{vanderBos2002,Ishihara2009,Wallace2010}. Direct numerical simulation (DNS) studies of HIT have revealed a characteristic teardrop-shaped contours of $P(Q,R)$, with a preferential alignment along the right branch of the Vieillefosse tail~\citep{Vieillefosse1982,Vieillefosse1984}. This shape reflects the dominance of enstrophy-producing and strain-producing regions, corresponding to vortex stretching and self-amplification of strain rate, respectively~\citep{Betchov1956,Carbone2020}. Nonetheless, reproducing this universal $Q$-$R$ topology within reduced-order or synthetic turbulence models remains a longstanding challenge.

Some modelling methods for the velocity gradient tensor concentrated on constructing random fields with specified spectra or fractal interpolations~\citep{Kraichnan1990,Zhang2011}. They can reproduce energy spectra and low-order statistics. 
Other methods employ closure models for the Lagrangian evolution of the velocity gradient tensor, which require approximations for the pressure Hessian and the viscous Laplacian. These include the linear relaxation model~\citep{Martin1998a}, the nonlinear relaxation model using the inverse Cauchy–Green tensor~\citep{Jeong2003}, the recent fluid deformation (RFD) approximation~\citep{Chevillard2006,Chevillard2008}, and the Gaussian field (GF) closure~\citep{Wilczek2014}. The GF closure computes conditional averages of the pressure Hessian and viscous Laplacian directly from Gaussian velocity fields, revealing that even in an isotropic Gaussian field the conditional pressure Hessian is not isotropic. The enhanced Gaussian field (EGF) closure~\citep{Wilczek2014} adjusts coefficients using DNS data to achieve statistical stationarity.
These models effectively prevent singularity and accurately reproduce $P(Q,R)$ at moderate Reynolds numbers.
Subsequently, the recent deformation of Gaussian field (RDGF) closure~\citep{Johnson2016} combines the RFD mapping with Gaussian field statistics for the upstream initial conditions, avoiding DNS-tuned coefficients while preserving the anisotropic structure of the pressure Hessian. A multiple-time-scale extension~\citep{Johnson2017} couples velocity gradients across scales, reproducing anomalous scaling, negative skewness, and topological features at high Reynolds numbers.
Recently, data-driven methods have also been applied to modelling the velocity gradient tensor in turbulence~\citep{Tian2021,Das2024,Tan2024}. 

However, some underlying mathematical closures obscure the physical mechanisms, such as vortex stretching and winding, which shape $P(Q,R)$. 
Some studies~\citep{Yang2020,Yang2023} seek to advance the modelling of velocity gradient tensor from a single-point statistical description to multi-point correlation models that incorporate local geometric constraints, linking it to physical processes such as vortex stretching and enstrophy generation.
\citet{Johnson2024} conducted a systematic review of multiscale velocity gradient dynamics, highlighting the necessity of a statistical theory that connects $P(Q,R)$ with elemental vortex structures.

In turbulent flows, vortices with complex topological characteristics constitute the inertial and fine-scale ranges through which energy is continuously accumulated and dissipated~\citep{Batchelor1953}. These structures are essential components that sustain the intricate dynamical state characteristic of turbulence~\citep{Betchov1957,Pullin1998}.
A modelling strategy grounded in fine-scale turbulence dynamics can be conceptualized as a building-block assembly process. The selection of flow elements parallels the choice of building blocks (analogous to cellular tissues), while their spatiotemporal configuration determines the emergent macroscale morphology (equivalent to skeletal architecture). This paradigm necessitates a rigorous coupling between the physical interpretation of elemental elements (e.g.~vortex tubes and sheets) and their multiscale organizational principles.

Among the various candidates, the Lundgren spiral vortex stands out as a compelling building block. 
\citet{Lundgren1982} introduced a spiral vortex ensemble model where stretched three-dimensional (3D) vortices are represented as unsteady two-dimensional (2D) flows through a spatiotemporal stretching transformation. Within this model, each spiral vortex resides in a strain field created by adjacent vortices and undergoes axial stretching, with asymmetric vorticity fluctuations (referred to as ``spiral arms'' contributing to the winding of vortices). The ensemble assumes random vortex ages and orientations, statistical equilibrium between generation, stretching, and annihilation, and ergodicity. It also provides an analytical framework for characterizing stretched and evolving vortices that encapsulate the essential features of fine-scale turbulence, illustrating how the dynamical processes of vortex stretching and scale reduction combine to produce Kolmogorov's -5/3 spectrum~\citep{Lundgren1982,Gilbert1993,Horiuti2008}. 
Subsequent studies extended the spiral vortex framework to higher-order statistics~\citep{Pullin1993,Saffman1996} and scalar transport~\citep{Pullin2001}, demonstrating its potential for capturing non-Gaussian features of turbulence.

However, spiral vortex ensembles have not been applied to modelling $P(Q,R)$. Although the spiral vortex model inherently includes vortex stretching and winding, which correspond to the physical processes underlying the different regions of the $Q$-$R$ diagram, a systematic investigation connecting ensemble statistics to $P(Q,R)$ has yet to be realized. Furthermore, the influence of essential ensemble parameters, such as the background stretching rate, vortex cutoff age and viscous decay factor, on $P(Q,R)$ remains unclear. 

``Woven turbulence'' (WT), a method for synthesizing turbulent fields, provides a platform for the study and control of turbulence statistics. 
\citet{Shen2024} introduced a bottom‑up approach that uses vortex filaments to construct turbulent fields consisting of intertwined Burgers vortex tubes. It successfully reproduces key metrics such as energy spectrum, energy flux, second-order structure functions and vorticity intermittency. 
Subsequently, \citet{Han2026} extended this framework by incorporating a fractal distribution of vortex tubes with scale-dependent parameters derived from dimensional analysis and geometric similarity. 
Moreover, \citet{Shen2026} adapted the weaving concept to wall‑bounded turbulent flows by constructing hierarchical hairpin vortex packets based on the attached‑eddy hypothesis. 
Collectively, these developments establish a versatile, structure-based framework for generating realistic turbulent fields with controllable statistics and coherent vortices. 
Despite these successes, the original WT produces a butterfly-shaped $P(Q,R)$, rather than the asymmetric teardrop characteristic of turbulence~\citep{Shen2024}. This discrepancy identifies a structural question beyond reproducing vortex intensities, sizes and spatial arrangements. Whereas $Q$ measures the balance between rotation and strain intensities, $R$ also depends on strain geometry and its coupling with vorticity. Recovering their joint distribution therefore places additional constraints on how vortical and straining motions are organized. In particular, the internal structure of vortex elements may contain information that is insufficiently represented by axisymmetric tube profiles.

\begin{figure}
	\centering
	\begin{overpic}
    [width=\linewidth]{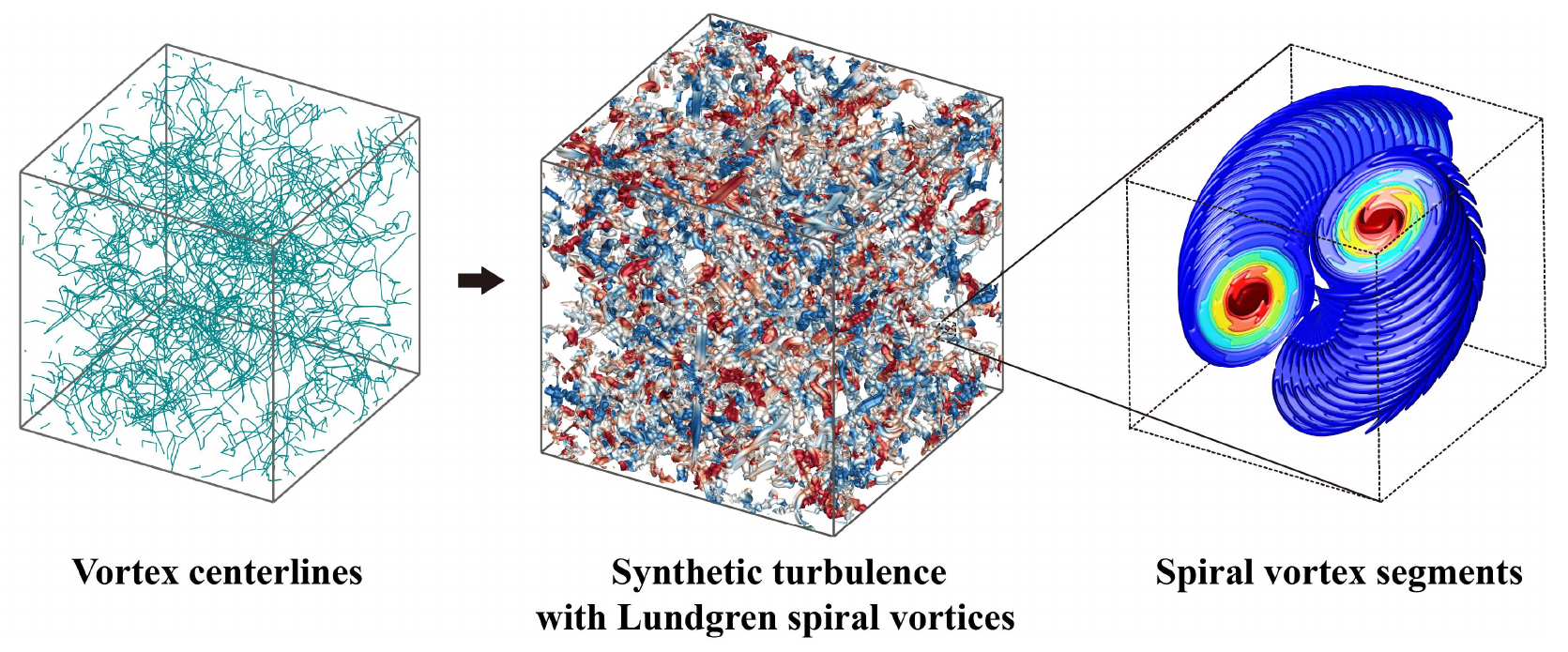}
    \put(0,38){\footnotesize{(a)}}
    \put(33,38){\footnotesize{(b)}}
    \put(70,38){\footnotesize{(c)}}
    \end{overpic}
    \caption{
    Schematic of the WLT based on Lundgren spiral vortices.
    (a) Vortex centerlines are obtained from the vortex filament simulation. 
    (b) Constructing vortex tubes with Lundgren spiral vortex profiles along the centerlines yields the WLT field. 
    The isosurface of the vorticity magnitude $|\boldsymbol{\omega}|$ is colored by the helicity density (inner product of the local velocity and vorticity). 
    (c) Lundgren vortex segments along the centerlines, which are represented by a set of isosurfaces of $|\boldsymbol{\omega}|$ with different colors.
    } \label{fig:WLT}
\end{figure}

Lundgren spiral vortices provide a representation of such internal organization. Differential rotation winds non-axisymmetric vorticity into spiral layers, whose development and persistence depend on stretching and viscous diffusion. An ensemble of such elements therefore represents varying degrees of internal organization, extending the description provided by smooth vortex profiles. Dependence on vortex age and viscous damping is particularly relevant because it determines the extent to which developed winding structures and their fine layers are represented in the ensemble.
As sketched in figure~\ref{fig:WLT}, we develop the ``woven Lundgren turbulence'' (WLT) framework by placing Lundgren spiral vortex profiles along centerlines obtained from vortex filament simulations, extending the woven turbulence framework established by \citet{Shen2024}.
Compared with previous studies which performed numerical simulations and theoretical analyses on the energy spectra of individual spiral vortex, we directly construct the spatial arrangement of the Lundgren spiral vortex ensemble through the multiscale coupling approach of woven turbulence, enabling direct statistical analysis of the turbulent field.

The paper is organized as follows. 
In \s\ref{sec:Weaving Turbulence}, we describe the synthesis of turbulent fields using an adjustable spiral vortex ensemble based on WLT.
In \s\ref{sec:Velocity Gradient Tensor}, we elucidate the dependence of $P(Q,R)$ on WLT parameters. 
In \s\ref{sec:model}, we present a model for $P(Q,R)$ in WLT, grounded in the statistical theory of the spiral vortex ensemble. 
Some conclusions are drawn in \s\ref{sec:conclusion}.

\section{Synthetic turbulence with Lundgren spiral vortices}\label{sec:Weaving Turbulence}
\subsection{Lundgren spiral vortex}\label{sec:profile}
We demonstrate the construction of a WLT field comprised of an ensemble of Lundgren spiral vortices, providing an adjustable platform for examining how vortex structures influence the statistics of velocity gradient invariants in HIT. 

The Lundgren spiral vortex offers an asymptotic solution to the Navier-Stokes (NS) equations for an axially stretched, temporally evolving vortex~\citep{Lundgren1982}. 
In the present study, following the Lundgren vortex framework, we assume that each segment of a curved, stretched vortex tube is locally subjected to an axisymmetric strain field, approximating the stretching effect induced by surrounding turbulence.
For a 3D vorticity field $\boldsymbol{\omega}$ under the axial strain rate $ \gamma $, the vorticity component field $\omega(r, \theta, t)$ along the vortex axis, denoted by the $Z$-axis, is derived from a 2D unsteady solution $\omega_{\mathrm{2D}}$ through the transformation
\begin{equation}\label{eq:2D}
    \omega(r, \theta, t) = \mathrm{e}^{\gamma t} \omega_{\mathrm{2D}} \left( \mathrm{e}^{\gamma t / 2} r, \theta, \frac{\mathrm{e}^{\gamma t} - 1}{\gamma} \right),
\end{equation}
where $r$ and $\theta$ are the radial and azimuthal coordinates in the $X$-$Y$ plane perpendicular to the $Z$-axis, respectively, and $t$ denotes time. 
For large times, the 2D spiral solution admits a Fourier series representation
\begin{equation}\label{eq:f2D}
    \omega_{\mathrm{2D}}(\xi, \theta, T) = \sum_{n=-\infty}^{\infty} f_n(\xi) \mathrm{e}^{\mathrm{i} n [\theta - \Omega(\xi) T]},
\end{equation}
where $\xi = \mathrm{e}^{\gamma t / 2} r$ represents the stretched radial coordinate, $T = (\mathrm{e}^{\gamma t} - 1) / \gamma$ denotes the straining time, $\Omega(\xi)$ is the angular velocity profile and the complex coefficients $f_n(\xi)$ encode the spiral structure.
Considering viscous effects by the factor $ \nu_L $, the Fourier modes of $\omega = \sum_n \omega_n \mathrm{e}^{\mathrm{i} n \theta}$ are
\begin{equation}\label{eq:Lundgren_spiral}
    \omega_n(r, t) = \mathrm{e}^{\gamma t} f_n\left(\mathrm{e}^{\gamma t/2} r\right) \exp \left[-\mathrm{i} n \Omega\left(\mathrm{e}^{\gamma t/2} r\right) T - \frac{1}{3} n^2\left(\frac{\mathrm{d} \Omega\left(\mathrm{e}^{\gamma t/2}r\right)}{\mathrm{d} \xi}\right)^2 \nu_L T^3\right].
\end{equation}

We adopt the method of \citet{Lundgren1993} to give adjustable profiles for vortex tubes. We consider a double-branched spiral, incorporating only even harmonics, and assume that all non-zero harmonics possess the same real amplitude $f(r)$ in the physical (unstretched) coordinate system. Furthermore, the angular velocity profile within the physical frame is represented by $\mu(r)$. For vortices of moderate age, the weak dependence of the radial coordinate on stretching is neglected. 
This assumption is justified within a statistical ensemble where the vortex age $\tau$ is considered as a parameter, and the stretching effect is incorporated into the definitions of $\mu(r)$ and $f(r)$. Under these assumptions, the vorticity of a single vortex, characterized by $\tau$, is expressed through the real Fourier series
\begin{equation}\label{eq:w}
\omega(r, \theta, \tau) = \omega_0(r) + 2f(r) \sum_{n=1}^{\infty} \cos \left[ 2n(\theta - \mu(r)\tau) \right] \exp \left( -\frac{4n^2 \mu'(r)^2 \nu_L \tau^3}{3} \right), 
\end{equation} 
where $\omega_0(r)$ is the axisymmetric component and $\mu'(r)$ is the radial derivative of angular velocity.
According to $ 2\pi r \omega_0 = \mathrm{d}(2\pi r^2 \mu) / \mathrm{d}r $, a direct consequence of Stokes' theorem applied to the axisymmetric vortex core, the kinematic compatibility relation
\begin{equation}\label{eq:mu}
\mu'(r) = \frac{\mathrm{d}\mu}{\mathrm{d}r} 
        = \frac{\omega_0(r) - 2\mu(r)}{r} 
\end{equation}
represents the differential rotation responsible for spiral winding, and $\mu'(r) \neq 0$ is a necessary condition for the spiral arms to tighten over time.

To obtain localized, smooth vorticity, we adopt the following dimensionless radial profiles~\citep{Lundgren1993} 
\begin{equation}\label{eq:w0_f}
    \omega_0(r) = \frac{\Gamma \exp \left[ - \left( {r}/{r_0} \right)^2 \right]}{(1+h)\pi r_0^2} + 2f(r)
\quad \textrm{and} \quad
    f(r) = \frac{\Gamma h \left( {r}/{r_0} \right)^2 \exp \left[ - \left( {r}/{r_0} \right)^2 \right]}{2(1+h)\pi r_0^2}
\end{equation}
for the axisymmetric vortex core and the asymmetric spiral arms, respectively, where $\Gamma$ denotes the total circulation, $r_0$ the vortex core radius, and $h$ the relative strength of the spiral arms.
Substituting \eqref{eq:w0_f} into \eqref{eq:mu} and performing the integration yields
\begin{equation}\label{eq:muu}
    \mu(r) = \frac{\Gamma \left[ 1 + h - \left( 1 + h + h \left( {r}/{r_0} \right)^2 \right) \exp \left( - \left( {r}/{r_0} \right)^2 \right) \right]}{2(1+h)\pi r^2},
\end{equation}
which implies that the vortex undergoes differential rotation across its radius, thereby forming leading-like spiral arms~\citep{Rowther2025}.
Specifically, for $ 0<h<1 $, \eqref{eq:muu} ensures that $\mu(r)$ decreases monotonically from a finite value $\Gamma/2(1+h)\pi r_0^2$ at $r = 0$ to zero as $r \to \infty$, and $f(r)$ reaches its maximum at a finite radius, thereby simulating the roll-up of vortex sheets. 

The viscous decay term in \eqref{eq:w} follows from the analysis of \citet{Lundgren1982}: the factor $4n^2$ originates from the harmonic index $2n$, while the exponent $-{4n^2 \mu'(r)^2 \nu_L \tau^3}/{3}$ represents the leading-order viscous damping of the $2n$-th harmonic. 
For positive values of $\nu_L$ and $\tau$, the contribution of the $2n$-th harmonic is proportional to $\exp(-4n^2)$; consequently, high-order harmonics are rapidly attenuated by viscous diffusion.
In all numerical calculations, the infinite series in \eqref{eq:w} is truncated after the first $N_\varepsilon$ harmonics.
The truncation order $N_\varepsilon = 100$ is determined through a convergence test in this work, which is sufficiently large to ensure that the neglected harmonics have a negligible effect on the vorticity field and the reported statistics.

\begin{figure} 
	\centering
	\begin{overpic}
    [width=1.0\textwidth]{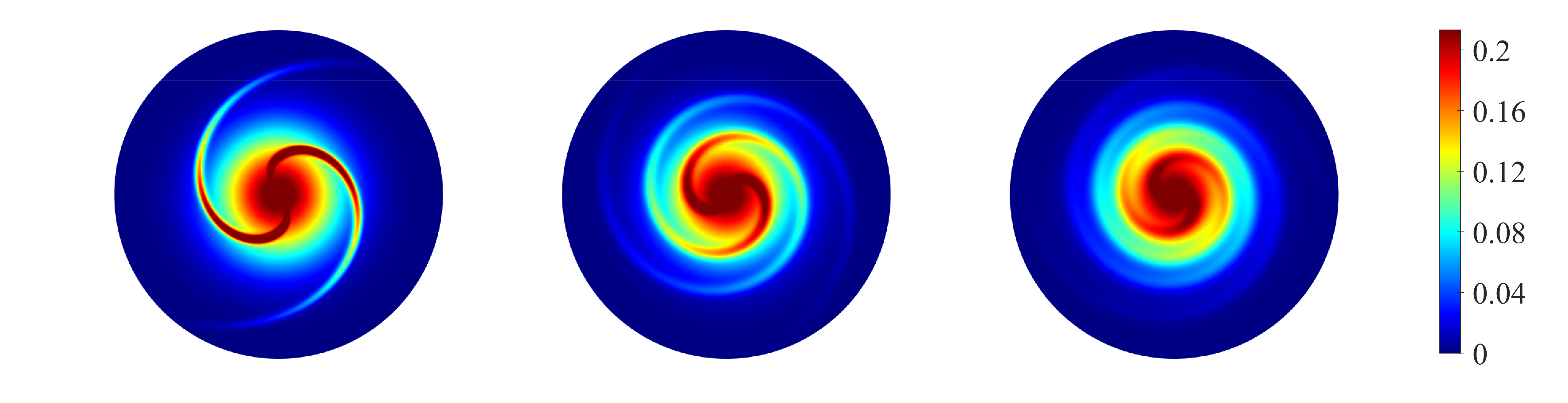}
\put(14,-0.5){\footnotesize{$\tau=50$}}
\put(43,-0.5){\footnotesize{$\tau=100$}}
\put(72,-0.5){\footnotesize{$\tau=150$}}
\put(91.7,0.3){\footnotesize{$\omega$}}
    \end{overpic}
    \caption{Vorticity contour of the spiral vortex defined in \eqref{eq:w} at different vortex ages $\tau$ for dimensionless parameters $r_0 = 1$, $\Gamma=1$, $\nu_L= 1 \times 10^{-4}$ and $ h = 0.5 $.} \label{fig:Profile}
\end{figure}

Hence, \eqref{eq:w} -- \eqref{eq:muu} form a specific, parametrized family of spiral vortices that preserve the fundamental characteristics of Lundgren's solution, yet remain sufficiently straightforward for both analytical and numerical analysis. 
Figure~\ref{fig:Profile} displays a series of spiral vortex profiles at different stages. As the vortex age $\tau$ increases, the two spiral arms progressively constrict, and the number of turns in the spiral increases due to differential rotation. 
To facilitate analysis, we take dimensionless forms $\Gamma=1$, $r_0 = 1$ and $ h = 0.5 $ below. 

\subsection{Adjustable Lundgren ensemble of vortex tubes}\label{sec:ensemble}
To construct a statistically stationary HIT field, we consider an ensemble of Lundgren vortices distributed in space, following the formulations of \citet{Lundgren1982} and \citet{Pullin1993}. 
The ensemble is based on four assumptions: ergodicity, allowing ensemble-averaged statistics to be represented by spatial averages over a sufficiently large realization of the vortex ensemble; locality of vorticity along each vortex axis, such that each vortex remains tube-like; random ages, orientations, and spatial positions of vortices, ensuring statistical homogeneity and isotropy; and dynamical balance between vortex generation, stretching, and annihilation, which maintains a steady-state ensemble.

The schematic of the woven turbulence generated by Lundgren spiral vortices is shown in Figure~\ref{fig:WLT}.
We conduct vortex filament simulations~\citep{Hanninen2014} to generate a statistically stationary, homogeneous and isotropic tangle of vortex filaments that serves as the set of centerlines of vortex tubes. 
The methodology and implementation of weaving turbulence have been thoroughly documented in \cite{Xiong2019,Xiong2020,Shen2023,Shen2024}. 
We focus here on the key innovation in WLT: the distribution of Lundgren spiral vortex profiles along these centerlines and the mapping of spatial coordinates to temporal parameters.

The central idea is to treat each point along a centerline as hosting a Lundgren spiral vortex element at a particular stage of its evolutionary life cycle. Rather than assigning independent vortices with random ages to each spatial location, we introduce a space-time mapping that continuously and cyclically connects vortex structures of different ages along the same centerline. 

This mapping requires that the spatial distribution of ages along the centerline be consistent with the stretching history of the vortices: a vortex element of age $\tau$ has an axial length $l(\tau) = l_0 S(\tau)$, where $l_0$ is its initial length and the stretch ratio is
\begin{equation}
    S(\tau) = \exp\left[\int_0^\tau \gamma(\varrho) \, d\varrho\right].
\end{equation} 
In WLT, the axial strain rate $\gamma$ is assumed to be an age-independent constant $\gamma_{L}$, yielding the simplified expression $l(\tau) = l_0 e^{\gamma_{L} \tau}$.
Therefore, we define a characteristic velocity $u_s = \gamma_{L} l_0 e^{\gamma_{L} \tau}$,
and obtain the relation of the arc-length coordinate $s$ along the centerline and $\tau$ through
\begin{equation}\label{eq:l_t}
\frac{ds}{d\tau} = u_s = \gamma_{L} l_0 e^{\gamma_{L} \tau}. 
\end{equation}
Equation~\eqref{eq:l_t}, together with \eqref{eq:w} -- \eqref{eq:muu}, determines the vorticity field within WLT and encapsulates the dynamical information contained in \eqref{eq:Lundgren_spiral}. 
We introduce a periodic age function, representing multiple cycles of vortex generation and decay, for a given closed centerline $\boldsymbol{\mathcal{C}}(s)$ with $s \in [0, \ell]$ and the total arc length $\ell$. 
A single complete life cycle consists of an increase in $\tau$ from $0$ to $\tau_{L}$ (corresponding to the growth and stretching phase), followed by a decrease from $\tau_{L}$ back to $0$ (corresponding to the decay and renewal phase). Here, the cutoff age $\tau_{L}$ of the vortices denotes the stage at which the spiral vortices in the ensemble undergo annihilation and subsequent reorganization. The arc length required for the increasing phase is obtained by integrating~\eqref{eq:l_t} as
\begin{equation}
    \ell_{\text{up}} =  \int_0^{\tau_{L}} u_s(\tau)  d\tau = {l_0} (e^{\gamma_{L} \tau_{L}} - 1).
\end{equation}
For the decreasing phase, a vortex element of age $\tau = \tau_{L} - \tau_0$, where $\tau_0$ increases from $0$ to $\tau_{L}$, still has the physical length change rate $u_s(\tau_{L} - \tau_0) =\gamma_{L} l_0 \mathrm{e}^{\gamma_{L}(\tau_{L} - \tau_0)}$. 
The arc length associated with this phase is therefore
\begin{equation}
\ell_{\text{down}} = \int_0^{\tau_{L}} u_s(\tau_{L} - \tau_0) d\tau_0 = {l_0} \left( \mathrm{e}^{\gamma_{L} \tau_{L}} - 1 \right). 
\end{equation}
Thus, a complete age cycle from $0$ to $\tau_{L}$ and back to $0$ corresponds to the total arc length
\begin{equation}
    \ell_{\text{cycle}} = \ell_{\text{up}} + \ell_{\text{down}} = {2l_0} \left( \mathrm{e}^{\gamma_{L} \tau_{L}} - 1 \right).
\end{equation}
For a closed centerline, $\mathscr{N} = \ell/\ell_{\text{cycle}}$ complete age cycles can be accommodated, where $\mathscr{N}$ is assumed to be an integer for simplicity. 
This yields a periodic age profile along the centerline, with each cycle comprising an increasing branch followed by a decreasing branch.

The explicit age function $\tau(s)$ along the centerline can be derived by inverting the relation $s(\tau)$. For the increasing branch with $0 \le s \le \ell_{\text{up}}$, we have
\begin{equation}
s(\tau) = \int_{0}^{\tau} \gamma_L l_0 \mathrm{e}^{\gamma_{L} \tau_0} d\tau_0 = {l_0} \left( \mathrm{e}^{\gamma_{L} \tau} - 1 \right),
\end{equation}
which gives
\begin{equation}\label{eq:t_up}
    \tau(s) = \frac{1}{\gamma_{L}} \ln \left( 1 + \frac{s}{l_0} \right), \quad 0 \le s \le \ell_{\text{up}}.
\end{equation}
For the decreasing branch with $\ell_{\text{up}} < s \le \ell_{\text{cycle}}$, we obtain
\begin{equation}\label{eq:t_down}
    \tau(s)= \tau(\ell_{\text{cycle}}-s)= \frac{1}{\gamma_{L}} \ln \left( 1 + \frac{\ell_{\text{cycle}}-s}{l_0} \right) , \quad \ell_{\text{up}} < s \le \ell_{\text{cycle}}.
\end{equation}
This age function satisfies $\tau(0) = 0$, $\tau(\ell_{\text{up}}) = \tau_{L}$ and $\tau(\ell_{\text{cycle}}) = 0$, and remains piecewise smooth on both the increasing and descending branches. 
Although the spatial derivative of $\tau(s)$ exhibits finite jumps at transitions between different branches, the constructed vorticity field remains finite and Hölder continuous. According to the singular kernel theory of the Biot–Savart integral~\citep{Majda2001}, the velocity field does not develop infinite gradients, and therefore WLT does not introduce nonphysical singularities.

\begin{figure}  
	\centering
	\begin{overpic}
		[width=1.0\textwidth]{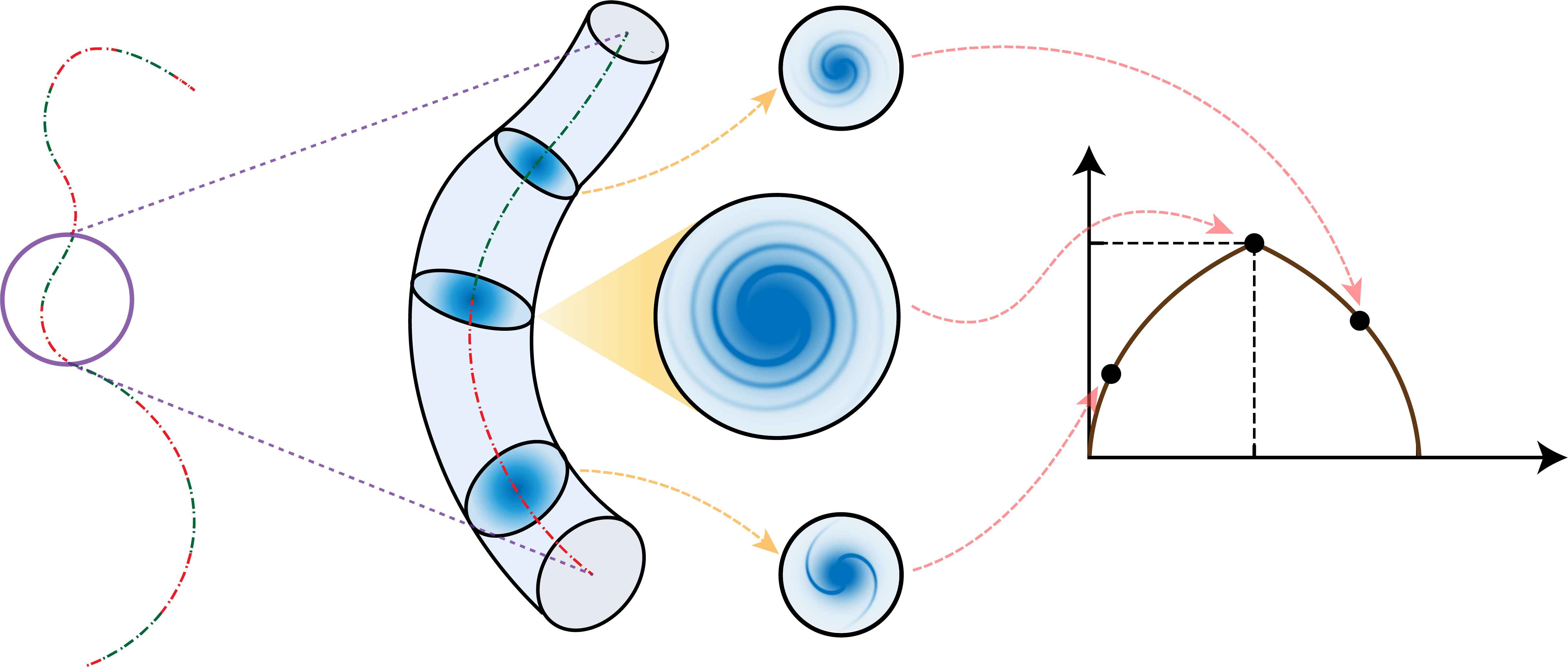}
        \put(3.5,25){\rotatebox{-24}{\textcolor{mygreen}{\tiny{$\ell_{\text{down}}$}}}}
		\put(3,21){\rotatebox{24}{\textcolor{red}{\tiny{$\ell_{\text{up}}$}}}}
        \put(76.5,11){\footnotesize{$\ell_{\text{cycle}}/2$}}
        \put(88,11){\footnotesize{$\ell_{\text{cycle}}$}}
        \put(98,11){\footnotesize{$s$}}
        \put(68.5,11.3){\footnotesize{$0$}}
        \put(67,31){\footnotesize{$\tau$}}
        \put(66.5,26.5){\footnotesize{$\tau_{L}$}}
        \put(-1,40){\footnotesize{(a)}}
        \put(25,40){\footnotesize{(b)}}
        \put(64.5,33.5){\footnotesize{(c)}}
        \put(12,38){\footnotesize{$\boldsymbol{\mathcal{C}}$}}
    \end{overpic}
    \caption{Schematic for constructing a periodic spiral vortex tube along the centerline in WLT. (a) The centerline $\boldsymbol{\mathcal{C}}$ is segmented with the period $\ell_{\text{cycle}}$, where red parts represent $\ell_{\text{up}}$ and green parts represent $\ell_{\text{down}}$. (b) Illustration of the spiral vortex tube constructed over $\ell_{\text{cycle}}$, together with the axial vorticity profiles at various positions, where darker/lighter blue indicates the stronger/weaker vorticity magnitude. (c) The space-time mapping in~\eqref{eq:t_up} and~\eqref{eq:t_down}, where the three points correspond to the ages of the spiral vortices at the three profiles shown in panel (b).}
    \label{fig:L_tube}
\end{figure}

Figure~\ref{fig:L_tube} illustrates how to construct the periodic spiral vortex tube along the centerline using the space-time mapping $\tau(s)$ in \eqref{eq:t_up} and \eqref{eq:t_down}. 
At a given $s$ on the centerline, the local axial vorticity $\omega$ is determined by the Lundgren spiral profile \eqref{eq:w} at $\tau(s)$. 
As one traverses the centerline, the observed vortex transitions continuously. On the increasing branch, starting from a weakly wound spiral ($\tau = 0$), the vortex progresses through increasingly developed states, reaching a fully wound, tightly coiled spiral at $\tau = \tau_{L}$; on the decreasing branch, it gradually unwinds and decays as $\tau$ decreases back to $0$. This vortex-tube segment is then extended periodically along the entire centerline by repeating this pattern every $\ell_{\text{cycle}}$. 
This construction ensures that the spatial distribution of $\tau$ along the centerline is fully consistent with the stretching history: vortex elements at different positions have physical lengths that exactly correspond to the arc-length increments over which they are distributed. 
The periodic structure naturally sustains a steady-state population without requiring explicit creation and annihilation events, 
and it captures the core concept in constructing the spiral vortex ensemble~\citep{Lundgren1982,Gilbert1993}: coding dynamical information directly into spatial geometry.

The cutoff age $\tau_{L}$ determines the maximum degree of spiral winding. A larger $\tau_{L}$ increases the total arc length required for a complete cycle, as $\ell_{\text{cycle}}$ grows exponentially with $\tau_{L}$, and allows the spiral arms to become more tightly wound before decaying. The number of cycles $\mathscr{N}$ along a given centerline provides an additional degree of freedom. For a fixed total arc length $\ell$, increasing $\mathscr{N}$ corresponds to a smaller $\ell_{\text{cycle}}$, which in turn imposes a relation among $l_0$, $\gamma_{L}$ and $\tau_{L}$ through $\ell_{\text{cycle}} = \ell/\mathscr{N}$. This provides flexibility for representing different spatial densities of vortex-generation events.

The background strain rate $\gamma_{L}$ also plays a central role in this mapping. A larger $\gamma_{L}$ renders the arc length distribution highly nonlinear. The increasing branch grows slowly for small $\tau$, and then accelerates rapidly as $\tau$ approaches $\tau_{L}$.
Consequently, for a given $\ell$, a larger $\gamma_{L}$ causes a larger fraction of the centerline to be occupied by vortices near the peak age and a smaller fraction by young vortices near $\tau = 0$, reflecting that stretched vortices become longer and therefore occupy more space. Conversely, a smaller $\gamma_{L}$ produces a more uniform distribution of $\tau$ along the centerline. This directly affects the statistical weight of different vortex morphologies in the ensemble and, consequently, the shape of $P(Q,R)$ discussed in \s\ref{sec:Velocity Gradient Tensor}.

\begin{figure}  
	\centering
	\begin{overpic}[width=1.0\textwidth]{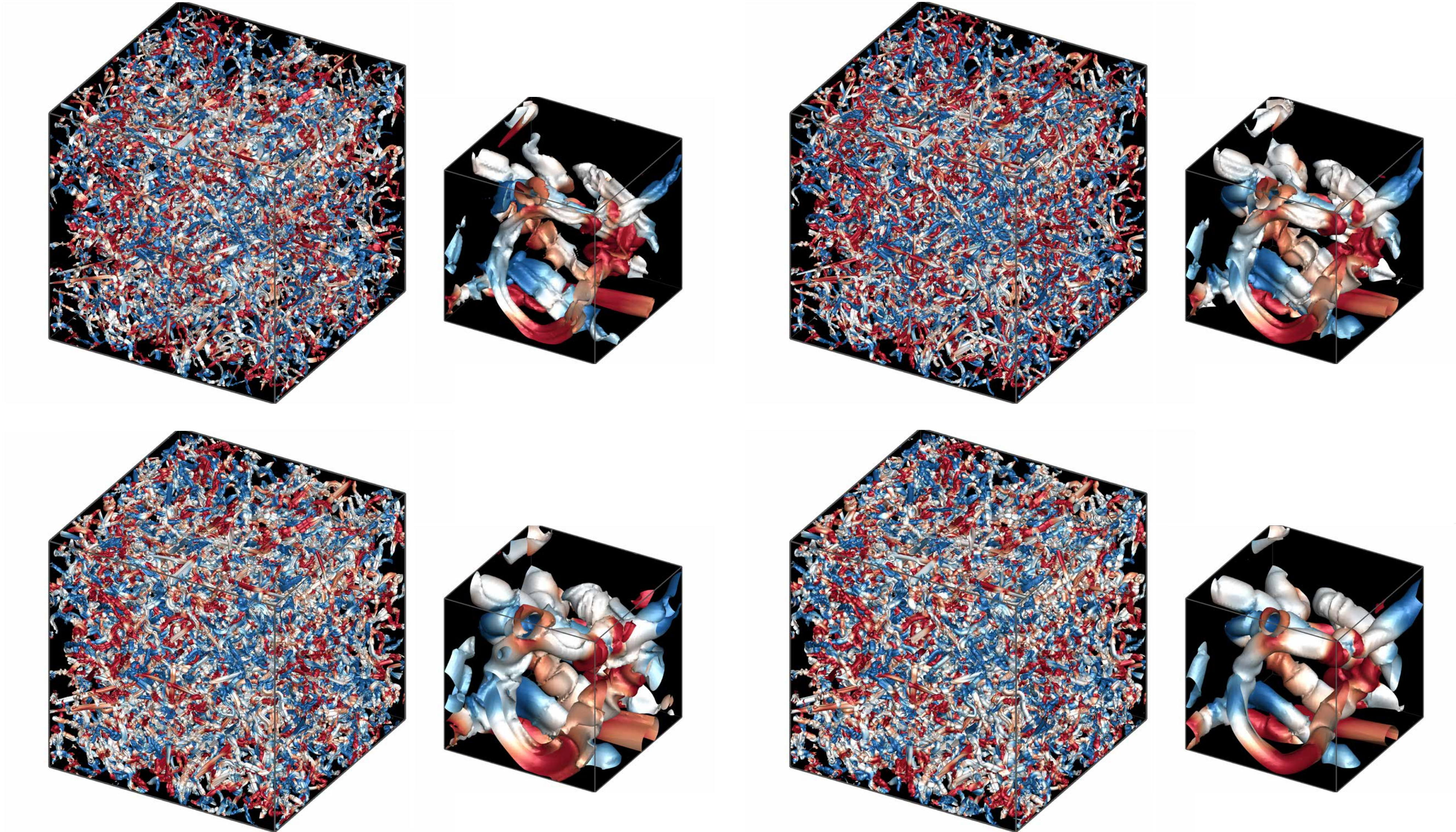}
        {\color{white}
        \linethickness{0.7pt}
        \put(0,57.1){\line(1,0){100}}
        \put(0,27.7){\line(1,0){100}}
        }
		{\color{white}
        \linethickness{0.7pt}
        \put(27.5,50.6){\line(1,0){25}}
        \put(27.5,21.2){\line(1,0){25}}
        }
        {\color{white}
        \linethickness{0.7pt}
        \put(79,50.6){\line(1,0){25}}
        \put(79,21.2){\line(1,0){25}}
        }
        \put(-2,56){\footnotesize{(a)}}
        \put(48.8,56){\footnotesize{(b)}}
        \put(-2,27){\footnotesize{(c)}}
        \put(48.8,27){\footnotesize{(d)}}
    \end{overpic}
    \caption{Isosurfaces of the vorticity magnitude colored by the helicity density $H = \boldsymbol u \cdot \boldsymbol \omega$ for the WLT (left in each panel) and a zoomed-in view (right in each panel) under different ensemble parameters: 
    (a) $\tau_{L}=50$, $\gamma_{L}=0.02$ and $\nu_{L}= 1 \times 10^{-4}$;
    (b) $\tau_{L}=90$, $\gamma_{L}=0.06$ and $\nu_{L}= 1 \times 10^{-4}$;
    (c) $\tau_{L}=130$, $\gamma_{L}=0.1$ and $\nu_{L}= 1 \times 10^{-4}$;
    (d) $\tau_{L}=130$, $\gamma_{L}=0.12$ and $\nu_{L}= 2 \times 10^{-3}$.
    }
    \label{fig:fields}
\end{figure}

As a synthesis of HIT, we generate the WLT in a periodic box of side $L=2\pi$ with $1024^{3}$ grid points.
Technical details of the numerical construction of WLT, including the divergence-free correction of the vorticity field and the transformation from the vorticity field to the velocity field, are provided in the supplementary materials.
We vary only the three core parameters, $\tau_{L}$, $\gamma_{L}$, and $\nu_{L}$, while keeping all other parameters, including the centerline configuration, unchanged. 
Figure~\ref{fig:fields} visualizes vortex structures in WLT under different ensemble parameters.
The large-scale arrangement of the spiral vortex tubes remains unchanged in all panels in figure~\ref{fig:fields}, whereas the primary differences are observed in the small-scale vortices. 
In figure~\ref{fig:fields}(a), the vortex structures are relatively flat, with pronounced curved vortex sheets emerging around the vortex core. 
In figures~\ref{fig:fields}(b) and (c), the vortex structures occupy a larger spatial volume, and the spiral vortices exhibit more compact, finer spiral arms along the vortex axis. 
In figure~\ref{fig:fields}(d), some small-scale spiral structures are smeared out, and the vortex tubes become smooth.
Therefore, the variations in $\tau_L$, $\gamma_L$ and $\nu_L$ alter the local geometry and topology of the vorticity field, including vortex tube coherence, bending, entanglement and fragmentation. 
By tuning these parameters, the WLT can systematically modify the statistical geometry of the velocity gradient tensor, thereby providing a physically interpretable platform for understanding $P(Q,R)$ in turbulence.

\section{Invariants of the velocity gradient tensor in WLT}\label{sec:Velocity Gradient Tensor}
\begin{figure}
	\centering
	\includegraphics[width=0.9\linewidth]{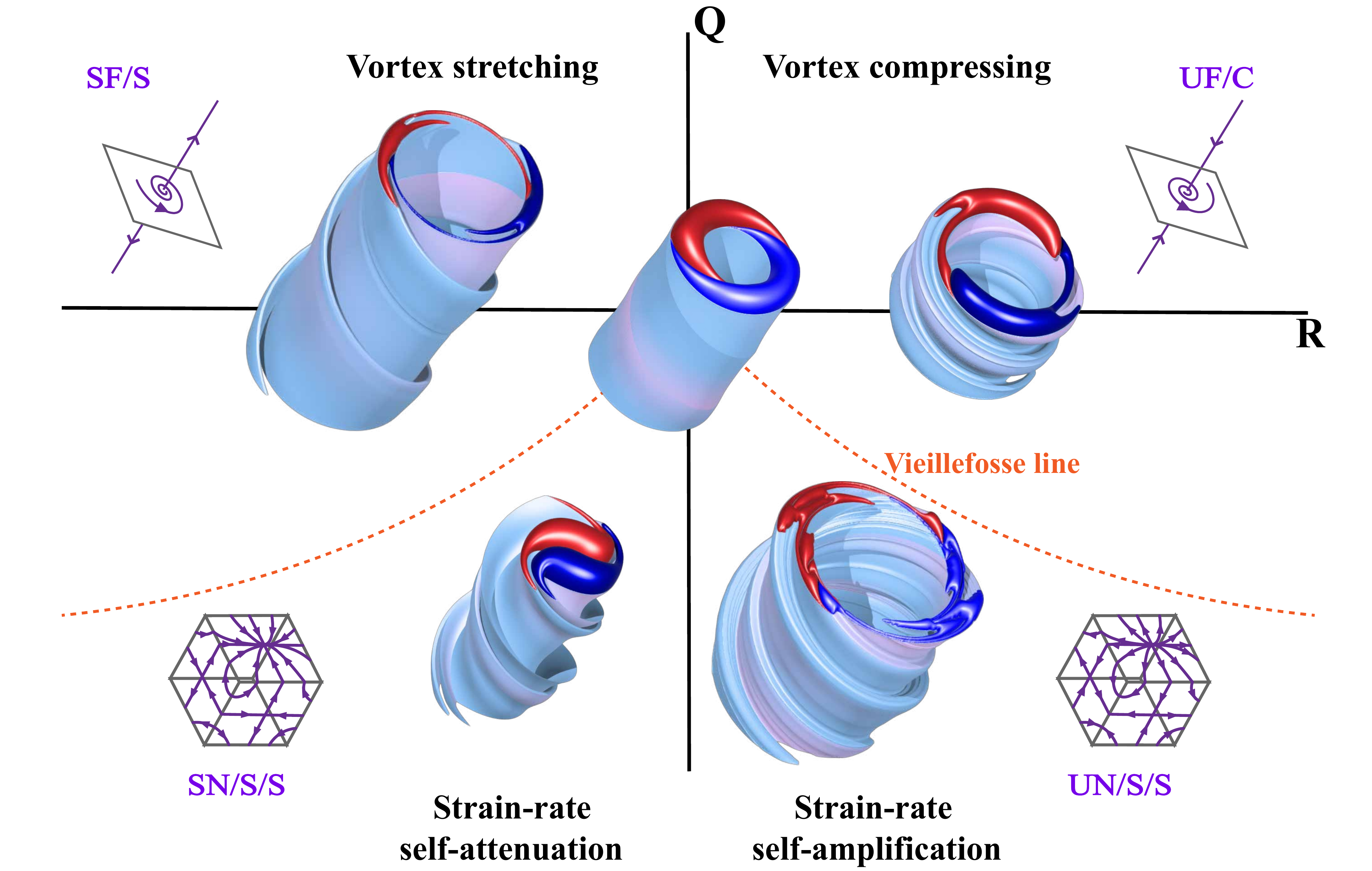}
    \caption{Schematic of the $Q$-$R$ plane in terms of the dominant gradient amplification or attenuation mechanism in each quadrant, together with sketches of the local flow topology and spiral vortex arms. 
    The translucent surfaces represent isosurfaces of vorticity magnitude, while the red and blue colors visually distinguish the two branches of the spiral-arm structure.} \label{fig:QR_mech}
\end{figure}

Considering an incompressible 3D velocity field 
$ \boldsymbol{u} (\boldsymbol{x},t) $, the velocity gradient tensor $\boldsymbol{A}$ or ${A}_{ij} = \partial u_i / \partial x_j=S_{ij}+\Omega_{ij}$ captures both the rate of deformation and rotation. Here, ${S}_{ij} = (A_{ij} + A_{ji})/2$ denotes the strain-rate tensor, while ${\Omega}_{ij} = (A_{ij} - A_{ji})/2$ denotes the rotation-rate tensor, related to the vorticity $\boldsymbol{\omega}$ by $\omega_i = -\varepsilon_{ijk} \Omega_{jk}$. 
Since the first invariant $P = -\text{tr} (\boldsymbol{A}) = 0$ in incompressible flow, we use the second and third invariants
\begin{equation}\label{eq:Q}
    Q = -\frac{1}{2} A_{ij} A_{ji} = \frac{1}{4} \omega_i \omega_i - \frac{1}{2} S_{ij} S_{ij}
\end{equation}
and
\begin{equation}\label{eq:R}
    R = -\frac{1}{3} A_{ij} A_{jk} A_{ki} = -\frac{1}{3} S_{ij} S_{jk} S_{ki} - \frac{1}{4} \omega_i S_{ij} \omega_j
\end{equation}
to analyze the local flow topology. 

Comprehensive background material on the topological methodology in the $Q$-$R$ plane can be found in \citet{Chong1990}, \citet{Perry1994}, \citet{Soria1994}, \citet{Blackburn1996}, \citet{Ooi1999} among others. An important tent-like curve in this plane is the Vieillefosse line $V_A = 27 R^2/4 + Q^3 = 0$. 
It plays a central role in the dynamics of the restricted Euler equation \citep{Vieillefosse1984,Cantwell1992} and marks the boundary between distinct topological regions. 
For $V_A > 0$, $A_{ij}$ has one real eigenvalue and two complex-conjugate ones, corresponding to focal (vorticity-dominated) topologies; for $V_A < 0$, all three eigenvalues are real, corresponding to nodal (strain-dominated) topologies.

As illustrated in figure~\ref{fig:QR_mech}, in the plane partitioned by the $Q$-axis and the Vieillefosse line, four non-degenerate regions of local flow topology exist in incompressible flows: 
stable-focus/stretching (SF/S), unstable-focus/contracting (UF/C), stable-node/saddle/saddle (SN/S/S) and unstable-node/saddle/saddle (UN/S/S) regions. 
For WLT, vortex structures are affected by the combined effects of asymmetric windup and axial stretching~\citep{Gilbert1993}, and we use a pair of columnar spiral arms as the reference (corresponding to the sketch in the central region of figure~\ref{fig:QR_mech} with $Q = 0 $ and $R = 0 $).
The distinct geometry of the spiral vortex arms reflects the dominant mechanism in each region.
For $V_A>0$ and $R<0$, rotation dominates, while vortex stretching elongates the structure in the axial direction, resulting in vorticity-magnitude isosurfaces with developed spiral branches.
For $V_A>0$ and $R>0$, vortex compression shortens and folds the tube, producing a compact structure with tightly wound and strongly curved spiral arms.
For $V_A<0$ and $R<0$, strain dominates but undergoes self-attenuation; consequently, the vortical structure remains small and compact, with shortened spiral branches and reduced axial coherence.
By contrast, the region for $V_A<0$ and $R>0$ is associated with strain-rate self-amplification. Persistent deformation stretches, folds and reorganizes the vorticity field, giving rise to broader, multilayered, and more extended spiral structures.
These structures with different local flow topologies determine the teardrop shape of $P(Q,R)$ for turbulence. 

\subsection{Ensemble parameters for vortex stretching, winding and dissipation}\label{sec:parameters}
\begin{figure}  
	\makebox[\textwidth][l]{%
        \hspace*{0.03\textwidth}%
        \begin{overpic}[width=0.85\textwidth]{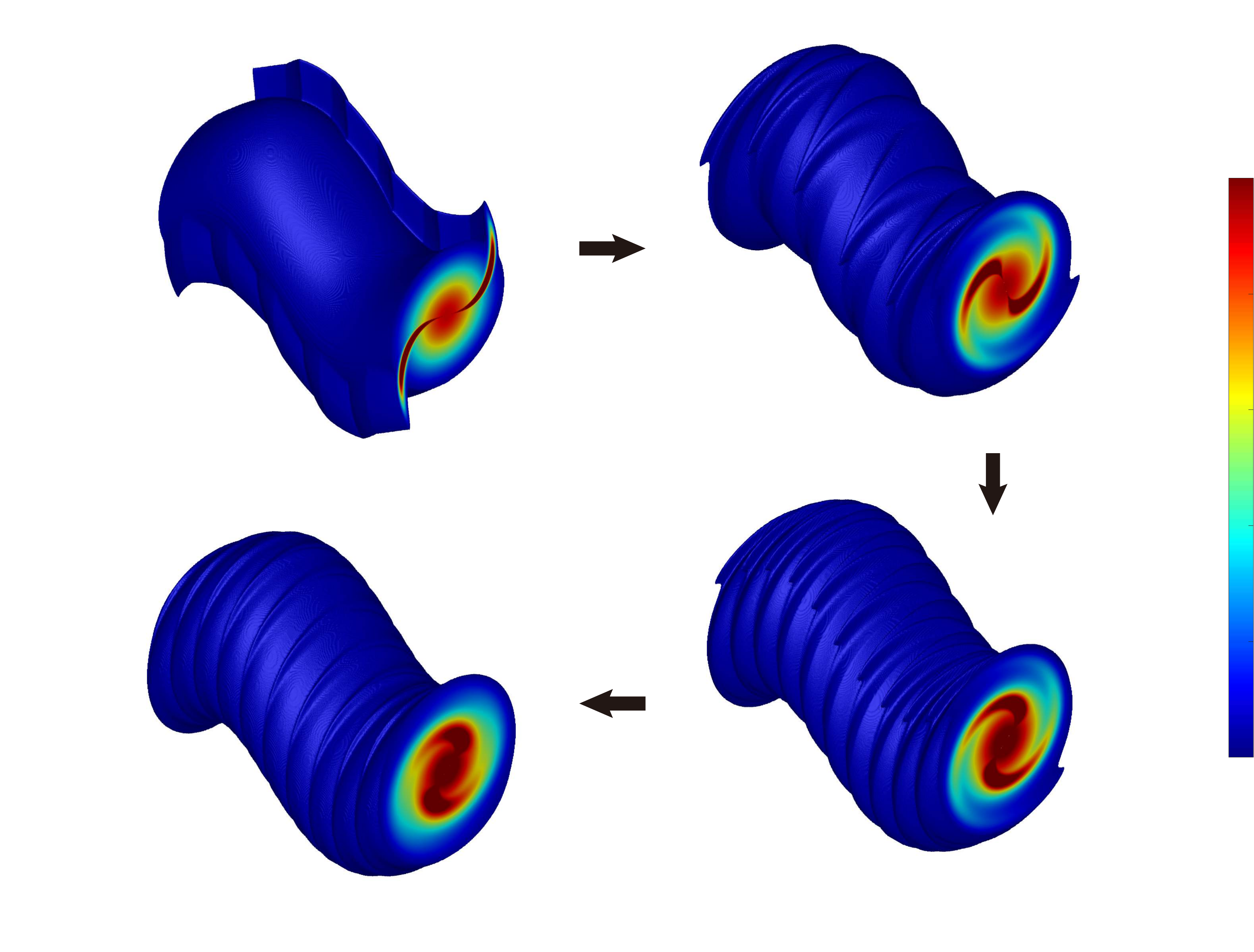}
        \put(45.5,58.5){\large{$ \tau_{L}\uparrow $}}
		\put(81,37){\large{$ \gamma_{L}\uparrow $}}
        \put(45.5,16){\large{$ \nu_{L}\uparrow $}}
        \put(97,12){\footnotesize{$ |\boldsymbol{\omega}| $}}
        \put(8,70){\footnotesize{(a)}}
        \put(52,70){\footnotesize{(b)}}
        \put(52,34){\footnotesize{(c)}}
        \put(8,34){\footnotesize{(d)}}
        \put(100.5,60.9){\footnotesize{$ 0.6 $}}
        \put(100.5,51.6){\footnotesize{$ 0.5 $}}
        \put(100.5,42.3){\footnotesize{$ 0.4 $}}
        \put(100.5,33.0){\footnotesize{$ 0.3 $}}
        \put(100.5,23.7){\footnotesize{$ 0.2 $}}
        \put(100.5,14.4){\footnotesize{$ 0.1 $}}
    \end{overpic}}
    \caption{The geometry of the Lundgren spiral vortex element varies with the ensemble parameters. Starting from the upper-left panel, the panels depict successive increases in $\tau_{L}$, $\gamma_{L}$ and $\nu_{L}$:
    (a) $\tau_{L}=30$, $\gamma_{L}=0.01$ and $\nu_{L}=1 \times 10^{-4}$;
    (b) $\tau_{L}=90$, $\gamma_{L}=0.01$ and $\nu_{L}=1 \times 10^{-4}$;
    (c) $\tau_{L}=90$, $\gamma_{L}=0.08$ and $\nu_{L}=1 \times 10^{-4}$;
    (d) $\tau_{L}=90$, $\gamma_{L}=0.08$ and $\nu_{L}=8 \times 10^{-4}$.
    The vortex element with a curved axis is represented by isosurfaces of the axial vorticity magnitude, and the cross-sectional distribution is colored by $ |\boldsymbol{\omega}| $.} \label{fig:struc_parameters}
\end{figure}

Within the WLT framework for synthesizing HIT, $P(Q,R)$ results from the superposition of numerous spiral vortices, each with distinct ages and orientations. The ensemble parameters, background strain rate $\gamma_{L}$, cutoff age $\tau_{L}$ and viscous decay factor $\nu_{L}$, play a crucial role in determining the balance between vortex stretching, winding and dissipation, respectively.
Figure~\ref{fig:struc_parameters} illustrates how each parameter affects the state of individual vortex elements. 

The cutoff age $\tau_{L}$ determines the extent to which the spiral winding process is permitted to advance before viscous diffusion ultimately eliminates fine-scale structures. 
In the spiral vortex solution~\eqref{eq:w}, the non-axisymmetric harmonics decay on a time scale $\tau_{\text{spiral}} \sim (\mu'(r)^2 \nu_{L})^{-1/3}$, which is substantially shorter than the diffusive decay time $\tau_{\text{core}} \sim r_0^2 / \nu_{L}$ of the axisymmetric core. For a single vortex of age $\tau$, the number of completed spiral turns at a given radius is proportional to $\mu(r)\tau$. Thus, a larger $\tau_{L}$ enables vortices to attain more tightly wound spiral configurations before being removed from the ensemble.
In figure~\ref{fig:struc_parameters}, the two branches of the spiral arms are widely separated and remain close to the symmetric vortex-sheet morphology at low $\tau_{L}$. As $\tau_{L}$ increases, the vortex element develops pronounced, asymmetric and compact spiral arms.
Meanwhile, increasing $\tau_{L}$ increases the number of older vortices in the population, which exhibit more tightly wound spirals and consequently contribute more strongly to non-Gaussian statistics and higher-order moments of the velocity gradient.

The background strain rate $\gamma_{L}$ enters the Lundgren transformation in \eqref{eq:2D} through the stretch ratio $S= \mathrm{e}^{\gamma_{L} \tau}$, the stretched radial coordinate $\xi= \mathrm{e}^{\gamma_{L} \tau / 2} r $ and the straining time $T = (\mathrm{e}^{\gamma_{L} \tau} - 1)/\gamma_{L}$. 
These quantities determine the extent of axial stretching experienced by the vortex cross-section and the rate at which the spiral arms tighten. 
In figure~\ref{fig:struc_parameters}, when $\gamma_{L}$ increases, the vortex element in the ensemble undergoes more rapid axial stretching, which has several consequences for the vortex morphology. 
First, the vorticity concentrates in more intense regions characterized by strong rotation. 
Second, the characteristic time scale for spiral winding, set by $\mu(r)$, is effectively accelerated because the same $\tau$ corresponds to a larger $T(\tau)$. 
Third, the alignment between the vortex axis and the local strain field becomes more pronounced, promoting the preferential orientation of vorticity with the intermediate eigenvector of $S_{ij}$, which will be discussed in \s\ref{sec:topology}. 
In accordance with~\eqref{eq:t_up} and~\eqref{eq:t_down}, $\gamma_{L}$ also governs the relative weighting of the younger and older vortices in WLT. 
When averaged over the ensemble, larger $\gamma_{L}$ amplifies the contribution of older vortices, which have had more lifetime to develop intense cores and tight spiral arms.

The viscous decay factor $\nu_{L}$ appears in \eqref{eq:w} through the exponential damping term $\exp(-4n^2\mu'(r)^2\nu_{L} \tau^3/3)$. This term represents the cumulative effect of viscous diffusion on the spiral arms and becomes increasingly significant at late times and small scales. 
For a fixed $\tau$, decreasing $\nu_{L}$ enables higher harmonics to persist for longer times and up to larger radii. 
Consequently, the spiral arms extend further from the vortex core, generating finer-scale vorticity variations and stronger velocity gradients.
%
As shown in figure~\ref{fig:struc_parameters}, at high $\nu_{L}$, the spiral arms diffuse rapidly, leaving a smoother core region; at low $\nu_{L}$, the spiral arms remain coherent down to much smaller scales, generating very large local velocity gradients.  
Thus, $\nu_{L}$ governs the degree of small-scale intermittency in our model.

A vortex with given circulation $\Gamma$ and core radius $r_0$ evolves through three overlapping phases: an initial stretching phase in which the core intensifies and the spiral begins to form; a winding phase in which differential rotation wraps the spiral arms into progressively tighter turns; a dissipative phase in which viscosity gradually destroys the fine-scale structure. 
Taken together, $\tau_{L}$, $\gamma_{L}$ and $\nu_{L}$ provide a complete characterization of the vortex life cycle within the spiral vortex ensemble. They modulate the state of individual vortex elements within the WLT, thereby altering their morphology, lifespan and contribution to $P(Q,R)$. 

Although these parameters are not entirely independent in real turbulence, we treat them as independent ones that separately govern distinct physical mechanisms within the WLT framework. This decoupling enables a systematic investigation of how each mechanism contributes to the statistical topology of the velocity gradient tensor.

\begin{figure}
	\centering
	\begin{overpic}
        [width=1.0\textwidth]{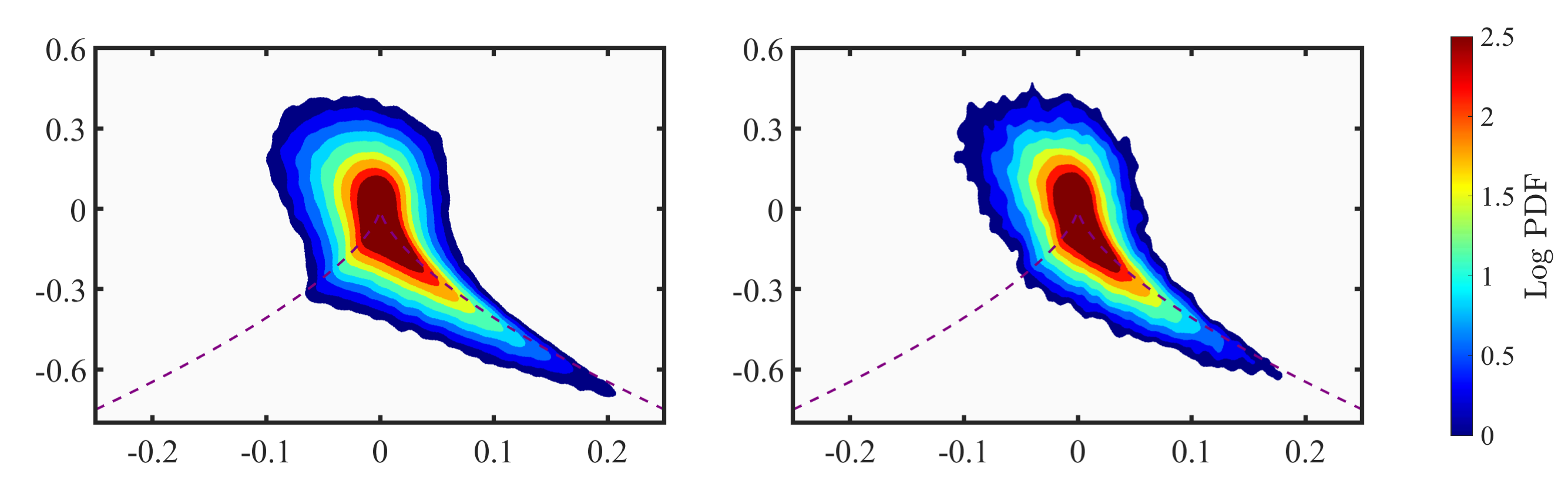}
        \put(0,29.3){\footnotesize{(a)}}
		\put(44.5,29.3){\footnotesize{(b)}}
        \put(1.3,16){\footnotesize{$ Q^{*} $}}
		\put(45.8,16){\footnotesize{$ Q^{*} $}}
        \put(23.3,-1){\footnotesize{$ R^{*} $}}
		\put(67.8,-1){\footnotesize{$ R^{*} $}}
    \end{overpic}
    \caption{Logarithmically scaled $P(Q^*,R^*)$ in (a) DNS and (b) WLT with $\tau_{L}=180$, $\gamma_{L}=0.1$ and $\nu_{L}=5 \times 10^{-5}$. The dashed line denotes the Vieillefosse line.}\label{fig:QR_DNS}
\end{figure}

\subsection{$P(Q,R)$ in WLT}\label{sec:topology}
To facilitate comparisons of $P(Q,R)$ across scales and Reynolds numbers, we adopt the normalized forms $Q^* = {Q}/{\langle S_{ij} S_{ij} \rangle}$ and $R^* = {R}/{\langle S_{ij} S_{ij} \rangle^{3/2}}$, based on the averaged strain-rate tensor ${\langle S_{ij} S_{ij} \rangle}$, where $\langle\cdot\rangle$ represents the volume average over the computational domain.  
Figure~\ref{fig:QR_DNS} compares $P(Q^*,R^*)$ from the DNS (detailed in Appendix \ref{sec:DNS}) and the WLT at $Re_\lambda=182.6$. 
The parameter values are selected to match the vortex dynamics in the DNS of HIT. 
Specifically, $\tau_L = 180$ defines a sufficiently large cutoff age to permit adequate development of spiral arms, $\gamma_{L} = 0.1$ is chosen to yield a realistic stretching rate consistent with enstrophy production, and $\nu_L = 5 \times 10^{-5}$ ensures an appropriate level of small-scale intermittency.
The WLT reproduces the major features of $P(Q^*,R^*)$ observed in the DNS results, including the teardrop-shape contours. 
In addition, Appendix \ref{sec:DNS} shows good agreement between the DNS and WLT results for the energy spectrum, the third-order velocity structure function and the PDFs of the enstrophy and dissipation rate.

\begin{figure}
	\centering
	\includegraphics[width=0.7\linewidth]{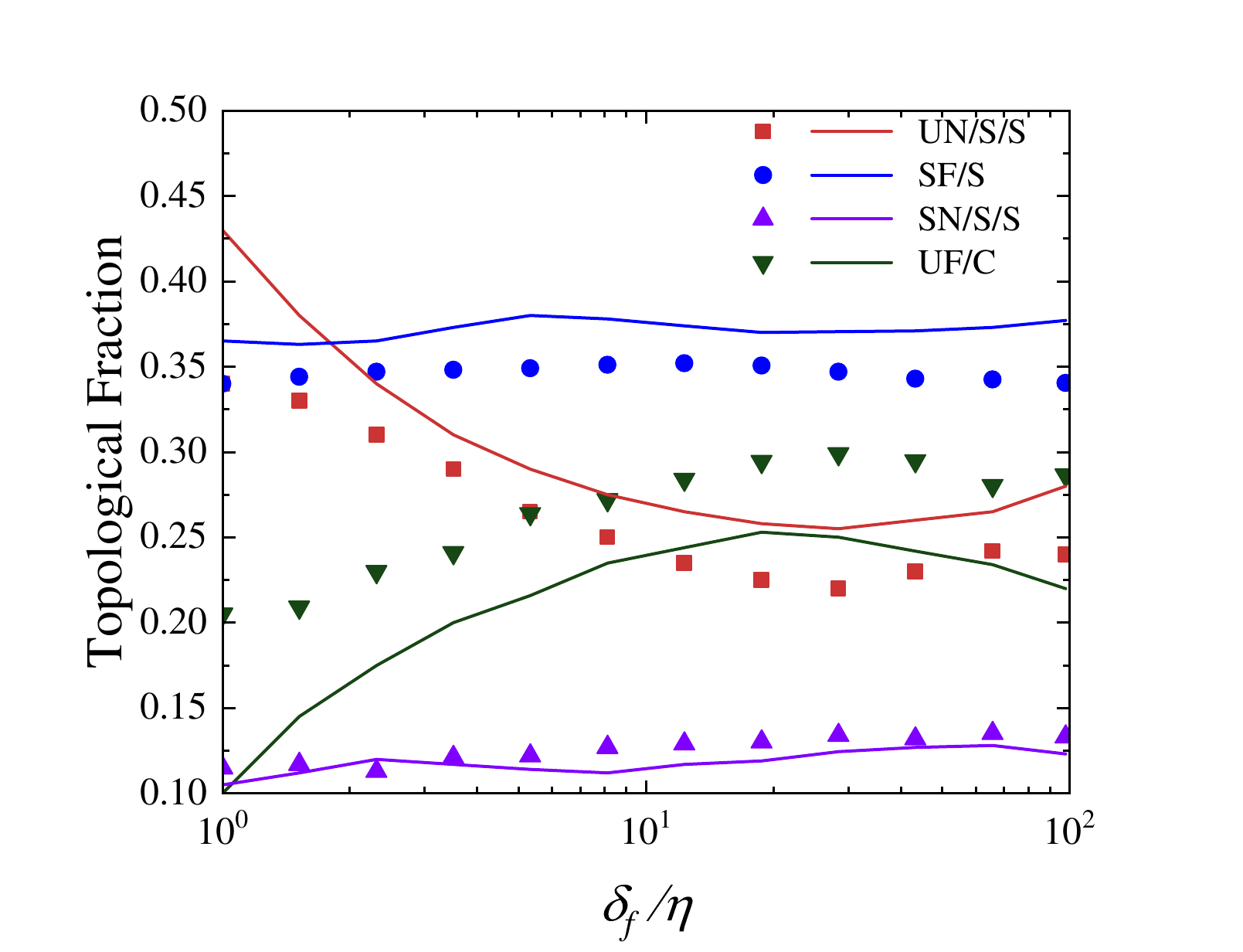}
    \caption{The fraction of different local flow topologies at varying scales $\delta_f/\eta$ (symbols: DNS; solid lines: WLT with $\tau_{L}=180$, $\gamma_{L}=0.1$ and $\nu_{L}=5 \times 10^{-5}$).}  \label{fig:Fraction}
\end{figure}

In figure~\ref{fig:Fraction}, we compare DNS and WLT through the population fractions of different topological regions \citep{Danish2018} in terms of $\delta_f/\eta$, with the filter width $\delta_f$ and the Kolmogorov scale $\eta$. Here, we apply the filter of \cite{Eyink2009}, which is smooth and compact, to the velocity fields obtained from DNS and WLT. By varying $\delta_f$ from the inertial range ($\delta_f \gg \eta$) down to the viscous range ($\delta_f \sim \eta$), we compute for each scale the population fractions of the four topological regions defined in figure~\ref{fig:QR_mech} and observe their transitions.  
Although specific percentage discrepancies are present, WLT reproduces the trends observed in DNS: the UN/S/S topology exhibits a marked increase in population from the inertial to the viscous range, the UF/C topology decreases, while the SF/S and SN/S/S topologies remain nearly constant.
This demonstrates that the multiscale spiral vortices hierarchy in WLT inherently captures the mechanism competition between vortex compressing and strain-rate self-amplification across scales.

\begin{figure}
	\centering
	\begin{overpic}
		[width=1.0\textwidth]{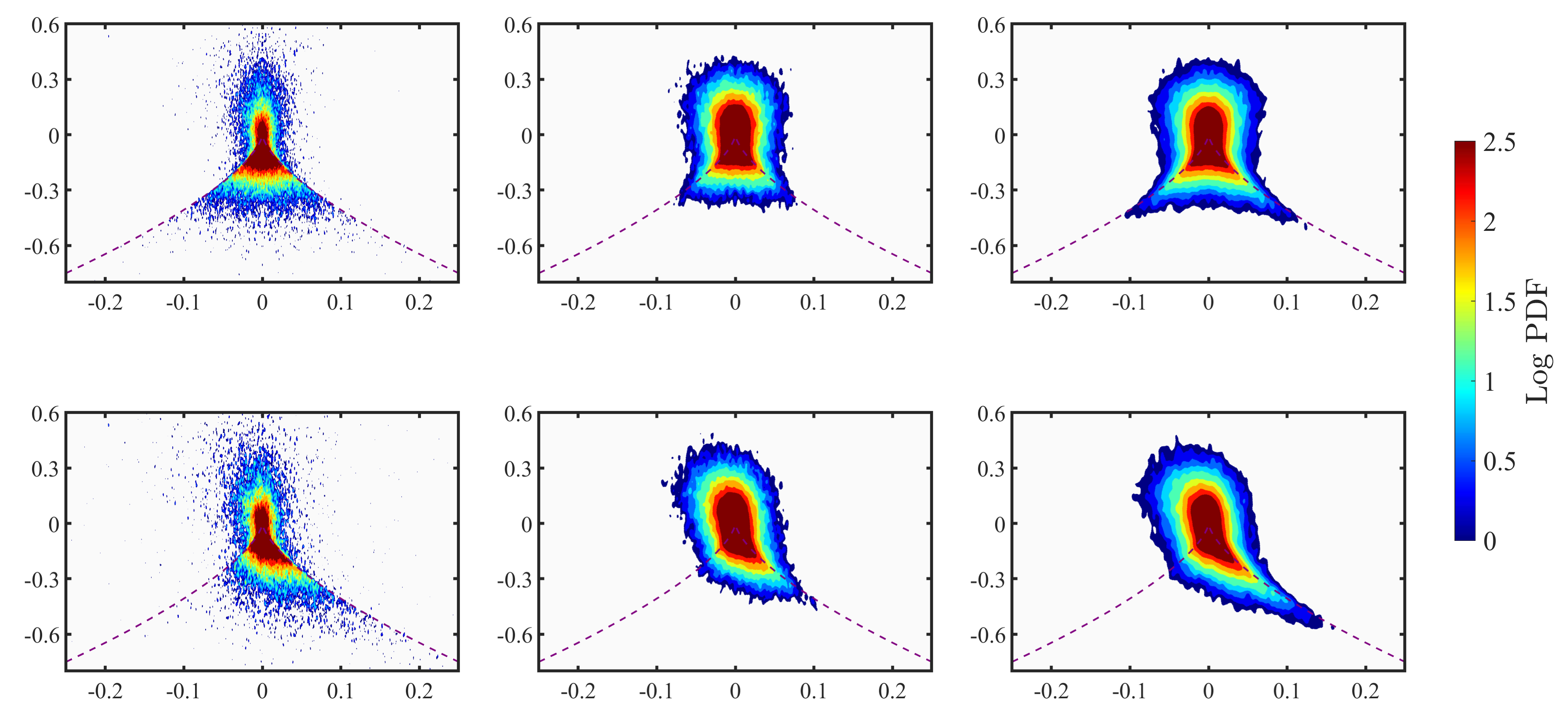}
		\put(-0.3,43.9){\scriptsize{(a)}}
		\put(29.8,43.9){\scriptsize{(b)}}
        \put(60.2,43.9){\scriptsize{(c)}}
    	\put(-0.3,19.1){\scriptsize{(d)}}
        \put(29.8,19.1){\scriptsize{(e)}}
        \put(60.2,19.1){\scriptsize{(f)}}
        \put(0.5,35.3){\scriptsize{$ Q^{*} $}}
		\put(30.7,35.3){\scriptsize{$ Q^{*} $}}
        \put(60.9,35.3){\scriptsize{$ Q^{*} $}}
    	\put(0.5,10.5){\scriptsize{$ Q^{*} $}}
        \put(30.7,10.5){\scriptsize{$ Q^{*} $}}
        \put(60.9,10.5){\scriptsize{$ Q^{*} $}}
        \put(15.9,23.6){\scriptsize{$ R^{*} $}}
        \put(46.1,23.6){\scriptsize{$ R^{*} $}}
        \put(76.3,23.6){\scriptsize{$ R^{*} $}}
        \put(15.9,-1.2){\scriptsize{$ R^{*} $}}
        \put(46.1,-1.2){\scriptsize{$ R^{*} $}}
        \put(76.3,-1.2){\scriptsize{$ R^{*} $}}
	\end{overpic}
    \caption{Logarithmically scaled $P(Q^*,R^*)$ under different ensemble parameters in WLT: (a) $\tau_{L}=10$, $\gamma_{L}=0.01$ and $\nu_{L}=1 \times 10^{-3}$; (b) $\tau_{L}=10$, $\gamma_{L}=0.08$ and $\nu_{L}=1 \times 10^{-3}$; (c) $\tau_{L}=10$, $\gamma_{L}=0.08$ and $\nu_{L}= 1 \times 10^{-4}$; (d) $\tau_{L}=160$, $\gamma_{L}=0.01$ and $\nu_{L}=1 \times 10^{-3}$; (e) $\tau_{L}=160$, $\gamma_{L}=0.08$ and $\nu_{L}=1 \times 10^{-3}$; (f) $\tau_{L}=160$, $\gamma_{L}=0.08$ and $\nu_{L}= 1 \times 10^{-4}$.
    The dashed line denotes the Vieillefosse line.} \label{fig:QR_parameters}
\end{figure}

As an adjustable turbulent field with entangled vortex tubes, WLT offers the advantage of relating the vortex elements to the $Q$‑$R$ statistics. 
The independent modulation of $\tau_{L}$, $\gamma_{L}$ and $\nu_{L}$ shows how vortex stretching, winding and dissipation influence the local flow topology of turbulence. 
Figure~\ref{fig:QR_parameters} demonstrates, through a series of cases, how the tuning of the three parameters affects the morphology of $P(Q^*,R^*)$: larger $\tau_{L}$ accentuates the asymmetry characterized by an enhanced inclination toward the right quadrant; increasing $\gamma_{L}$ broadens the head region; decreasing $\nu_{L}$ extends the tail along the Vieillefosse line while preserving the characteristic tear-drop shape. 

The parametric influence on the shape of $P(Q^*,R^*)$ in figure~\ref{fig:QR_parameters} can be interpreted in terms of variations in the topological flow structures within the ensemble. 
For the WLT, increasing $\tau_{L}$ leads to the emergence of vortices with more fully developed spiral arms. These arms consist of thin, alternating layers of positive and negative vorticity, which generate intense, localized velocity gradients. At a point passing through a spiral arm, the velocity gradient undergoes rapid oscillations crossing alternating vorticity layers, thereby contributing to extreme values of both $Q$ and $R$. \citet{Lundgren1982} shows that $ S\sim k^{2/3} $ in the spiral arms, where $k$ denotes the wavenumber, so the strain-rate self-amplification dominates. In the $Q$-$R$ plane, as illustrated by the transitions from figures~\ref{fig:QR_parameters}(a) to (d), (b) to (e) and (c) to (f), this manifests as an increased probability in the strain-dominated region, characterized by a strong tendency toward the UN/S/S region.

Increasing $\gamma_{L}$ shifts the vortex population toward states that develop more intense cores and tighter arms.
Physically, fluid elements located near the cores of stretched vortices undergo strong rotation, thereby populating the SF/S and UF/C topologies. In figures~\ref{fig:QR_parameters}(a) to (b) and (d) to (e), this appears as an expansion of the vorticity-dominated head region at $V_A > 0$.

Decreasing $\nu_{L}$ sharpens the spirals and extends the tail region in the $Q$-$R$ plane toward increasingly extreme values of $Q$ and $R$ in figures~\ref{fig:QR_parameters}(b) to (c) and (e) to (f). Physically, this enhances intermittency at high Reynolds numbers, where rare but intense events, such as those associated with tightly wound spiral structures, occur with greater probability. 
The tail extension is asymmetric, as the sign of $R$ depends on whether the spiraling motion is converging ($R<0$, SN/S/S) or diverging ($R>0$, UN/S/S). In our model, fully developed spiral arms typically produce UN/S/S topology~\citep{Lundgren1993,Gilbert1993}, so the tail extends toward the right branch of the Vieillefosse line.

\subsection{Vorticity and strain rate alignment in WLT}\label{sec:strain analysis}
The symmetric and trace-free strain-rate tensor ${S}_{ij}$ in incompressible turbulent flows has three orthogonal eigenvectors $\boldsymbol{e}_i$, $i=1,2,3$ with corresponding eigenvalues ordered as $\lambda_{1} \ge \lambda_{2} \ge \lambda_{3}$, where $\boldsymbol{e}_1$ represents the most positive strain (stretching) direction, $\boldsymbol{e}_3$ the most negative strain (compressing) direction, and $\boldsymbol{e}_2$ the intermediate strain direction.

\begin{figure}
	\centering
	\includegraphics[width=0.7\linewidth]{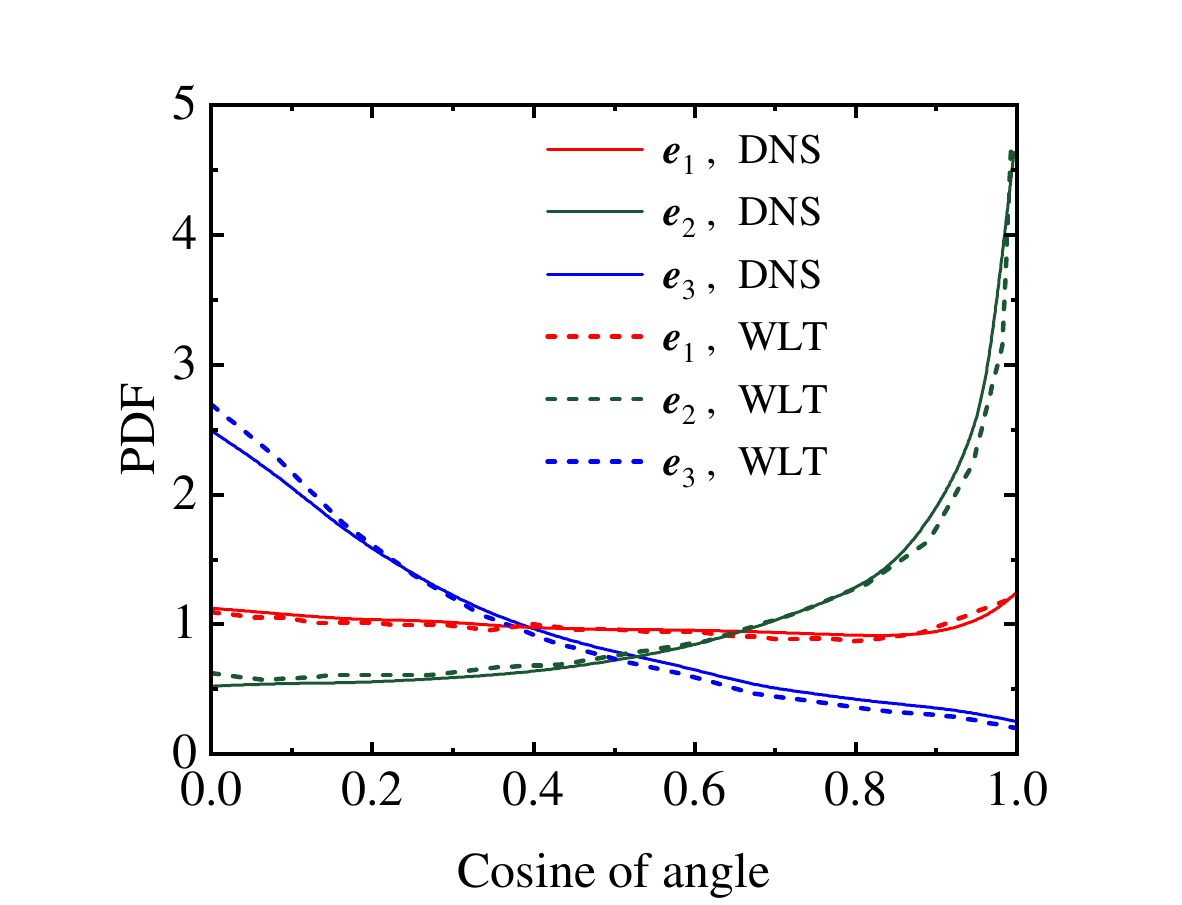}
   \caption{PDFs for the cosine of the angle between the vorticity and the strain-rate eigenvectors $\boldsymbol{e}_i$ from DNS and WLT with $\tau_{L}=180$, $\gamma_{L}=0.1$ and $\nu_{L}=5 \times 10^{-5}$.} \label{fig:COS_1}
\end{figure}

Figure~\ref{fig:COS_1} compares the PDFs of cosine of angle between $\boldsymbol{\omega}$ and $\boldsymbol{e}_i$ in DNS and WLT. The WLT shows a good agreement with DNS: 
$\boldsymbol{\omega}$ preferentially aligns with the intermediate eigenvector $\boldsymbol{e}_2$ of $S_{ij}$, tends to be nearly perpendicular to the most compressive eigenvector $\boldsymbol{e}_3$, and shows relatively uniform alignment with the most stretching eigenvector $\boldsymbol{e}_1$. 
The restricted Euler model explains these trends as an intrinsic consequence of the nonlinear self-amplification of the velocity gradient tensor~\citep{Ashurst1987,Cantwell1992}, while pressure-Hessian and nonlocal effects shape the quantitative form of the PDFs in real turbulence~\citep{Chevillard2008,Hamlington2008}. 
This agreement is dynamically significant because these preferential orientations directly reflect the coupling between the vortex structure and the local strain field in WLT. 
The bending and twisting of the vortex centerline control the reorientation of the surrounding principal strain axes, while the axial vorticity distribution along the spiral vortex tube further constrains the orientation of the vorticity relative to the principal strain directions.
The ability of WLT to reproduce the full alignment distribution, rather than only its mean trend, indicates that it captures the local geometry governing vortex stretching and the intensification of coherent vortex filaments, which are central to the energy cascade and the small-scale turbulence structure~\citep{Johnson2020,Johnson2021,Musci2025}.

\begin{figure}
	\centering
	\includegraphics[width=0.7\linewidth]{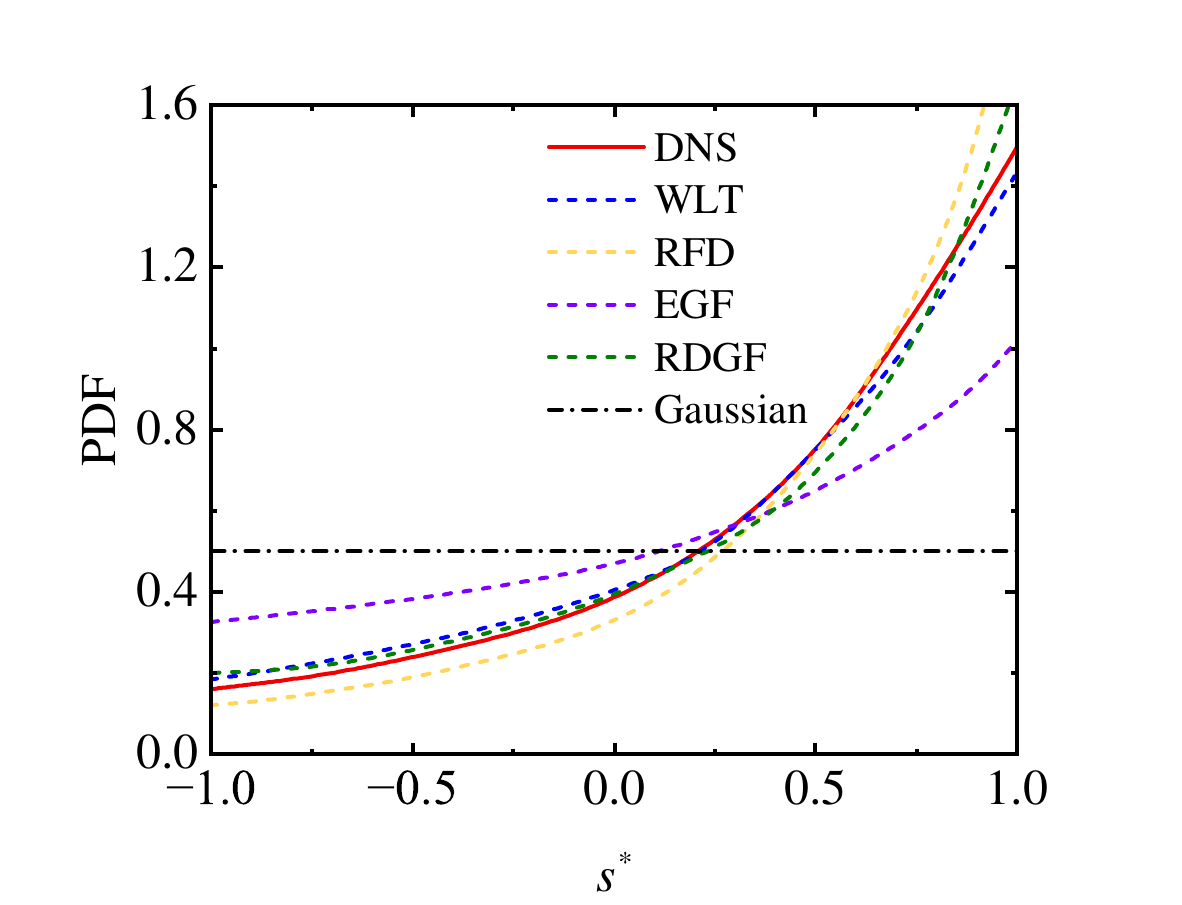}
    \caption{PDFs for $ s^{*} $ in different models compared with DNS and Gaussian field statistics. The Lagrangian models RFD, EGF and RDGF can be found in \citet{Johnson2016}. The parameters in the WLT are $\tau_{L}=180$, $\gamma_{L}=0.1$ and $\nu_{L}=5 \times 10^{-5}$.} \label{fig:s_model}
\end{figure}

Figure~\ref{fig:s_model} shows that the WLT also captures the preferential bias toward the biaxial strain observed in turbulent flows, and the distribution of $s^*$ is in good agreement with DNS. 
Here,
\begin{equation}\label{eq:s*}
    s^* = - \frac{3 \sqrt{6} \lambda_1 \lambda_2 \lambda_3}{(\lambda_1^2 + \lambda_2^2 + \lambda_3^2)^{3/2}} 
\end{equation}
with $-1 \le s^* \le 1$ characterizes the strain topology \citep{Lundgren1994}: $s^* < 0$ denotes axial strain with $\lambda_1 > 0$ and $\lambda_3 < \lambda_2 < 0$; $s^* > 0$ denotes biaxial strain with $\lambda_1 > \lambda_2 > 0$ and $\lambda_3 < 0$; $s^* = 0$ corresponds to plane strain with $\lambda_2 = 0$. Extensive experimental and DNS studies have shown that biaxial strain occurs more frequently than axial strain in turbulent flows~\citep{Lundgren1994,Tsinober2009}. Consequently, the bias toward $s^* > 0$ indicates a greater tendency for disk-like fluid-element deformation~\citep{Meneveau2011,Johnson2024}. 
%
The WLT and the Lagrangian model RDGF yield results of $s^*$ that are most consistent with DNS in figure~\ref{fig:s_model}.

\section{Modelling the joint PDF of $Q$ and $R$ in WLT}\label{sec:model}
\subsection{Decomposition based on local dynamics}
We develop an explicit model of $P(Q,R)$ in terms of Lundgren spiral vortex ensemble parameters in WLT. 
We start from the definition of
\begin{equation}
    P(Q,R)
    =
    \left\langle
    \delta\left(Q+\frac{1}{2}\mathrm{tr}(\boldsymbol{A}^2)\right)
    \delta\left(R+\frac{1}{3}\mathrm{tr}(\boldsymbol{A}^3)\right)
    \right\rangle,
    \label{eq:QR_pdf_delta}
\end{equation}
where \(\langle\cdot\rangle\) denotes the ensemble average, equivalently a
spatial or temporal average under statistical stationarity, and $\delta(\cdot)$ is the Dirac delta function.

Within the WLT framework, $\boldsymbol{A}$ is assumed to be composed of three statistically distinct contributions: the background straining that stretches the vortex tube axially and compresses it radially, the inertial-range spiral arms generated by the differential rotation of the Lundgren vortex and the viscous dissipation where small-scale gradients are strongly smoothed by viscosity. 
In actual HIT fields, these processes are not strictly separated in space, but in WLT, as mentioned in \s\ref{sec:parameters}, they are dominated by independent parameters. 
We therefore assume that each local vortex element is assigned to the process that gives the leading contribution to its velocity gradient statistics. 
Accordingly, we decompose domain into stretching, spiraling and dissipation subregions as
\begin{equation}
    V
    =
    V_{\mathrm{stretch}}
    +
    V_{\mathrm{spiral}}
    +
    V_{\mathrm{diss}}
    =V_1 +V_2 +V_3,
\end{equation}
with corresponding volume fractions
\begin{equation}
    w_i=\frac{|V_i|}{L^3},
    \quad
    i=1,2,3,
    \quad
    w_1+w_2+w_3=1.
    \label{eq:volume_fractions_def}
\end{equation}
The three mechanisms may overlap physically, but it is resolved through the dominance criterion and is not double-counted in the volume decomposition. 
Here, \(V_{\rm stretch}\), \(V_{\rm spiral}\) and \(V_{\rm diss}\) should be understood as statistical subdomains rather than strictly material regions with physical boundaries.

Applying the law of total probability 
$P(Q, R) = \sum P(Q, R |V_i) P(V_i) = \sum w_i P_i(Q, R)$,
we obtain 
\begin{equation}
    P(Q,R)
    =
    w_1 P_{\mathrm{stretch}}(Q,R)
    +
    w_2 P_{\mathrm{spiral}}(Q,R)
    +
    w_3 P_{\mathrm{diss}}(Q,R),
    \label{eq:QR_mixture}
\end{equation}
where $P_{\mathrm{stretch}}(Q,R)$, $P_{\mathrm{spiral}}(Q,R)$ and $P_{\mathrm{diss}}(Q,R)$ denote the stretching, spiraling and dissipation contributions, respectively. 
This decomposition is a conditional mixture over the local dynamical mechanisms represented in the Lundgren ensemble.

\subsection{Stretching contribution}
In this subsection, we consider only the contribution of the stretching strain field to the velocity gradient. The contribution associated with the spiral vortex structure generated by the differential-rotation mechanism is treated separately in \s\ref{sec:spiral-arm}. 

The background stretching in the Lundgren vortex model is locally approximated by an axisymmetric strain tensor
\begin{equation}
    S_{ij}
    =
    \begin{pmatrix}
        -\gamma_L/2 & 0 & 0 \\
        0 & -\gamma_L/2 & 0 \\
        0 & 0 & \gamma_L
    \end{pmatrix},
    \label{eq:axisymmetric_strain_tensor}
\end{equation}
with $\gamma_L>0$. 
If the velocity gradient were given only by this deterministic strain field, the corresponding invariants would be
\begin{equation}
    Q_0
    =
    -\frac{1}{2}\mathrm{tr}(S^2)
    =
    -\frac{3}{4}\gamma_L^2
    \quad \textrm{and} \quad
    R_0
    =
    -\frac{1}{3}\mathrm{tr}(S^3)
    =
    -\frac{1}{4}\gamma_L^3.
\end{equation}
In a turbulent field, however, this idealized background strain can be perturbed by neighboring vortices, finite-time vortex evolution and random orientations of the vortex axes. We therefore model
\begin{equation}
    P_{\mathrm{stretch}}(Q,R)
    =
    \frac{1}{2\pi\sqrt{\det\boldsymbol{\Sigma}}}
    \exp
    \left[
    -\frac{1}{2}
    \left(
    \boldsymbol{X}-\boldsymbol{X}_0
    \right)^T
    \boldsymbol{\Sigma}^{-1}
    \left(
    \boldsymbol{X}-\boldsymbol{X}_0
    \right)
    \right]
    \label{eq:P_stretching}
\end{equation}
as a bivariate Gaussian distribution centered at $(Q_0,R_0)$ with 
$\boldsymbol{X}=(Q,R)^T$ and
$\boldsymbol{X}_0=(Q_0,R_0)^T$.
Here, the covariance matrix
\begin{equation}
    \boldsymbol{\Sigma}
    =
    \begin{pmatrix}
        \sigma_Q^2 & \rho_{QR}\sigma_Q\sigma_R\\
        \rho_{QR}\sigma_Q\sigma_R & \sigma_R^2
    \end{pmatrix}
    \label{eq:covariance_stretching}
\end{equation}
contains the variances and covariance of the local strain-induced fluctuations. 
In WLT, the vortices are assigned random positions and orientations, and the ensemble is constructed to recover statistical homogeneity and isotropy at the flow-field level. 
To close $\boldsymbol{\Sigma}$, we assume that the leading-order stretching perturbations are distributed axisymmetrically about the local vortex axis.
Under this approximation, no preferred signed correlation between the $Q$- and $R$-fluctuations is retained.
We therefore set $\rho_{QR}=0$, so that
\begin{equation}
    \boldsymbol{\Sigma}=
\begin{pmatrix}
\sigma_Q^2 & 0\\
0 & \sigma_R^2
\end{pmatrix}.
\end{equation}
Substituting this into \eqref{eq:P_stretching} yields the stretching contribution
\begin{equation}
    P_{\rm stretch}(Q,R)=
\frac{1}{2\pi \sigma_Q \sigma_R}
\exp\left[
-\frac{(Q-Q_0)^2}{2\sigma_Q^2}
-\frac{(R-R_0)^2}{2\sigma_R^2}
\right].
\end{equation}

\subsection{Spiral-arm contribution}\label{sec:spiral-arm}
We obtain the spiral-arm contribution $P_{\mathrm{spiral}}(Q,R)$ by conditioning the local velocity-gradient statistics on the vortex properties $\tau$, $k$, $\theta_k$ and $n$.  
For fixed $n$, $k$, $\tau$ and $\theta_k$, we decompose
\begin{equation}
    A_{ij}
    =
    \overline{A}_{ij}^{(n)}(k,\tau,\theta_k)
    +
    A_{ij}^{\prime(n)}(k,\tau,\theta_k),
    \label{eq:A_decomposition_mode}
\end{equation}
where $\overline{A}_{ij}^{(n)}$ is the contribution of the $n$-th spiral mode and $A_{ij}^{\prime(n)}$ denotes fluctuations due to random phases, interactions with other vortices and neglected higher-order modes. Therefore, the conditional mean invariants are defined as 
$\overline{Q}_n(k,\tau)=\left\langle Q |n,k,\tau \right\rangle$ and $\overline{R}_n(k,\tau)=\left\langle R |n,k,\tau \right\rangle$.

The local spiral-arm gradients are therefore controlled by $\gamma_L$, $n$, $k$ and the radial variation of the angular velocity $\Omega$. 
Retaining the leading-order dependence obtained from the stationary-phase scaling, we write the modal centers of the spiral-arm contribution as
\begin{equation}
    \overline{Q}_n(k)
    =
    -\frac{\gamma_L^2}{8}
    \left(\frac{n}{k}\right)^{4/3}
    \left|
    \frac{\Omega''}{\Omega'}
    \right|
\quad \textrm{and} \quad
    \overline{R}_n(k)
    =
    \frac{\gamma_L^3}{16}
    L\,n\,k^{-2}
    \left|
    \frac{\Omega''}{\Omega'}
    \right|^{1/2}.
\end{equation}
The negative sign in $\overline{Q}_n(k)$ indicates that the spiral arms contribute predominantly to strain-dominated regions of the $Q$--$R$ plane, whereas $\overline{R}_n(k)$ describes the skewed distribution induced by the spiral winding of the vortex arms.

For fixed $n$, $k$, $\tau$ and $\theta_k$, we approximate the conditional PDF of $(Q,R)$ by a bivariate Gaussian PDF centered at $(\overline{Q}_n,\overline{R}_n)$:
\begin{equation}
    P(Q,R|n,k,\tau,\theta_k)
    =
    \frac{1}{2\pi\sqrt{\det \boldsymbol{C}_n(k,\tau)}}
    \exp
    \left[
    -\frac{1}{2}
    \left(
    \boldsymbol{X}-\overline{\boldsymbol{X}}_n
    \right)^T
    \boldsymbol{C}_n^{-1}(k,\tau)
    \left(
    \boldsymbol{X}-\overline{\boldsymbol{X}}_n
    \right)
    \right],
    \label{eq:conditional_spiral_general}
\end{equation}
with $\overline{\boldsymbol{X}}_n = (\overline{Q}_n(k), \overline{R}_n(k))^T$. 
Here, the covariance matrix $\boldsymbol{C}_n(k,\tau)$ generally contains the conditional covariance between the fluctuations of $Q$ and $R$ about their modal centers $(\overline{Q}_n,\overline{R}_n)$. In the WLT ensemble, however, the spiral vortices are randomly positioned and oriented, and the phase of each spiral arm is likewise randomized through ensemble averaging over the vortex age and orientation. After conditioning on given $n$, $k$ and $\tau$, these random orientations and phases do not imply a preferred signed correlation between the residual fluctuations of $Q$ and $R$. 
We therefore adopt the leading-order diagonal closure for the conditional covariance as
\begin{equation}
    \boldsymbol{C}_n(k,\tau)
    =
    \begin{pmatrix}
        \sigma_{Q,n}^2(k,\tau) & 0\\
        0 & \sigma_{R,n}^2(k,\tau)
    \end{pmatrix},
\end{equation}
and \eqref{eq:conditional_spiral_general} becomes
\begin{equation}
    P(Q,R|n,k,\tau,\theta_k)
    =
    \frac{1}{2\pi\sigma_{Q,n}(k,\tau)\sigma_{R,n}(k,\tau)}
    \exp
    \left[
    -\frac{\left(Q-\overline{Q}_n(k)\right)^2}
    {2\sigma_{Q,n}^2(k,\tau)}
    -
    \frac{\left(R-\overline{R}_n(k)\right)^2}
    {2\sigma_{R,n}^2(k,\tau)}
    \right].
    \label{eq:conditional_spiral_diagonal}
\end{equation}
This approximation provides a simple normalizable closure once the modal centers already encode the dominant $Q$-$R$ correlation induced by the spiral-arm dynamics.

Next, we determine the statistical weight for each spiral mode. This modal weight quantifies the relative contribution of the \(n\)-th azimuthal mode to $P_{\mathrm{spiral}}(Q,R)$. 
Its form is obtained from the third-order vorticity correlation in Appendix \ref{sec:suppression_LEV}. 
Specifically, substituting the Fourier representation of the Lundgren spiral vortex into the third-order vorticity correlation, and then applying the Fourier-Bessel transform together with the stationary-phase approximation, yields the modal contribution $Z_n$ in \eqref{eq:Z*}.
This contribution contains the product of three azimuthal Fourier components, \(f_m f_{n-m} f_{-n}\), which represents the triadic coupling among the modes \(m\), \(n-m\), and \(-n\).
Based on the exponential factor in \eqref{eq:Z*} for
describing the viscous decay of the coupled modal contribution,  
we define the effective viscous weight of the $n$-th spiral mode by summing over the coupled azimuthal modes as 
\begin{equation}
F_{\mathrm{visc}}(n,k)
=
\sum_{m=-m_{\max}}^{m_{\max}}
a_{n m}
\exp\left[
-\frac{2\nu_L}{3\gamma_L}
\frac{n^2+m^2+nm}{n^2}k^2
\right],
\end{equation}
where \(a_{nm}\) contains \(f_m f_{n-m}f_{-n}\). 
The cutoff \(m_{\max}\) denotes the largest retained azimuthal mode in the finite-mode representation of the spiral arms, as the modes with sufficiently large \(|m|\) are exponentially suppressed by viscosity. 

In the absence of detailed information on the mode amplitudes, we assume that \(a_{nm}\) varys slowly over the retained modes compared with the exponential viscous factor. Then, the relative importance of each spiral mode is controlled primarily by viscous attenuation. The normalized modal probability is therefore defined as
\begin{equation}
p_n(k)=
\frac{F_{\mathrm{visc}}(n,k)}
{\sum_{j=1}^{n_{\max}(k)}F_{\mathrm{visc}}(j,k)} .
\end{equation}

WLT is a statistically stationary state of an ensemble of spiral vortices. Consistent with \citet{Lundgren1982}, when the spiral vortex tube constructed in \s\ref{sec:ensemble} is treated as a sequence of discrete spiral vortex elements, the repeated birth-decay process yields a time-independent age distribution, despite vortices of different ages occupying different axial lengths.
Thus, the vortex age $\tau$ is uniformly sampled over \(0\leq\tau\leq\tau_L\) under the vortex-number measure. 
The orientation of $\boldsymbol{k}$ is also statistically isotropic.
Therefore, we obtain the spiral contribution
\begin{equation}
    P_{\mathrm{spiral}}(Q,R)
    =
    \frac{1}{Z_s}
    \int_0^{\tau_L}
    \frac{d\tau}{\tau_L}
    \int_{k_{\min}}^{k_{\max}}
    p_k(k)\,dk
    \int_0^{2\pi}
    \frac{d\theta_k}{2\pi}
    \sum_{n=1}^{n_{\max}(k)}
    p_n(k)
    P(Q,R|n,k,\tau,\theta_k)
    \label{eq:P_spiral_general}
\end{equation}
by averaging the conditional PDF in \eqref{eq:conditional_spiral_diagonal} over $\tau$, $k$, $\theta_k$ and $n$, where the probability density of active inertial-range wavenumbers is approximated by a uniform distribution
$p_k(k)=1/(k_{\max}-k_{\min})$ for $k_{\min}\leq k\leq k_{\max}$. 
The normalization factor $Z_s$ in \eqref{eq:P_spiral_general} is chosen such that
\begin{equation}
    \int_{-\infty}^{\infty}
    \int_{-\infty}^{\infty}
    P_{\mathrm{spiral}}(Q,R)\,dQ\,dR
    =
    1.
    \label{eq:Zs_normalization}
\end{equation}
If $p_k(k)$ and $p_n(k)$ are normalized and the conditional density \eqref{eq:conditional_spiral_general} is normalized, then $Z_s=1$.
Applying the approximation \eqref{eq:conditional_spiral_diagonal} to \eqref{eq:P_spiral_general} yields the spiral-arm contribution 
\begin{equation}
\begin{aligned}
    P_{\text{spiral}}(Q,R)
    =
    \int_0^{\tau_L}
    \frac{d\tau}{\tau_L}
    \int_{k_{\min}}^{k_{\max}}
    p_k(k)\,dk
    \int_0^{2\pi}
    \frac{d\theta_k}{2\pi}
    \sum_{n=1}^{n_{\max}(k)}
    p_n(k)
    \times
    \\    
    \frac{1}
    {2\pi\sigma_{Q,n}(k,\tau)\sigma_{R,n}(k,\tau)}    
    \exp
    \left[
    -\frac{\left(Q-\overline{Q}_n(k)\right)^2}
    {2\sigma_{Q,n}^2(k,\tau)}
    -
    \frac{\left(R-\overline{R}_n(k)\right)^2}
    {2\sigma_{R,n}^2(k,\tau)}
    \right].
\end{aligned}
    \label{eq:P_spiral_final}
\end{equation}

\subsection{Dissipation contribution}
In the dissipation range, the velocity-gradient statistics are strongly affected by viscosity. 
Instead of deriving the full dissipative distribution from the inertial-range spiral-arm asymptotics, we model it through a conditional decomposition as 
\begin{equation}
    P_{\mathrm{diss}}(Q,R)
    =
    P_{\mathrm{diss}}(Q|R)P_{\mathrm{diss}}(R).
    \label{eq:Pdiss_decomposition}
\end{equation}
This modelling choice is motivated by the particular sensitivity of \(R\) to nonlinear self-amplification within the dissipation range \citep{Cantwell1992,Ooi1999}. In \eqref{eq:Pdiss_decomposition}, $P_{\rm diss}(R)$ is employed to represent the primary dissipative modification of the topological asymmetry, whereas $P_{\rm diss}(Q|R)$ describes the distribution of the second invariant at a prescribed value of this local topological state.

Since $P(Q,R)$ exhibits a characteristic teardrop shape in the $Q$--$R$ plane, the dominant branch can be approximated by a curve of the form
$Q\sim -\kappa |R|^{2/3}$,
where $\kappa>0$ is a model parameter. We therefore model the conditional distribution of $Q$ for fixed $R$ as
\begin{equation}
    P_{\mathrm{diss}}(Q|R)
    =
    \frac{1}{\sqrt{2\pi}\sigma_\varepsilon(R)}
    \exp
    \left[
    -
    \frac{
    \left(
    Q+\kappa |R|^{2/3}
    \right)^2
    }
    {2\sigma_\varepsilon^2(R)}
    \right],
    \label{eq:Pdiss_Q_given_R}
\end{equation}
where the conditional width \(\sigma_\varepsilon(R)\) induced by fluctuations is determined from the ensemble of WLT-generated velocity fields obtained for different model parameters. Specifically, after computing $Q_L$ and $R_L$ from each velocity-gradient field, we estimate the residual $\Delta Q_L = Q_L+\kappa |R_L|^{2/3}$ and compute $\sigma_\varepsilon^2(R)=\left\langle \Delta Q_L^2|R_L\right\rangle_{\mathrm{WLT}}$. This conditional average is evaluated in narrow \(R\)-bins across all sampled WLT realizations and parameter sets.

In the dissipative range, the most intense velocity-gradient events are constrained by viscous smoothing. Accordingly, $P_{\rm diss}(R)$ is expected to decay more rapidly than an algebraic law. At the same time, $P(Q,R)$ in HIT is not symmetric with respect to \(R=0\), reflecting the preferential sampling of distinct strain-production and rotation-production events along the Vieillefosse branches. To capture these features with a minimal parametrization, we therefore adopt a minimal skewed exponential form for
\begin{equation}
P_{\rm diss}(R)=
C_R
\frac{\exp(-\beta_0 |R|)}
{1+\exp(-R/R_a)}.
\label{eq:Pdiss_R}
\end{equation}
Here, \(C_R\) is determined by the normalization condition 
$\int_{-\infty}^{\infty} P_{\rm diss}(R)\,dR=1$; 
\(\beta_0\) and \(R_a\) are empirical parameters obtained from the WLT ensemble across the sampled model parameter sets. 
In \eqref{eq:Pdiss_R}, \(\beta_0\) governs the far-tail decay rate of the distribution, whereas \(R_a\) controls the asymmetry degree between positive and negative values of \(R\). 
Combining \eqref{eq:Pdiss_Q_given_R} and \eqref{eq:Pdiss_R}, we obtain
\begin{equation}
    P_{\mathrm{diss}}(Q,R)
    =
    C_R
    \frac{
    \exp(-\beta_0 |R|)
    }
    {
    1+\exp(-R/R_a)
    }
    \frac{1}{\sqrt{2\pi}\sigma_\varepsilon(R)}
    \exp
    \left[
    -
    \frac{
    \left(
    Q+\kappa |R|^{2/3}
    \right)^2
    }
    {2\sigma_\varepsilon^2(R)}
    \right].
    \label{eq:P_diss_final}
\end{equation}

\subsection{Volume-fraction weights}
We specify the weights $w_1$, $w_2$ and $w_3$ in \eqref{eq:QR_mixture} for the mean volume fractions occupied by the background stretching, spiral-arm and dissipative regions, respectively.
For a Lundgren vortex subjected to an axial stretching rate $\gamma_L$, the radial length scale decreases as $r(\tau)=r_0\exp(-\gamma_L \tau/2)$. The stretching-dominated core volume is approximated by $V_{\mathrm{stretch}}(\tau)=
\pi r^2(\tau)l_0=\pi r_0^2l_0\exp(-\gamma_L \tau)$. Taking $N_c$ vortices in a domain of volume $L^3$, and averaging $V_{\mathrm{stretch}}(\tau)$ uniformly over the vortex lifetime $0<\tau<\tau_L$ yields the weight for stretching
\begin{equation}
    w_1
    =
    \frac{N_c}{\tau_L L^3}\int_0^{\tau_L}
    V_{\mathrm{stretch}}(\tau)\,d\tau
    =
    \frac{N_c}{L^3}
    \pi r_0^2l_0
    \frac{1-\exp(-\gamma_L\tau_L)}
    {\gamma_L\tau_L}.
    \label{eq:w1_final}
\end{equation}

The spiral-arm region is associated with the axial elongation and winding of the vortex tube. Its leading-order volume is modeled as $V_{\mathrm{spiral}}^{(0)}(\tau)=\pi r_0^2 l_0 \exp(\gamma_L \tau)$. 
The effective lifetime-averaged volume occupied by the spiral arms is further enlarged by the combined effect of circulation-induced winding and viscous diffusion. We represent this correction by the dimensionless factor $\Xi_V = 1+\Gamma\gamma_L\tau_L/(2\nu_L)$, and then yields the weight for spiral-arm
\begin{equation}
    w_2
    =
    \frac{N_c\Xi_V}{\tau_L L^3}
    \int_0^{\tau_L}
    V_{\mathrm{spiral}}^{(0)}(\tau)\,d\tau
    =
    \frac{N_c}{L^3}
    \pi r_0^2l_0
    \frac{\exp(\gamma_L\tau_L)-1}
    {\gamma_L\tau_L}
    \left(
    1+\frac{\Gamma\gamma_L\tau_L}{2\nu_L}
    \right).
    \label{eq:w2_final}
\end{equation}

Finally, the dissipative contribution is assigned the remaining volume fraction as 
\begin{equation}
\begin{aligned}
    w_3
    =1-
    \frac{N_c}{L^3}
    \pi r_0^2l_0
    \bigg[
    \frac{1-\exp(-\gamma_L\tau_L)}
    {\gamma_L\tau_L}
    +
    \frac{\exp(\gamma_L\tau_L)-1}
    {\gamma_L\tau_L}
    \left(
    1+\frac{\Gamma\gamma_L\tau_L}{2\nu_L}
    \right)
    \bigg].
    \label{eq:w3_final}
\end{aligned}
\end{equation}
The parameters must satisfy 
\begin{equation}
    0\leq w_1\leq 1,
    \quad
    0\leq w_2\leq 1,
    \quad
    0\leq w_3\leq 1,
    \label{eq:weight_constraints}
\end{equation}
which imposes the realizability condition
\begin{equation}
    \frac{N_c}{L^3}
    \pi r_0^2l_0
    \bigg[
    \frac{1-\exp(-\gamma_L\tau_L)}
    {\gamma_L\tau_L}
    +
    \frac{\exp(\gamma_L\tau_L)-1}
    {\gamma_L\tau_L}
    \left(
    1+\frac{\Gamma\gamma_L\tau_L}{2\nu_L}
    \right)
    \bigg]
    \leq 1.
    \label{eq:weight_consistency}
\end{equation}
It constrains the dimensionless stretching time \(T_L=\gamma_L\tau_L\). 
For positive parameters, the weights are realizable provided
\begin{equation}
F(T_L)=
\frac{1-\exp(-T_L)}{T_L}
+
\frac{\exp(T_L)-1}{T_L}
\left(1+\frac{\Gamma T_L}{2\nu_L}\right) \leq \frac{L^3}{N_c\pi r_0^2l_0}.
\end{equation}
Equivalently, upon fixing the ensemble constants $N_c, L, r_0, l_0, \Gamma$ and the parameter $\nu_L$, our model of $P(Q,R)$ requires
$0\leq \gamma_L\tau_L\leq T_L^{m}$,
where \(T_L^{m}\) is determined by $F(T_L^{m})={L^3}/{N_c\pi r_0^2l_0}$.

\subsection{Model assessment}
Substituting the three contributions \eqref{eq:P_stretching}, \eqref{eq:P_spiral_final} and \eqref{eq:P_diss_final} into \eqref{eq:QR_mixture}, we finally obtain the WLT-based model for $P(Q,R)$: 
\begin{equation}\label{eq:QR_pdf_final}
\begin{gathered}
\displaystyle 
P(Q,R)=
\frac{w_1}{2\pi \sigma_Q \sigma_R}
\exp\left[
-\frac{(Q-Q_0)^2}{2\sigma_Q^2}
-\frac{(R-R_0)^2}{2\sigma_R^2}
\right]+w_2\int_0^{\tau_L}
    \frac{d\tau}{\tau_L}
    \int_{k_{\min}}^{k_{\max}}
    p_k(k)\,dk \times
    \\
    \displaystyle 
    \int_0^{2\pi}
    \frac{d\theta_k}{2\pi}
    \sum_{n=1}^{n_{\max}(k)}
    p_n(k)
 \displaystyle 
\frac{1}
    {2\pi\sigma_{Q,n}(k,\tau)\sigma_{R,n}(k,\tau)}    
    \exp
    \left[
    -\frac{\left(Q-\overline{Q}_n(k)\right)^2}
    {2\sigma_{Q,n}^2(k,\tau)}
    -
    \frac{\left(R-\overline{R}_n(k)\right)^2}
    {2\sigma_{R,n}^2(k,\tau)}
    \right] \\\\
\displaystyle 
+w_3\frac{C_R}{\sqrt{2\pi}\sigma_{\epsilon}(R)} \frac{\exp(-\beta_0|R|)}{1 + \exp(-R/R_a)} \exp \left[ -\frac{(Q + \kappa |R|^{2/3})^2}{2\sigma_{\epsilon}^2(R)} \right],   
\end{gathered}
\end{equation}
with three weights in \eqref{eq:w1_final}, \eqref{eq:w2_final} and \eqref{eq:w3_final}.

\begin{figure}
	\centering
	\begin{overpic}
		[width=1.0\textwidth]{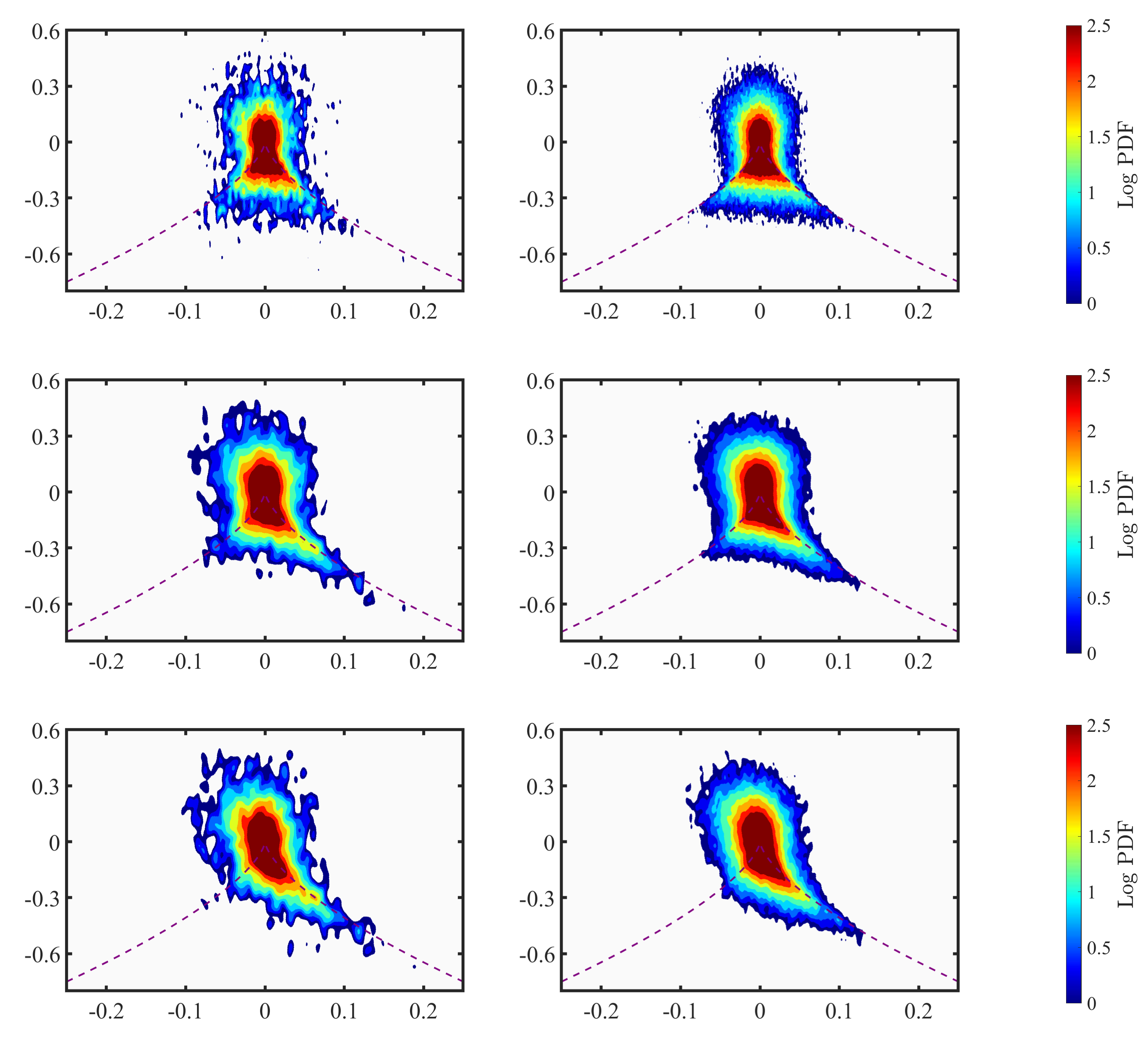}
		\put(-0.3,89.2){\footnotesize{(a)}}
		\put(-0.3,58.7){\footnotesize{(b)}}
        \put(-0.3,28.2){\footnotesize{(c)}}
        \put(1,76.8){\footnotesize{$ Q^{*} $}}
		\put(1,46.3){\footnotesize{$ Q^{*} $}}
        \put(1,15.8){\footnotesize{$ Q^{*} $}}
    	\put(44.1,76.8){\footnotesize{$ Q^{*} $}}
        \put(44.1,46.3){\footnotesize{$ Q^{*} $}}
        \put(44.1,15.8){\footnotesize{$ Q^{*} $}}
        \put(22.3,61){\footnotesize{$ R^{*} $}}
        \put(65.4,61){\footnotesize{$ R^{*} $}}
        \put(22.3,30.5){\footnotesize{$ R^{*} $}}
        \put(65.4,30.5){\footnotesize{$ R^{*} $}}
        \put(22.3,0){\footnotesize{$ R^{*} $}}
        \put(65.4,0){\footnotesize{$ R^{*} $}}
	\end{overpic}
    \caption{Logarithmically scaled $P(Q^*,R^*)$ under different ensemble parameters:
    (a) $\tau_{L}=30$, $\gamma_{L}=0.06$ and $\nu_{L}=1 \times 10^{-3}$;
    (b) $\tau_{L}=80$, $\gamma_{L}=0.08$ and $\nu_{L}=2 \times 10^{-4}$;
    (c) $\tau_{L}=140$, $\gamma_{L}=0.09$ and $\nu_{L}=2 \times 10^{-4}$.
    The left column presents the results obtained from our model in \eqref{eq:QR_pdf_final}. The right column shows the WLT results. 
    The dashed line denotes the Vieillefosse line. } \label{fig:QR_models}
\end{figure}

Figure~\ref{fig:QR_models} compares the logarithmically scaled $P(Q^*,R^*)$ obtained from our model in \eqref{eq:QR_pdf_final} with those from the WLT for three representative parameter sets of $\tau_{L}$, $\gamma_{L}$ and $\nu_{L}$. Overall, our model reproduces the WLT results for $P(Q^*, R^*)$ well in terms of both magnitude and morphology. Across all three parameter sets, the modeled distribution captures the characteristic teardrop-shaped contours of $P(Q^*, R^*)$, including the broad head near the strain-dominated region and the elongated tail along the Vieillefosse line. This agreement indicates that our model in \eqref{eq:QR_pdf_final} retains the dominant physical mechanisms controlling the $Q$-$R$ joint PDF in WLT. 

The agreement is particularly strong in the head and tail regions of $P(Q^*,R^*)$. 
The head is governed primarily by the background stretching contribution in \eqref{eq:P_stretching} and \eqref{eq:covariance_stretching}, whose center and width are directly determined by $\gamma_L$ and its local fluctuations. 
The tail is controlled mainly by the dissipation contribution in \eqref{eq:P_diss_final}, where the conditional relation $Q\sim-\kappa |R|^{2/3}$ in \eqref{eq:Pdiss_Q_given_R} and the skewed marginal distribution of $R$ in \eqref{eq:Pdiss_R} explicitly encode the geometry and asymmetry of high-gradient events in WLT. As a result, the model provides a robust approximation of the most probable core and the viscosity-regularized events.

By contrast, the agreement is weaker in the asymmetric body and the surrounding spiral-arm region. In \eqref{eq:P_spiral_final}, the spiral contribution is represented by a discrete sum over spiral modes and an integral over the inertial-range wavenumbers. This term is therefore more fragmented than the stretching and dissipative terms, as it retains only the leading modal contributions after the stationary-phase approximation. Moreover, the model captures only the dominant influence of the first few active spiral modes, while the weaker contributions from higher modes, phase correlations, and vortex-vortex interactions are absorbed into the conditional covariance. Finally, the spiral contribution in \eqref{eq:P_spiral_final} is restricted to the prescribed inertial-range wavenumber interval and therefore cannot fully represent the transition regions connecting the strain-dominated core, the spiral-arm body, and the dissipative tail. These approximations result in a locally fragmented distribution in the modelled body region.

The good agreements in figure~\ref{fig:QR_models} support the central assumption of our model in \eqref{eq:QR_pdf_final}: $P(Q,R)$ can be decomposed into stretching, spiral-arm and dissipation contributions. The explicit stretching and dissipative terms faithfully determine the head and tail, whereas the spiral term provides a leading-order approximation of the asymmetric body region. Further improvement of the body region requires retaining additional spiral modes, including their phase correlations, and incorporating interactions among neighbouring vortices.

\section{Conclusions}\label{sec:conclusion}
We develop the WLT method, which embeds Lundgren spiral vortices within the woven turbulence framework of \cite{Shen2024}. By assigning a periodic age function along the centerlines obtained from vortex filament simulations and relating the arc-length coordinate to the vortex age through the stretching history, we construct an ensemble of stretched, winding vortices for WLT, as a synthesis of HIT fields. The construction introduces three independently adjustable parameters: the background strain rate $\gamma_L$, the cutoff age $\tau_L$ and the viscous decay factor $\nu_L$, which respectively govern the rates of vortex stretching, spiral winding and dissipation. 

The statistical effect of the vortex element depends on the organization of its internal layers and associated strain, alongside its intensity and characteristic size.
By systematically varying $\gamma_L$, $\tau_L$ and $\nu_L$ and monitoring the evolution of the joint PDF $P(Q, R)$ of the velocity-gradient invariants in WLT, we demonstrate that the teardrop shape of $P(Q, R)$ can be partitioned into three distinct regions: the head region governed by $\gamma_L$, the body region controlled by $\tau_L$ and the tail region regulated by $\nu_L$. Increasing $\gamma_L$ broadens the head, increasing $\tau_L$ enhances the inclination of the body towards positive $R$ and decreasing $\nu_L$ extends the tail along the right Vieillefosse branch. 
This parametric control establishes a direct link between vortex dynamics and the statistical topology of the velocity-gradient tensor, suggesting that the characteristic \(Q\)-\(R\) topology could emerge from the geometry and evolution of the underlying vortex structures.

A series of assessments is also conducted on WLT. 
First, the model's ability to capture the local flow topology of the velocity gradient tensor is validated by comparing the teardrop shape of the normalized joint PDF $P(Q^*, R^*)$ with DNS results. 
Second, a multiscale analysis is conducted to compute the population fractions of the four principal topologies, UN/S/S, UF/C, SN/S/S and SF/S, as functions of the ratio of the filter scale to the Kolmogorov scale $\delta_f/\eta$, thereby assessing the scale-dependent transition from the inertial to the viscous range.
Third, the alignment of vorticity with the strain-rate eigenvectors is quantified, and the strain-type measure $s^*$ is employed to examine the preferential occurrence of biaxial strain. 
Thus, these good agreements confirm the WLT captures velocity gradient invariants in synthetic turbulence. 

Furthermore, based on the statistical theory of WLT, we develop an explicit model of $P(Q,R)$ in \eqref{eq:QR_pdf_final}. It decomposes $P(Q,R)$ into three contributions: stretching within the vortex core, winding of the spiral vortex arms and dissipation associated with small-scale structures. These contributions are weighted and normalized by the corresponding volume fractions occupied by each vortical structure. 
The controlling parameters $\gamma_L$, $\tau_L$ and $\nu_L$ of our model are the same with those of WLT. 
The model is validated for characterizing the teardrop-shaped contours of $P(Q,R)$ in HIT. 


In summary, WLT offers a physically interpretable framework in which the structural effects of stretching, winding and viscous smoothing can be varied. It therefore provides a testbed for identifying the vortex-scale mechanisms essential for reproducing the velocity-gradient statistics in homogeneous isotropic turbulence. 
A broader implication is that phenomenological descriptions of the fine structure of turbulence should account for the joint organization of vortex cores, the surrounding vorticity layers, and the associated strain field. The improvement over the original WT suggests that the development of spiral layers around vortex cores contributes to the \(Q\)-\(R\) asymmetry, whereas their viscous attenuation limits the extreme tail. This motivates a picture of real turbulence in which velocity-gradient statistics reflect an ensemble of vortex structures with varying degrees of internal winding and viscous smoothing.
Future work can include incorporating multifractal vortex hierarchies~\citep{Horiuti2008,Stegmayer2024}, extending the model to wall-bounded flows, and introducing vortex-vortex interactions to further improve the tails of $P(Q, R)$ and the representation of energy transfer across scales.

\backsection[Acknowledgements]{Numerical simulations were carried out on the TH-2A supercomputer in Guangzhou, China.}

\backsection[Funding]{This work has been supported in part by the National Natural Science Foundation of China (Grants No.~12432012, No.~12525201 and No.~12588201).}

\backsection[Declaration of interests]{The authors report no conflict of interest.}

\backsection[Author contributions]
{Y.Y., Z.Z. and W.S. contributed to the conception of this research. 
Z.Z., W.S. and Y.Y. contributed to the development of the research methodology. 
Z.Z. performed the research. 
Z.Z. and J.-H.X. contributed to the mathematical modelling. 
Y.Y. supervised the project. 
All the authors discussed the results and wrote the manuscript. 
All the authors have given approval for the manuscript.}

\appendix
\section{Validation of WLT using DNS}\label{sec:DNS}
Three WLT cases with different ensemble parameters are selected in Table~\ref{tab:WLT_simulations}, and their flow statistics are compared with DNS for validation. 
We performed DNS of HIT for solving the incompressible NS equations with an external forcing
in a periodic box using the open-source pseudospectral code ``HIT3D'' \citep{Chumakov2008}. 
The computational domain size is $L^3 = (2\pi)^3$ with $1024^3$ grid points. 
Correspondingly, we performed three DNS runs at $Re_\lambda=227.5$, $182.6$ and $137.3$ in Table~\ref{tab:DNS_simulations}. 
They are chosen to match $Re_\lambda$ of the WLT cases in Table~\ref{tab:WLT_simulations}. 
\begin{table}
	\centering\tabcolsep = 10pt
	\begin{tabular}{c c c c c}
		\toprule 
		   $Re_{\lambda}$ & $\tau_L$ & $\gamma_L$ & $\nu_L$ \\
		\midrule 
        227.5 & 210  & 0.12 & \num{1e-6}  \\
        182.6 & 180  & 0.10 & \num{5e-5}  \\
        137.3 & 120 & 0.07 & \num{8e-4}  \\
		\bottomrule 
	\end{tabular} 
    \caption{Three WLT cases at different $Re_\lambda$ and their corresponding ensemble parameters.}
    \label{tab:WLT_simulations}
\end{table}

\begin{table}
	\centering\tabcolsep = 10pt
	\begin{tabular}{c c c c c}
		\toprule 
		   $Re_{\lambda}$ & $k_{t}$ & $\langle \varepsilon \rangle$ & $\nu$ \\
		\midrule 
        227.5 & 3.77  & 4.03 & \num{4.54e-4}  \\
        182.6 & 3.31  & 3.75 & \num{5.84e-4}  \\
        137.3 & 2.98  & 3.54 & \num{8.87e-4}  \\
		\bottomrule 
	\end{tabular} 
    \caption{Three DNS cases of HIT for matching the WLT cases at different $Re_\lambda$, where $k_{t}$ is the turbulent kinetic energy, $\langle \varepsilon \rangle$ is the mean dissipation rate and $\nu$ is the kinematic viscosity.}
    \label{tab:DNS_simulations}
\end{table}

\begin{figure}
	\centering
	\includegraphics[width=0.6\linewidth]{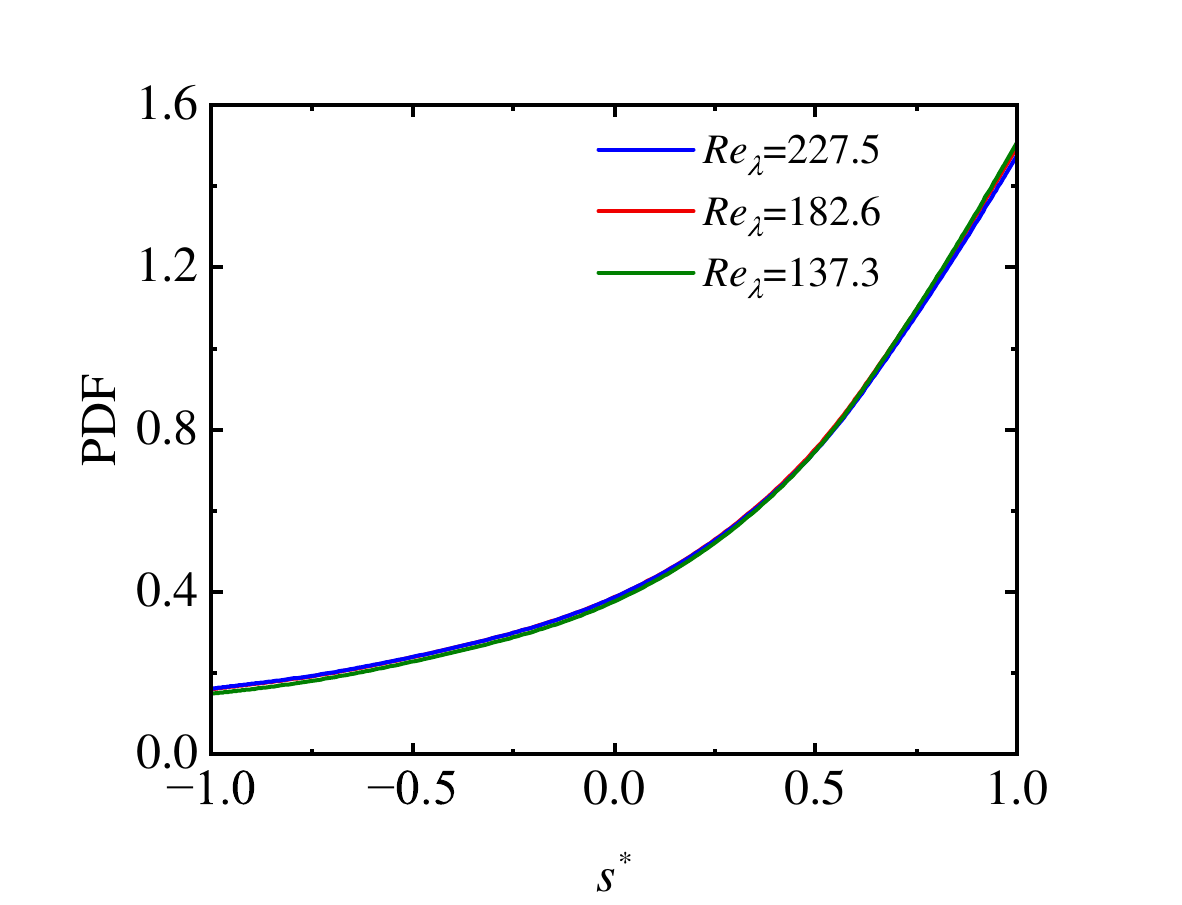}
    \caption{PDFs for $ s^{*} $ in DNS results at different $Re_\lambda$.} \label{fig:s_DNS2}
\end{figure}

On the velocity gradient tensor modelling, we use the alignment of $\boldsymbol{\omega}$ with $\boldsymbol{e}_i$ and the strain measure $s^*$ defined in \eqref{eq:s*} to assess WLT.  
Figure~\ref{fig:s_DNS2} shows the PDFs for $ s^{*} $ in different DNS cases show no dependence on $Re_\lambda$, consistent with the conclusion in \citet{Johnson2016}. 
In \s\ref{sec:Velocity Gradient Tensor}, we choose the DNS results at $Re_\lambda=182.6$ to compare with WLT.

\begin{figure}
	\centering
	\begin{overpic}
        [width=1.0\textwidth]{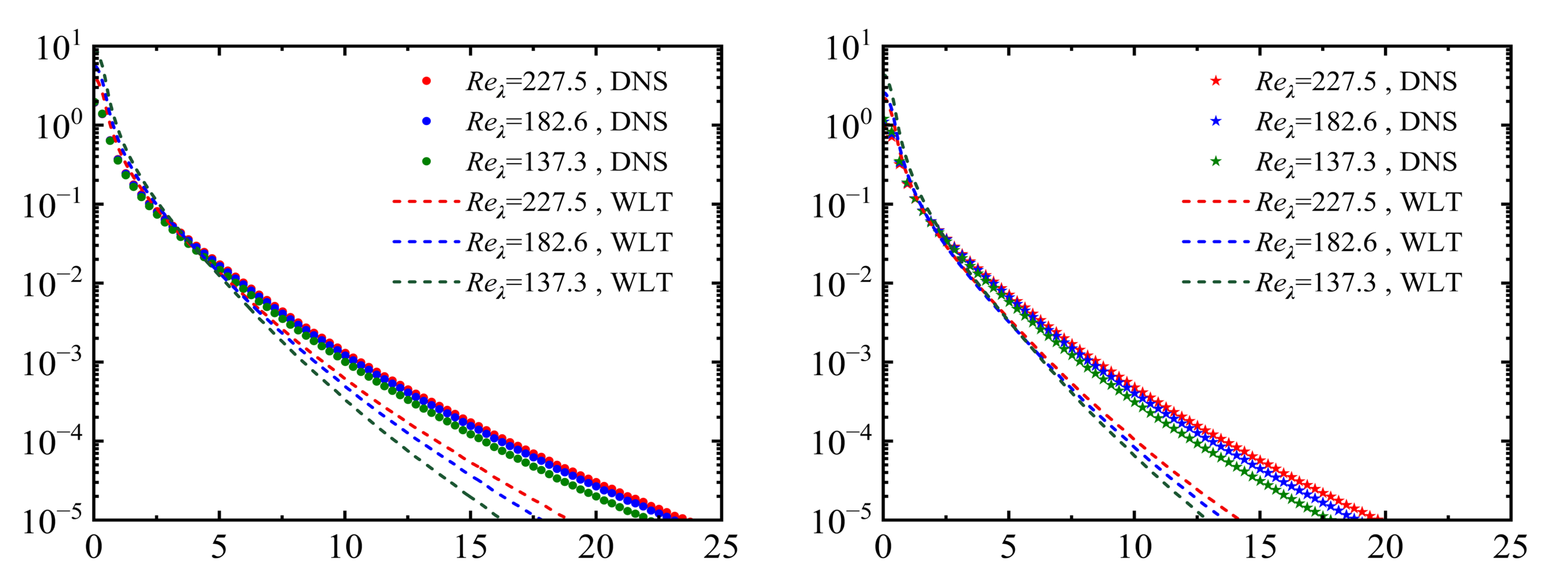}
        \put(-1,36){\footnotesize{(a)}}
		\put(49.5,36){\footnotesize{(b)}}
        \put(-1.5,12){\rotatebox{90}{\footnotesize{Enstrophy PDF}}}
        \put(49.3,12){\rotatebox{90}{\footnotesize{Dissipation PDF}}}
        \put(23,-1){\footnotesize{$\chi /\langle \chi \rangle$}}
        \put(74.2,-1){\footnotesize{$\varepsilon / \langle \varepsilon \rangle$}}
       
    \end{overpic}
    \caption{PDFs of the (a) enstrophy $\chi$ and (b) dissipation rate $\varepsilon$ normalized by their mean values for DNS and WLT at different $Re_\lambda$.}
    \label{fig:w_PDF}
\end{figure}

\begin{figure}
	\centering
	\begin{overpic}
        [width=1.0\textwidth]{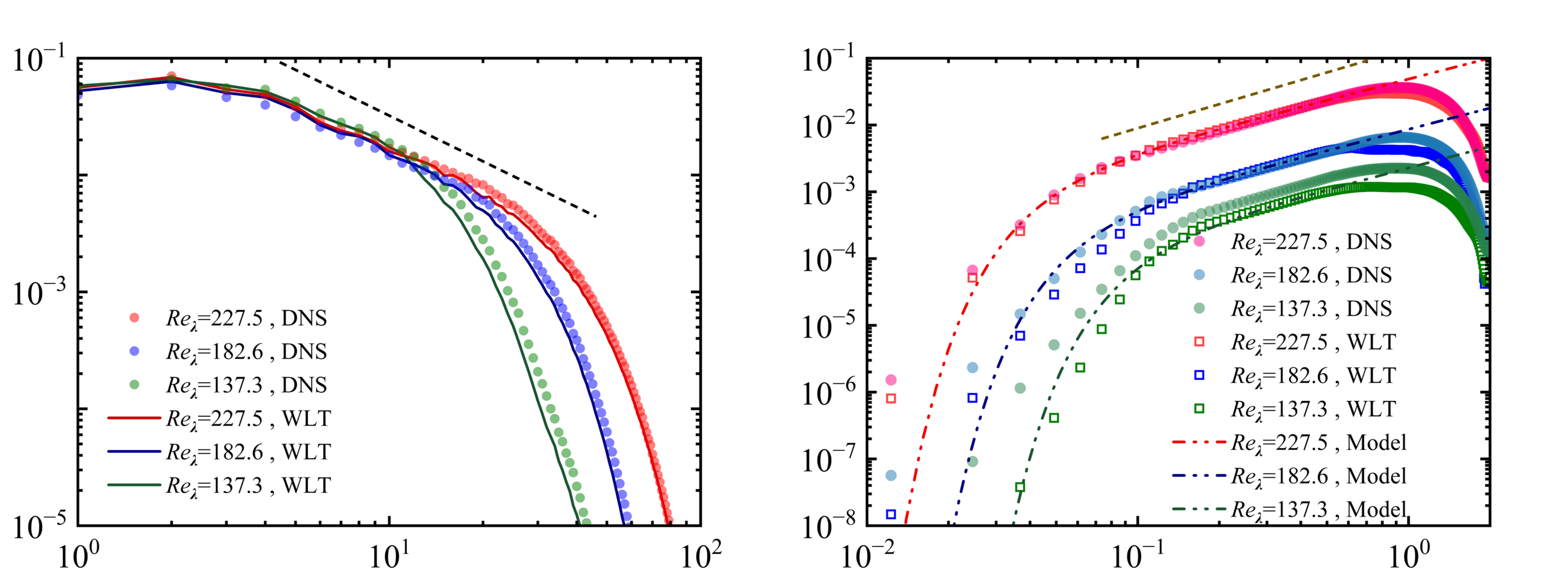}
        \put(-2,35.4){\footnotesize{(a)}}
		\put(48.1,35.4){\footnotesize{(b)}}
        \put(27,29.6){\rotatebox{338}{\scriptsize{$k^{-5/3}$}}}
        \put(77,31){\rotatebox{25}{\textcolor{brown}{\scriptsize{$r_s$}}}}
        \put(-2,16.5){\rotatebox{90}{\footnotesize{$E(k)$}}}
        \put(48.1,16.5){\rotatebox{90}{\footnotesize{$\mathit{-D_3(r_s)}$}}}
        \put(24.2,-1){\footnotesize{$k$}}
        \put(74.5,-1){\footnotesize{$r_s$}}

    \end{overpic}
    \caption{(a) Comparison of the energy spectra between the DNS and WLT at different $Re_\lambda$. (b) Comparison of the third-order velocity structure functions among the DNS, WLT and theoretical model in \eqref{eq:WLT_3D} at different $Re_\lambda$.}
    \label{fig:statistics}
\end{figure}

Moreover, figure~\ref{fig:w_PDF} shows that the WLT provides good agreement with both dissipation and enstrophy distributions, displaying the stretched-exponential shapes. 
Figure~\ref{fig:statistics}(a) plots the energy spectra for three WLT cases and corresponding DNS cases at the same $Re_{\lambda}$. The results demonstrate nearly identical spectral shapes between the WLT and DNS, in which the inertial range width increases with $Re_{\lambda}$. 
Figure~\ref{fig:statistics}(b) compares the third-order velocity structure functions in WLT and DNS, showing good agreement. 
Discrepancies between the synthetic WLT data, the analytical model in \eqref{eq:WLT_3D} and the DNS results diminish as $Re_{\lambda}$ increases. 

\section{Modelling the third-order vorticity correlation in WLT} \label{sec:third-order vorticity}
\subsection{Vorticity correlation}
Based on Lundgren's model in terms of vorticity, we link the vorticity and velocity correlations.
The velocity field in turbulence under well-posed boundary conditions is fully determined by the vorticity field through the Biot--Savart integral operator
\begin{equation}\label{eq:ui}
    u_{i}(\boldsymbol{x})= \frac{1}{4\pi}\int \epsilon_{i j k} \nabla_{j}\left(\frac{1}{y}\right) \omega_{k}(\boldsymbol{x}+\boldsymbol{y}) d \boldsymbol{y},
\end{equation}
where $\boldsymbol{x}$ denotes the observation point, and $\boldsymbol{y}$ is the relative displacement from $\boldsymbol{x}$ to the source point $\boldsymbol{x}+\boldsymbol{y}$, with $y=|\boldsymbol{y}|$.

When the Lundgren ensemble contains a sufficiently large number of spiral vortices, each exhibiting random variation in both their age and orientation, the resultant WLT tends to be statistically homogeneous and isotropic.
In physical $\boldsymbol{r}_s$-space, this indicates that the velocity or vorticity correlation depends exclusively on the separation distance ${r_s}$=$|\boldsymbol{r}_s|$ between the points.
We define a three-point, third-order vorticity correlation in WLT as
\begin{equation}\label{eq:r_ijk}
R_{i j k}^{\omega}\left(r_{s1}, r_{s2}\right)\equiv\left\langle\omega_{i}(\boldsymbol{x}) \omega_{j}\left(\boldsymbol{x}+\boldsymbol{r}_{s1}\right) \omega_{k}\left(\boldsymbol{x}+\boldsymbol{r}_{s2}\right)\right\rangle,
\end{equation}
where \(\boldsymbol {r}_{s1}\) and \(\boldsymbol {r}_{s2}\) denote the separation vectors from the reference point \(\boldsymbol x\) to the second and third points, respectively, with $r_{s1}=|\boldsymbol {r}_{s1}|$ and $r_{s2}=|\boldsymbol {r}_{s2}|$. 
It can be reduced to two-point vorticity correlations, i.e.~$R_{i j k}^{\omega}(0, r_s)=R_{i j k}^{\omega}(r_s)=\left\langle \omega_{i}(\boldsymbol{x}) \omega_{j}(\boldsymbol{x}) \omega_{k}(\boldsymbol{x}+\boldsymbol{r}_s)\right\rangle$.  
%
Combining \eqref{eq:ui} and \eqref{eq:r_ijk} yields
\begin{equation}
   \begin{split}\label{eq:R_ijk}
R_{i j k}(r_s) &\equiv\left\langle u_{i}(\boldsymbol{x}) u_{j}(\boldsymbol{x}) u_{k}(\boldsymbol{x}+\boldsymbol{r}_s)\right\rangle\\
&=\frac{1}{(4 \pi)^{3}} \int \frac{d^{3} y d^{3} y_{1} d^{3} y_{2}}{y^{3} y_{1}^{3} y_{2}^{3}} \epsilon_{i l m} \epsilon_{j n p} \epsilon_{k q s} y_{l} y_{1_{n}} y_{2_{q}}\\
&\times R_{m p s}^{\omega}\left(\left|\boldsymbol{y}-\boldsymbol{y}_{1}\right|,\left|\boldsymbol{r}_s+\boldsymbol{y}-\boldsymbol{y}_{2}\right|\right).
\end{split}
\end{equation}

In order to maintain dimensional consistency in the subsequent analysis, we assume that the vorticity and velocity correlations have scaling invariance. Namely, if the $ \beta $th-order two-point vorticity correlation scales as 
\begin{equation}\label{eq:scalelaw1}
    R_{\beta}^{\omega}(r_s)\sim r_s^{\alpha},
\end{equation}
then the $ \beta $th-order two-point velocity correlation scales as
\begin{equation}\label{eq:scalelaw2}
    R_{\beta}(r_s)\sim r_s^{\alpha+\beta}.
\end{equation}
This bridges the vorticity correlations within Lundgren's spiral vortex framework to Kolmogorov's scaling laws for velocity correlations. 

Considering the form of the Lundgren spiral vortex solution \eqref{eq:Lundgren_spiral}, we transform the vorticity correlation from the Cartesian coordinates $\boldsymbol{x}$ to the cylindrical coordinates $\boldsymbol{r}$. 
For a fixed Lundgren vortex with $\tau$ and $\boldsymbol{r}$, the two-point vorticity correlation becomes
\begin{equation}
    {R}_{l m p}^{\omega}(\boldsymbol{r}, \boldsymbol{r}_s, \tau)=\omega_{l}(\boldsymbol{r}, \tau) \omega_{m}(\boldsymbol{r}, \tau) \omega_{p}(\boldsymbol{r}+\boldsymbol{r}_s, \tau). 
\end{equation}
Within the cube of side $L$, we first make a average over $\boldsymbol r$ as
\begin{equation}\label{eq:spatavg_vor}
    \overline{{R}_{l m p}^{\omega}}(\boldsymbol{r}_s, \tau)=\frac{1}{L^{3}} \int \omega_{l}(\boldsymbol{r}, \tau) \omega_{m}(\boldsymbol{r}, \tau) \omega_{p}(\boldsymbol{r}+\boldsymbol{r}_s, \tau) d \boldsymbol{r},
\end{equation}
yielding the correlation depending only on $\boldsymbol{r}_s$, along with the vortex age and orientation. 

Taking into account \eqref{eq:Lundgren_spiral} and the computational feasibility in Fourier space, we re-express~\eqref{eq:spatavg_vor} in terms of $\boldsymbol{k}$ as 
\begin{equation}
     \begin{split}
\overline{{G}_{l m p}^{\omega}}(\boldsymbol{k}, \tau)& \equiv \frac{1}{(2 \pi)^{3}} \int \exp (-i \boldsymbol{k} \cdot \boldsymbol{r}_s) \overline{{R}_{l m p}^{\omega}}(\boldsymbol{r}_s, \tau) d \boldsymbol{r}_s\\
&=\int d \boldsymbol{k}^{\prime} \frac{1}{L^{3}} \frac{1}{(2 \pi)^{3}} \widetilde{\omega}_{l}\left(-\boldsymbol{k}-\boldsymbol{k}^{\prime},\tau\right) \widetilde{\omega}_{m}\left(\boldsymbol{k}^{\prime},\tau\right) \widetilde{\omega}_{p}(\boldsymbol{k},\tau).
\end{split}
\end{equation}
Here,
\(\tilde{\omega}_l(-\boldsymbol{k}-\boldsymbol{k}')\),
\(\tilde{\omega}_m(\boldsymbol{k}')\) and
\(\tilde{\omega}_p(\boldsymbol{k})\) are the Fourier amplitudes of the three
vorticity components entering the third-order correlation.

In the statistical ensemble of \cite{Lundgren1982}, where spiral vortices are randomly distributed across spatial scales (parameterized by $ \boldsymbol{k} $) and temporal maturation states (parameterized by $\tau$), the ensemble average is thus performed simply by integrating the spiral solution in time and over all orientations
\begin{equation}
G_{l m p}^{\omega}(k)=\left\langle\overline{{G}_{l m p}^{\omega}}\right\rangle \equiv \int d \tau \int d \widehat{\boldsymbol{k}} \overline{{G}_{l m p}^{\omega}}(\boldsymbol{k}, \tau),
\end{equation}
where $\widehat{\boldsymbol{k}}$ is the unit vector in the $\boldsymbol{k}$-direction and $k=|\boldsymbol{k}|$. 
Notably, under the statistically homogeneity and ergodicity of WLT, the volume average over the computational domain and the ensemble average derived above are equivalent~\citep{Lundgren1993}. Therefore, we use the same operator $ \langle \cdot \rangle $ to represent them.

In line with Lundgren's framework, we employ a lower-dimensional approximation where the vorticity vector ($Z$-axis) of the spiral vortex strictly adheres to the principal stretching axis. This alignment decouples the dynamics from the $X$-$Y$ plane. Consequently, with $ \mathit{l=m=p}=3 $, tensors such as ${R}_{l m p}^{\omega}$ and ${G}_{l m p}^{\omega}$ simplify to the unidirectional forms ${R}^{\omega}$ and ${G}^{\omega}$, so that the subscripts $lmp$ are omitted in all subsequent analyses. 

In WLT, the preceding assumption posits that the majority of the vorticity is oriented along the curve's axis as $ \boldsymbol{r}=\boldsymbol{r}_{0}(s) $. 
The local tangent vector is defined by $ \hat{\boldsymbol{t}}(s)=d\boldsymbol{{r}}_{0}(s)/ds $. The position of a point in space in proximity to the vortex can be expressed as $\boldsymbol{r}=\boldsymbol{{r}}_{0}(s)+x\boldsymbol{\hat{i}}(s)+y\boldsymbol{\hat{j}}(s)$, 
where $\boldsymbol{\hat{i}}$ and $\boldsymbol{\hat{j}}$ constitute a pair of orthogonal direction vectors that lie in the plane orthogonal to $\hat{\boldsymbol{t}}(s)$. 
The vorticity field can be expressed by $\bm{\omega}=\boldsymbol{\hat{t}}(s)\bm{\omega}(x,y,s,\tau)$. 
As the result of \citet{Lundgren1982}, when $\bm{r}_s$ is small enough, the vorticity at the point $\boldsymbol{r}+\bm{r}_s$ is approximated by
$\bm{\omega}(\boldsymbol{r}+\bm{r}_s,\tau) 
   \approx \hat{\boldsymbol{t}}(s)\omega(x+\Delta x,y+\Delta y,s,\tau)$, with $ \Delta x=\bm{r}_s\cdot\boldsymbol{\hat{i}}(s)$ and $\Delta y=\bm{r}_s\cdot\boldsymbol{\hat{j}}(s)$.  

This linear approximation suggests the reformulation of \eqref{eq:spatavg_vor}: 
\begin{equation}
    \overline{{R}^{\omega}}\left(r_s,\tau\right)=\frac{1}{L^{3}}\int\omega^{2}\left(x,y,s,\tau\right)\omega(x+\Delta x,y+\Delta y,s,\tau)dxdyds.
\end{equation}
To enhance computational efficiency, we write its spectral density integral expression and restructure it from the $X$-$Y$ plane to the $r$-$\theta$ plane for polar coordinates, which allows the angular part of the integration to be reformulated as a Bessel function with the Fourier-Bessel transform. 
Its Fourier integral results in
\begin{equation}
    \begin{split}
        \overline{{G}^{\omega}}\left(\boldsymbol{k},\tau\right)&=\frac{1}{(2 \pi)^{3}} \int \exp (-i \boldsymbol{k} \cdot \boldsymbol{r}_s) \overline{{R}^{\omega}}(\boldsymbol{r}_s, \tau) d \boldsymbol{r}_s\\&=\frac{1}{\left(2\pi L\right)^{3}}\int ds\delta\left(\boldsymbol{k}\cdot\hat{\boldsymbol{t}}\left(s\right)\right)
        \int\exp\left[ikr\cos\left(\theta_k-\theta\right)\right]{\omega}\left(r,\theta,s,\tau\right)rdrd\theta\\&\times\int\exp\left[-ikr\cos\left(\theta_k-\theta\right)\right]{\omega}^2\left(r,\theta,s,\tau\right)rdrd\theta.
    \end{split}
\end{equation}
Here, $\theta_k$ is the azimuthal angle of the projection of $\boldsymbol{k}$ onto this plane;  
the Dirac delta function $\delta(\boldsymbol{k}\cdot\hat{\boldsymbol{t}})$ enforces the constraint that only wavevectors perpendicular to the local tangent direction contribute to the local 2D spiral-arm transform. 

Using $ \delta\left[\boldsymbol{k}\cdot\hat{\boldsymbol{t}}(s)\right]=\delta(\cos\varphi_{k})/k $ and $l_s(\tau)=\int ds$, which denotes the vortex-axis length at age $\tau$, and integrating over 
$d\boldsymbol{\hat{k}}=\sin\varphi_kd\varphi_kd\theta_k $ and over $\tau$, we have
\begin{equation}\label{eq:third-orderintegral}
    \begin{gathered}
G^{\omega}\left(k\right)=\frac{1}{k\left(2\pi L\right)^3}\int l_s(\tau) d\tau \int_0^{2\pi}d\theta_k\int\exp\left[ikr\cos\left(\theta_k-\theta\right)\right]{\omega}\left(r,\theta,s,\tau\right)rdrd\theta \\
\times\int\exp{[-ikr\cos{(\theta_k-\theta)}]}{\omega}^2{(r,\theta,s,\tau)}rdrd\theta.
    \end{gathered}
\end{equation}
It is the general form of the third‑order vorticity correlation in the Lundgren spiral vortex ensemble. 
We make a comparison between the third-order integral presented in \eqref{eq:third-orderintegral} and the second-order integral of Lundgren's vorticity energy spectrum \citep{Lundgren1982}.
Their mathematical formulations exhibit a similar structure, with the principal difference being the substitution of $\omega$ by $\omega^2$.
Compared with the $\int \omega \int \omega$ integral in the second-order spectrum from \citet{Lundgren1982}, the present $\int \omega \int \omega^2$ integral in \eqref{eq:third-orderintegral} gives preferential weight to intense and asymmetric vorticity structures. 
Therefore, the spiral arms, whose vorticity distribution is spatially skewed and asymmetric, make a stronger contribution to the third-order correlation in \eqref{eq:third-orderintegral}. 
This enhanced weighting provides a mechanism by which the spiral-arm geometry produces stronger non-Gaussian vorticity statistics.

\subsection{Calculation of vorticity integrals}\label{sec:suppression_LEV}
Based on the vorticity expression for the spiral vortex and the construction conditions of the Lundgren ensemble, we derive a more specific form of $G^{\omega}\left(k\right)$.
The construction of the WLT in \s\ref{sec:ensemble} captures the following key processes of the Lundgren vortex ensemble: vortices are assumed to be generated continuously at a constant rate \(\alpha_L\), with an initial axial length \(l_0\), and to be removed from the ensemble after a cutoff age \(\tau_L\).

For a spiral vortex element with initial axial length $l_0$, its length at age $\tau$, following the stretching history, is $l(\tau)=l_0 S(\tau)$. The total vortex length in ensemble over all ages is $\alpha_L \int_{0}^{\tau_{L}} l_0 S(\tau)d\tau$.
Therefore, \eqref{eq:third-orderintegral} becomes
\begin{equation}
    \begin{gathered}\label{eq:ensemble1}
G^{\omega}\left(k\right)=\frac{\alpha_Ll_{0}}{kL^3}\frac{1}{\left(2\pi\right)^3}\int_{0}^{\tau_{L}} S(\tau) d\tau \int_0^{2\pi}d\theta_k \\
\times\int\exp\left[ikr\cos\left(\theta_k-\theta\right)\right]{\omega}\left(r,\theta,s,t\right)rdrd\theta \\
\times\int\exp{[-ikr\cos{(\theta_k-\theta)}]}{\omega}^2{(r,\theta,s,t)}rdrd\theta.
    \end{gathered}
\end{equation}
%
For a Fourier-Bessel integral expressed in polar coordinates, we have
\begin{equation}\label{eq:bessel}
    \int_{0}^{2\pi}\exp[in\theta-ikr\cos(\theta-\theta_{k})]d\theta=(-i)^{n}2\pi J_{n}(kr)\mathrm{exp}(in\theta_{k}).
\end{equation}
Then, we represent the vorticity terms $\omega$ and $\omega^2$ as
\begin{equation}
    \omega\left(r,\theta\right)=\sum_{n=-\infty}^{\infty}\omega_{n}\left(r\right)\exp\left(in\theta\right)
\end{equation}
and 
\begin{equation}
    \begin{gathered}
    \omega^{2}\left(r,\theta\right)=\sum_{n=-\infty}^{\infty}\left(\omega^{2}\right)_{n}\exp\left(in\theta\right)=\sum_{n=-\infty}^{\infty}\sum_{m=-\infty}^{\infty}\omega_{n-m}\omega_{m}\exp\left(in\theta\right).
    \end{gathered}
\end{equation}

By applying \eqref{eq:bessel} and integrating over $\theta_{k}$, \eqref{eq:ensemble1} becomes
\begin{equation}\label{eq:ensemble2}
    G^\omega(k)=\int d\tau\frac{\alpha_Ll_{0}}{kL^3}\frac{1}{(2\pi)^3}S(\tau)M(\theta_k),
\end{equation}
with 
\begin{equation}
    \begin{gathered}\label{eq:ensembleM}
M\left(\theta_{k}\right)=4\pi^2\sum_{n,n'=-\infty}^{\infty}\int d\theta_k(-i)^{n-n^{\prime}}J_n\left(kr\right)\iint J_{n^{\prime}}\left(k\zeta\right)(\omega^2)_n\left(r\right)\omega_{n^{\prime}}^*\left(\zeta\right)e^{i\theta_k(n-n^{\prime})}drd\zeta. 
\end{gathered}
\end{equation}
Here, $n$ denotes the azimuthal Fourier mode associated with the
vorticity factor $\omega^{2}$ in the correlation; $n'$ denotes the azimuthal Fourier mode associated with $\omega$; both $n$ and $n'$ are integer modal indices. 
The integration over $\theta_k$ uses the orthogonality of the azimuthal Fourier modes. 
Therefore, only identical azimuthal modes remain after the orientation average. 
Subsequently, employing $J_{-n}(x) = (-1)^n J_n(x)$ for the Bessel function, along with $\omega_n^*(\zeta) = \omega_{-n}(\zeta)$ for the real-valued vorticity field, yields
\begin{equation}\label{eq:ensembleM2}
    M\left(k,\tau\right) 
= 8\pi^3 \sum \iint J_n\left(kr\right) J_{-n}\left(k\zeta\right) (\omega^2)_n\left(r\right) \omega_{-n}\left(\zeta\right) dr d\zeta.
\end{equation}

The formulation now involves a summation over products consisting of two distinct radial integrals. By denoting the integral for $\omega_{-n}$ as $I_{-n}^{1}$ and the one for $(\omega^2)_n$ as $I_{n}^{2}$, we define 
\begin{equation}\label{eq:I1}
    I_{-n}^{1}(k,\tau)=\int_{0}^{\infty}J_{-n}(kr)\omega_{-n}(r,\tau)rdr
\end{equation} 
and
\begin{equation}\label{eq:I2}
    I_{n}^{2}(k,\tau)=\int_{0}^{\infty}J_{n}(kr)(\omega^2)_n(r,\tau)rdr.
\end{equation}
Incorporating \eqref{eq:ensembleM2}, \eqref{eq:I1} and \eqref{eq:I2} into \eqref{eq:ensemble2} yields
\begin{equation}\label{eq:G_Bessel}
   G^\omega(k)=\int d\tau\frac{2\alpha_Ll_{0}}{kL^3}S(\tau)\sum_{n=0}^{\infty} I_{-n}^{1}(k,\tau)I_{n}^{2}(k,\tau).
\end{equation}

We now specify the vorticity functional in the spiral vortex solution, incorporating both axial stretching and viscous diffusion effects. We rewrite the Fourier mode $\omega_n$ in \eqref{eq:Lundgren_spiral} as
\begin{equation}\label{eq:Lun_wn}
    \omega_{n}=Sf_{n}\left(S^{1/2}r\right)\exp\left[-\operatorname{in}\Omega\left(S^{1/2}r\right)T-\frac{1}{3}\nu_L n^2{\Omega}'\left(S^{1/2}r\right)^2T^3\right].
\end{equation} 
In the same way, we rewrite 
\begin{equation}
\begin{gathered}
(\omega^2)_n=S^{2}\sum_{m}f_{n-m}\left(S^{1/2}r\right)f_{m}\left(S^{1/2}r\right) \\
\times\exp\left[-i\left(n-m\right)\Omega\left(S^{1/2}r\right)T-\frac{1}{3}(n-m)^{2}{\Omega}'\left(S^{1/2}r\right)^{2}\nu_L T^{3}\right] \\
\times\exp\left[-im\Omega\left(S^{1/2}r\right)T-\frac{1}{3}m^{2}{\Omega}'\left(S^{1/2}r\right)^{2}\nu_L T^{3}\right],
\end{gathered}
\end{equation}
with ${\Omega}'\left(S^{1/2}r\right)= \mathrm{d}\Omega\left(S^{1/2}r\right)/\mathrm{d}\xi.$

Then, we use the stationary phase approximation~\citep{Bender1999} to calculate the Bessel integrals in~\eqref{eq:G_Bessel}. For large $kr$, we approximate the Bessel function
\begin{equation}
    J_n\left(kr\right)\sim\left(\frac{1}{2\pi kr}\right)^{1/2}i^{n+1/2}\exp\left(-ikr\right)+\mathrm{c.c.},
\end{equation}
where ``c.c'' refers to complex conjugate terms. 
We rewrite the $\omega$-integral in \eqref{eq:I1} as $Ae^{i\Phi}+\textrm{c.c.}$ with the phase
\begin{equation}\label{eq:Besselw1}
   \Phi=-kr-n\Omega\left(S^{1/2}r\right)T.
\end{equation}
Similarly, the phase for the $\omega^2$-integral in \eqref{eq:I2} is
\begin{equation}\label{eq:Besselw2}
\Phi_{\omega2}=-kr-m\Omega\left(S^{1/2}r\right)T-(n-m)\Omega\left(S^{1/2}r\right)T=\Phi.
\end{equation}
For large \(k\) and \(\tau\), we consider \(T/k = (\mathrm{e}^{\gamma \tau} - 1) / \gamma k \sim \mathcal{O}(1)\); thus, the integral can be evaluated asymptotically using the method of stationary phase. Since the phase varies rapidly with respect to \(r\), contributions from regions away from the stationary points cancel due to oscillation.
Consequently, the dominant contribution to the integral arises from the critical points $r_{n}$ determined by
\begin{equation}\label{eq:zeropoint}
    0={\Phi}'=\frac{\mathrm{d} \Phi}{\mathrm{d} r}
    =-k-nS^{1/2}\Omega^{\prime}\left(S^{1/2}r_n\right)T.
\end{equation}
This approximation for a general integral of this type is
\begin{equation}\label{eq:WKBJ}
I = \int B(r) e^{i \Phi(r)} d r \sim \sum_{r_{n}}\left[\frac{2 \pi}{\left|\Phi^{\prime \prime}\left(r_{n}\right)\right|}\right]^{1 / 2} B\left(r_{n}\right) \exp \left[i \Phi\left(r_{n}\right)+\frac{i \pi}{4} \operatorname{sgn} \Phi^{\prime \prime}\left(r_{n}\right)\right],
\end{equation}
where the sum is taken over all interior, non-degenerate stationary points
$r_n$ satisfying
$\Phi'(r_n)=0$ and $\Phi''(r_n)\neq 0$.

Both Bessel integrals in \eqref{eq:I1} and \eqref{eq:I2} are evaluated on the same asymptotic branch; consequently, their product carries the phase $e^{2i\Phi}$. 
In the following scaling analysis, we retain the modulus of this branch contribution; its overall phase produces oscillations in Fourier space but does not affect the inertial-range power law of interest.
Combining \eqref{eq:Lun_wn} -- \eqref{eq:zeropoint} and using \eqref{eq:WKBJ} yields 
\begin{equation}
\begin{gathered}\label{eq:I111}
    I_{-n}^{1}=\pi\left(\frac{r_{n}\Omega^{\prime}}{k^{2}\Omega^{\prime\prime}}\right)^{1/2}i^{n+1}Sf_{n}\left(S^{1/2}r_{n}\right)
    \exp\left[-ikr_{n}+in\Omega T -\nu_L n^{2}T^{3}\left(\Omega^{\prime}\right)^{2}/3\right].
    \end{gathered}
\end{equation}
Similarly, we have 
\begin{equation}
    \begin{gathered}\label{eq:I222}
I_{n}^{2}=\sum_{m=-\infty}^{m=\infty}\pi\left(\frac{r_{n}\Omega^{\prime}}{k^{2}\Omega^{\prime\prime}}\right)^{1/2}i^{n}S^{2}f_{m}\left(S^{1/2}r_{n}\right)f_{n-m}\left(S^{1/2}r_{n}\right) \\
\times\exp\left[-ikr_{n}-in\Omega T\right.-\nu_L\left(m^{2}+\left(n-m\right)^{2}\right)T^{3}\left(\Omega^{\prime}\right)^{2}/3].
\end{gathered}
\end{equation}

\citet{Lundgren1982} decomposed the vorticity integral terms into contributions from the core $(n=0)$ and the spiral arms $(n>0)$, and derived the Kolmogorov energy spectrum from the contributions of the spiral arms, thereby demonstrating their predominance in the inertial range. Here, we aim to investigate the scaling properties of the third-order correlation within the inertial range and, accordingly, restrict our calculations to the contributions from the spiral arms $(n>0)$.

Now we change the integration variables from $\tau$ to $\xi_n=S^{1/2}r_{n}(\tau)$ to simplify the calculation. For large $\tau$, we have 
\begin{equation}
    S/T=\frac{\exp(\gamma_L \tau)}{[\exp(\gamma_L \tau)-1]/\gamma_L}\approx \gamma_L
\end{equation}
and using the definition of $r_{n}$, the strain rate function integration $S(\tau)$ could be expressed as
\begin{equation}\label{eq:sgamma}
    S = \gamma_L T = \left[ -\frac{\gamma_L k}{n \Omega'(\xi_n)} \right]^{2/3} .
\end{equation}
Differentiating \eqref{eq:sgamma} with respect to \(\xi_n\), and using
\(d\ln S/d\tau = \gamma_L\), we can get
\begin{equation}\label{eq:tauxin}
    d\tau=-\frac{2}{3\gamma_L}\frac{\Omega^{\prime\prime}}{\Omega^{\prime}}d\xi_n.
\end{equation}

By substituting \eqref{eq:I111} and \eqref{eq:I222}, using \eqref{eq:sgamma} to replace $T^3(\Omega')^2$ with $k^2 / (n^2 \gamma_L)$ and applying \eqref{eq:tauxin} to change the integration variable from $\tau$
to $\xi_n$,
the integral in \eqref{eq:G_Bessel} becomes
\begin{equation}\label{eq:G*}
    G^\omega(k)=\frac{1}{k}\frac{2\gamma_L\pi^{2} \alpha_Ll_0}{L^3}\sum_{n=1}^\infty Z_n
\end{equation}
with
\begin{equation}\label{eq:Z*}
    \begin{split}
Z_{n}=-\sum_{m=-\infty}^{\infty}\int\frac{1}{\Omega^{\prime}\left(\xi_n\right)^{2}n}f_{m}\left(\xi_n\right)f_{n-m}\left(\xi_n\right)f_{-n}\left(\xi_n\right)  \exp\left[\frac{-2\nu_L\left(n^2+m^2+nm\right)k^2}{3\gamma_L n^2}\right]\xi_n d\xi_n.
\end{split}
\end{equation}

Equations \eqref{eq:G*} and \eqref{eq:Z*} give the explicit $k$-space parameter dependence of the third-order vorticity correlation in the WLT. 
The prefactor in \eqref{eq:G*} is controlled by \(\gamma_L\), \(\nu_L\) and computational constants \(\alpha_L\), \(\l_0\) and \(L\).  
The factor $1/k$ in \eqref{eq:G*} originates from the Fourier transform of the localized spiral structure.
The dimensionless coefficient \(Z_n\) in \eqref{eq:Z*} encapsulates the dependence on the spiral-vortex profile through \(f_n(\xi_n)\) and \(\Omega'(\xi_n)\), while the exponential term accounts for the viscous damping of the fine-scale azimuthal structure.  
The triple product $f_m f_{n-m} f_{-n}$ represents the nonlinear coupling among different spiral harmonics, analogous to the triadic interactions that govern energy transfer in turbulence~\citep{Kinjangi2023}.
Equation \eqref{eq:Z*} is used in \s\ref{sec:spiral-arm} to determine the modal weight entering the spiral-arm contribution in the model of $P(Q,R)$. 

\section{Modelling the third-order velocity structure function in WLT}
\label{sec:third-order velocity structure function}
To demonstrate that the WLT captures the intermittency statistics, we derive a model for the third-order velocity structure function $D_3$ in WLT, based on the inertial-range scaling law from the vorticity correlation \eqref{eq:G*} with \eqref{eq:Z*}.

The exponential factor in \eqref{eq:Z*} governs the dissipation-range behavior, where viscous effects smooth the finest spiral turns. 
Although the dissipation-range form is more complex than the Gaussian decay $\exp[-(2\nu_L/3\gamma_L)k^{2}]$ observed in second-order correlations~\citep{Lundgren1982}, this complexity does not affect the inertial-range scaling law that constitutes the primary focus of our analysis. 
For any finite and non-zero $f_n$, Gaussian integrals yield a similar exponential cutoff factor, though it does not prevail within the relevant range of $k$.
In the WLT, all $f_n(\xi_n)$ are equal (introduced in~\s\ref{sec:profile}), so we replace the sum over $m$ by
\begin{equation}
\begin{split}
    I_n&=-\int_{-\infty}^\infty\exp\left[-\frac{2\nu_L(n^2+m^2+nm)k^2}{3\gamma_L n^2}\right]dm \\&
    =-\exp\left(-\frac{2\nu_L k^{2}}{3\gamma_L n^{2}}\frac{3n^{2}}{4}\right)\int_{-\infty}^{\infty}\exp\left[-\frac{2\nu_L k^{2}}{3\gamma_L n^{2}}\left(m+\frac{n}{2}\right)^{2}\right]dm.
    \end{split}
\end{equation}
For the inertial range with $k\ll\sqrt{\gamma_L/\nu_L}$, the dissipation factor $\exp[-(\nu_L/2\gamma_L)k^2]\sim \mathcal{O}(1)$, so that the dominated factor is $1/k$ in \eqref{eq:G*} from the Fourier integral.

Therefore, within the inertial range, the third-order vorticity correlation in \eqref{eq:G*} conforms
\begin{equation}
    G^{\omega}(k)\sim k^{-1}.
\end{equation}
If we now assume the relation between vorticity and velocity scalings expected from the dimensional arguments in \eqref{eq:scalelaw1} and \eqref{eq:scalelaw2}, the third-order velocity correlation conforms
\begin{equation}\label{eq:scalingV}
    G(k)\sim k^{-4}.
\end{equation}

Finally, to establish a connection between $D_3(r_s)$ and $G(k)$, we utilize the relation~\citep{Saffman1996}
\begin{equation}\label{eq:3strc}
    \frac{D_{3}}{6}=16\pi\int_0^\infty\left(\frac{\sin(kr_s)}{k^2r_s^2}+\frac{3\cos(kr_s)}{k^3r_s^3}-\frac{3\sin(kr_s)}{k^4r_s^4}\right)k^2G(k)dk.
\end{equation}
Changing the integration variables from $k$ to $kr_s$, and substitute $G(k)=B k^{-4}$ suggested by \eqref{eq:scalingV} into \eqref{eq:3strc} yields
\begin{equation}\label{eq:3strc_D}
\frac{D_{3}}{6}=16B\pi r_s\int_0^\infty\frac{1}{(kr_s)^4}\left(\sin(kr_s)+\frac{3\cos(kr_s)}{kr_s}-\frac{3\sin(kr_s)}{(kr_s)^2}\right) d(kr_s).
\end{equation}
It shows $D_3(r_s)\sim r_s$ in the inertial range, consistent with the scaling law of \cite{K41}. 
Incorporating the viscous correction in the dissipation range into \eqref{eq:3strc_D}, we obtain a WLT-based model for the third‑order velocity structure function
\begin{equation}\label{eq:WLT_3D}
    D_3(r_s) = C_L  r_s  \exp\left( -\frac{\nu_{L}}{2\gamma_{L} r_s^2}\right),
\end{equation}
where $C_L$ is an empirical constant. 
This model is further validated with DNS in figure~\ref{fig:statistics}(b). 
It accurately predicts the inertial-range scaling laws and the extent of the inertial range in HIT. 

\bibliographystyle{jfm}
\bibliography{jfm}
\end{document}


\begin{center}

{\fontsize{15}{20}\selectfont
\textbf{Supplementary Materials for}\par
\textbf{``Modelling velocity gradient invariants in turbulence with Lundgren spiral vortex ensemble''}}

\vspace{0.25in}

{\fontsize{12}{14}\selectfont
\textbf{Zhi Zheng}\textsuperscript{1},
\textbf{Weiyu Shen}\textsuperscript{1,2},
\textbf{Jin-Han Xie}\textsuperscript{1},
\textbf{Yue Yang}\textsuperscript{1,3,\textsuperscript{*}}\par}

\vspace{0.15in}

{\small\itshape

\textsuperscript{1}State Key Laboratory for Turbulence and Complex Systems,\\
School of Mechanics and Engineering Science, Peking University,
Beijing 100871, PR China


\textsuperscript{2}Max Planck Institute for Solar System Research, 
G\"ottingen 37077, Germany


\textsuperscript{3}HEDPS-CAPT, Peking University, 
Beijing 100871, PR China
}


\footnotetext[1]{yyg@pku.edu.cn}



\end{center}





\section{Divergence-free construction for the Lundgren spiral vortex tube}

In this section, we derive an expression for the divergence-free vorticity within a closed vortex tube with an arbitrary smooth centerline. The tube’s internal structure is determined by the Lundgren spiral vortex profile, whose local age varies along the centerline. 
This construction method, based on~\citet{Xiong2019,Xiong2020,Shen2023,Shen2024}, extends to arbitrary vortex tube structures with asymmetric internal vorticity distributions.

\subsection{Frenet--Serret frame and tube coordinates}
\begin{figure}  
	\centering
	\begin{overpic}
		[width=0.8\textwidth]{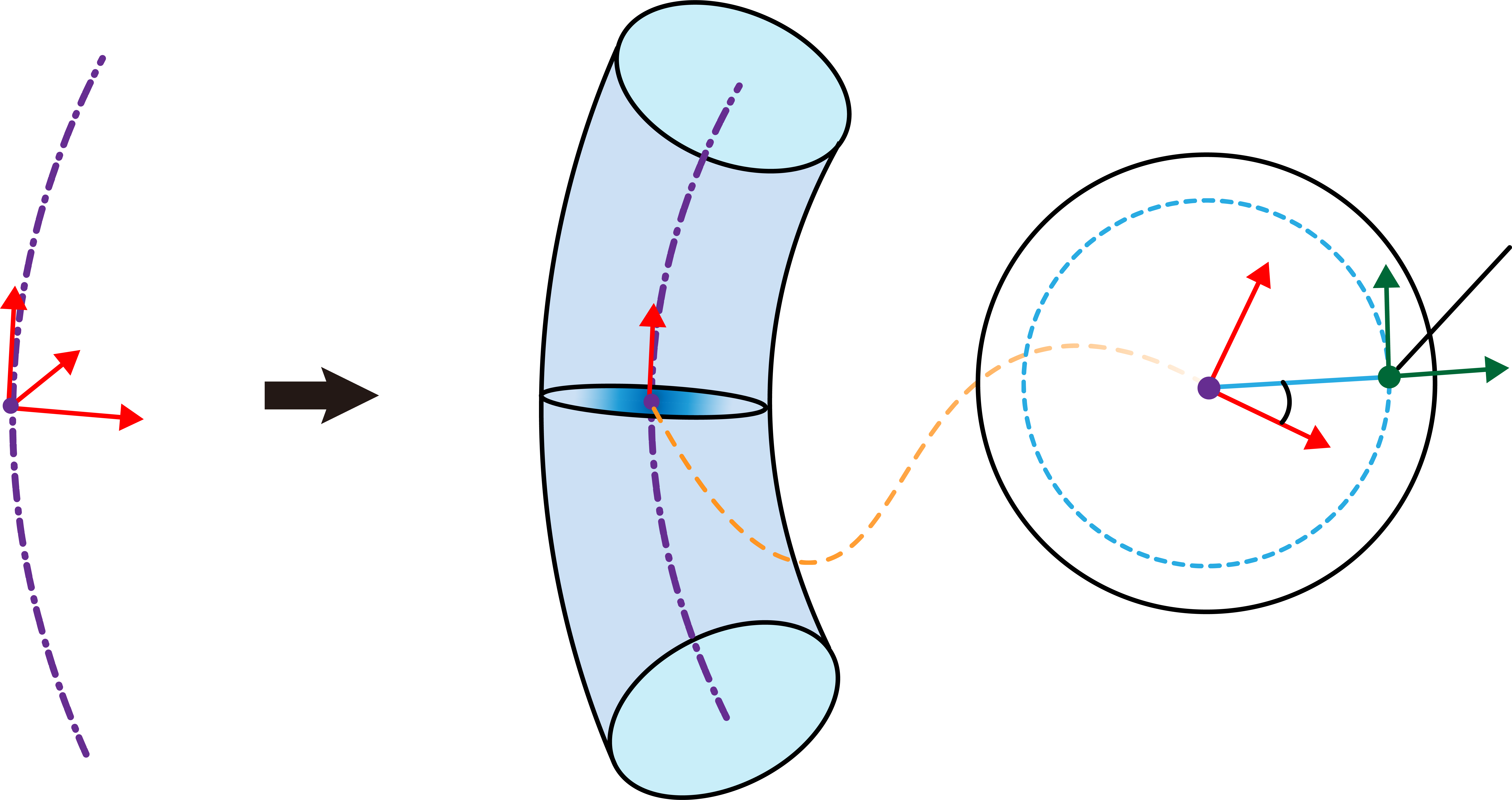}
        \put(-5,49){\small{(a)}}
        \put(32,49){\small{(b)}}
        \put(7,44){\textcolor{myPurple}{\footnotesize{$\boldsymbol{\mathcal{C}}$}}}
        \put(-2,32){\rotatebox{-1}{\textcolor{red}{\footnotesize{$\boldsymbol{T}$}}}}
        \put(8,26.5){\rotatebox{-1}{\textcolor{red}{\footnotesize{$\boldsymbol{N}$}}}}
        \put(3,29.8){\rotatebox{-1}{\textcolor{red}{\footnotesize{$\boldsymbol{B}$}}}}
        \put(40.5,31){\rotatebox{-1}{\textcolor{red}{\footnotesize{$\boldsymbol{T}$}}}}
        \put(79,32){\rotatebox{-1}{\textcolor{red}{\footnotesize{$\boldsymbol{B}$}}}}
        \put(84,21){\rotatebox{-1}{\textcolor{red}{\footnotesize{$\boldsymbol{N}$}}}}
        \put(97,37.2){{{\small{$\boldsymbol{x}(s,r,\theta)$}}}}
        \put(88.7,35.6){\rotatebox{5}{\textcolor{mygreen}{\small{$\boldsymbol{e}_{\theta}$}}}}
        \put(98.5,26.3){\rotatebox{5}{\textcolor{mygreen}{\small{$\boldsymbol{e}_{r}$}}}}
        \put(86.5,28.2){\rotatebox{5}{{\small{$r$}}}}
        \put(85.7,25){{{\small{$\theta$}}}}
    \end{overpic}
    \caption{Construction and coordinate systems of the vortex tube.
    (a) The curve $\boldsymbol{\mathcal{C}}$ described by the Frenet--Serret frame $(\boldsymbol{T}, \boldsymbol{N}, \boldsymbol{B})$.
    (b) Construction of the tubular region around the vortex centerline and the transformation between the Frenet--Serret frame $(\boldsymbol{T}, \boldsymbol{N}, \boldsymbol{B})$ and the local frame $(\boldsymbol{e}_s, \boldsymbol{e}_{r}, \boldsymbol{e}_{\theta})$.}
    \label{fig:L_tube}
\end{figure}

In ``woven Lundgren turbulence'' (WLT), we use the Frenet--Serret frame $(\boldsymbol{T}, \boldsymbol{N}, \boldsymbol{B})$ to characterize the vortex centerlines, as illustrated in figure~\ref{fig:L_tube}(a).
We represent a single vortex centerline $\boldsymbol{\mathcal{C}}$ as $\boldsymbol{\mathcal{C}}(s)$ with $s\in[0,\ell]$, where $s$ denotes the physical arc-length coordinate and $\ell$ is the total length.
The Frenet–Serret formulas on $\boldsymbol{\mathcal{C}}(s)$ are
\begin{equation}
\begin{cases}
\begin{aligned}
    \frac{\mathrm{d}\boldsymbol{T}(s)}{\mathrm{d}s}
    &=\kappa_c(s)\boldsymbol{N}(s),\\
    \frac{\mathrm{d}\boldsymbol{N}(s)}{\mathrm{d}s}
    &=-\kappa_c(s)\boldsymbol{T}(s)
      +\tau_c(s)\boldsymbol{B}(s),\\
    \frac{\mathrm{d}\boldsymbol{B}(s)}{\mathrm{d}s}
    &=-\tau_c(s)\boldsymbol{N}(s).
\end{aligned} 
\end{cases}
\label{eq:FS_equations}
\end{equation}
Here, $\boldsymbol{T}(s) \equiv \mathrm{d} \boldsymbol{\mathcal{C}} / \mathrm{d} s$ denotes the unit tangent, $\boldsymbol{N}(s) \equiv(\mathrm{d} \boldsymbol{T} / \mathrm{d} s) /|\mathrm{d} \boldsymbol{T} / \mathrm{d} s|$ the unit normal, $\boldsymbol{B}(s) \equiv \boldsymbol{T} \times \boldsymbol{N}$ the unit binormal, $\kappa_c(s)$ the curvature of $\boldsymbol{\mathcal{C}}$ and $\tau_c(s)$ the torsion of $\boldsymbol{\mathcal{C}}$. 

Figure~\ref{fig:L_tube}(b) illustrates the use of the curved cylindrical coordinate system $(s, r,\theta)$ to describe the tubular region surrounding the curve $\boldsymbol{\mathcal{C}}$, in which a point $\boldsymbol{x}$ is parameterized by
\begin{equation}
    \boldsymbol{x}(s,r,\theta)
    =
    \boldsymbol{\mathcal{C}}(s)
    +
    r\cos\theta\,\boldsymbol{N}(s)
    +
    r\sin\theta\,\boldsymbol{B}(s).
\label{eq:Stube}
\end{equation}
Here, $r$ is the radial distance from the centerline and $\theta$ is the polar angle measured in the local Frenet cross-section.
By defining the local unit vectors frame $(\boldsymbol{e}_s, \boldsymbol{e}_{r}, \boldsymbol{e}_{\theta})$ by
\begin{equation}
	\begin{cases}
	  \boldsymbol{e}_s = \boldsymbol{T}, \\
	  \boldsymbol{e}_{r} = \cos \theta \boldsymbol{N} + \sin \theta \boldsymbol{B}, \\
	  \boldsymbol{e}_{\theta} = - \sin \theta \boldsymbol{N} + \cos \theta \boldsymbol{B},
	\end{cases}
\end{equation}
the vorticity $\boldsymbol{\omega}$ in the tubular region could be expressed as
\begin{equation}\label{eq:Sw}
    \boldsymbol{\omega}(s,r,\theta)
    = \omega_s(s,r,\theta)\boldsymbol{e}_s
    + \omega_r(s,r,\theta)\boldsymbol{e}_r
    + \omega_\theta(s,r,\theta)\boldsymbol{e}_\theta.
\end{equation}
Using the Frenet--Serret equations \eqref{eq:FS_equations}, the derivatives of the coordinate basis mapping with respect to $s$, $r$ and $\theta$ are
\begin{equation}\label{eq:S_mapping}
    \begin{cases}
\begin{aligned}
\frac{\partial \boldsymbol{x}}{\partial s} &= (1-\kappa_cr\cos\theta)\boldsymbol{e}_s + \tau_c r\boldsymbol{e}_\theta, \\
\frac{\partial \boldsymbol{x}}{\partial r} &= \boldsymbol{e}_r, \\
\frac{\partial \boldsymbol{x}}{\partial \theta} &= r\boldsymbol{e}_\theta.
\end{aligned}
\end{cases}
\end{equation}
In the following derivation, we use $J_c=1-\kappa_c r\cos\theta$ to represent the geometric factor associated with the centerline curvature.
\vspace{0.1in}

\subsection{Radial correction from the divergence-free condition}
Next, we calculate the divergence of the vorticity $\boldsymbol{\omega}$ within the tubular region and impose the divergence-free condition.
To apply the divergence formula in general curvilinear coordinates~\citep{Aris1962}, we rewrite $\boldsymbol{\omega}$ in terms of the coordinate basis vectors $\partial\boldsymbol{x}/\partial s$, $\partial\boldsymbol{x}/\partial r$, and $\partial\boldsymbol{x}/\partial\theta$:
\begin{equation}\label{eq:S_contravariant}
    \boldsymbol{\omega}
    = \omega^s \frac{\partial\boldsymbol{x}}{\partial s}
    + \omega^r \frac{\partial\boldsymbol{x}}{\partial r}
    + \omega^\theta \frac{\partial\boldsymbol{x}}{\partial\theta}.
\end{equation}
Here, $\omega^s$, $\omega^r$, and $\omega^\theta$ are the contravariant components in the coordinate basis and differ from the components in the local unit-vector frame.
Using \eqref{eq:S_mapping}, we obtain the following relations between these coordinates:
\begin{equation}\label{eq:S_J}
    \omega^s = \frac{\omega_s}{J_c},
    \qquad
    \omega^r = \omega_r,
    \qquad
    \omega^\theta
    = \frac{\omega_\theta}{r}
      - \frac{\tau_c\omega_s}{J_c}.
\end{equation}
The volume Jacobian factor of the curved cylindrical coordinate system is
\begin{equation}\label{eq:S_g}
    \sqrt{g}
    =
    \frac{\partial\boldsymbol{x}}{\partial s}
    \cdot
    \left(
      \frac{\partial\boldsymbol{x}}{\partial r}
      \times
      \frac{\partial\boldsymbol{x}}{\partial\theta}
    \right)
    = J_c r.
\end{equation}
Combing \eqref{eq:S_J} and \eqref{eq:S_g}, the divergence of $\boldsymbol{\omega}$ is 
\begin{equation}\label{eq:S_div_w}
    \begin{aligned}
\nabla \cdot \boldsymbol{\omega} &= \frac{1}{\sqrt{g}} \left[ \frac{\partial(\sqrt{g} \omega^{s})}{\partial s} + \frac{\partial(\sqrt{g} \omega^{r})}{\partial r} + \frac{\partial(\sqrt{g} \omega^{\theta})}{\partial \theta} \right]\\
&=\frac{1}{J_c r}\left[\frac{\partial}{\partial s}(r\omega_s)+ \frac{\partial}{\partial r}(J_c r\omega_r)+ \frac{\partial}{\partial\theta}(J_c\omega_\theta-\tau_c r\omega_s)\right]\\
&=\frac{1}{J_c r}\left[r\frac{\partial\omega_s}{\partial s}+ \frac{\partial}{\partial r}(J_c r\omega_r)+ \frac{\partial}{\partial\theta}(J_c\omega_\theta)- \tau_c r\frac{\partial\omega_s}{\partial\theta}\right].
\end{aligned}
\end{equation}
In \eqref{eq:S_div_w}, $\omega_s$ denotes the axial component of vorticity along the tubular region (the vorticity flux), which is governed by the Lundgren spiral vortex structure~\citep{Lundgren1993}. $\omega_r$ and $\omega_{\theta}$ are the radial and azimuthal vorticity components introduced to ensure the divergence-free condition $\nabla \cdot \boldsymbol{\omega}=0$.

We adopt a concise yet general approach to constructing the vortex tubes, setting the azimuthal component $\omega_{\theta}= 0$ while retaining the dominant axial component $\omega_s$ and the radial correction component $\omega_r$.
Therefore, \eqref{eq:Sw} simplifies to
\begin{equation}\label{eq:S_www}
    \boldsymbol{\omega}(s,r,\theta)
    = \omega_s(s,r,\theta)\boldsymbol{e}_s
    + \omega_r(s,r,\theta)\boldsymbol{e}_r.
\end{equation}
and \eqref{eq:S_div_w} becomes
\begin{equation}\label{eq:S_w2}
    \begin{aligned}
\nabla\cdot\boldsymbol{\omega}&=\frac{1}{J_cr} \left[r\frac{\partial \omega_s}{\partial s}+ \frac{\partial}{\partial r}(J_cr\omega_r)-\tau_c r\frac{\partial \omega_s}{\partial\theta}\right]\\
&=\frac{1}{J_c}\left[\frac{\partial \omega_s}{\partial s}-\tau_c\frac{\partial \omega_s}{\partial\theta}\right] +\frac{1}{J_cr}\frac{\partial}{\partial r}(J_cr\omega_r).
\end{aligned}
\end{equation}
Imposing the divergence-free condition $\nabla \cdot \boldsymbol{\omega}=0$ into \eqref{eq:S_w2} gives 
\begin{equation}
    \frac{\partial}{\partial r}\left[J_cr\omega_r\right]=-r \left[\frac{\partial \omega_s}{\partial s}-\tau_c\frac{\partial \omega_s}{\partial\theta}\right].
\label{eq:S_radial_correction}
\end{equation}
Integrating \eqref{eq:S_radial_correction} from $0$ to the radial position $r$ and imposing the centerline regularity condition $\lim_{r \to 0}r\omega_r(s,r,\theta)=0$, we obtain the general radial correction for arbitrary vortex tubes as
\begin{equation}\label{eq:S_wr_c}
    \omega_r(s,r,\theta)=-\frac{1}{J_cr}\int_0^r \mathcal{R}\left[\frac{\partial \omega_s(s,\mathcal{R},\theta)}{\partial s}-\tau_c\frac{\partial \omega_s(s,\mathcal{R},\theta)}{\partial\theta}\right]d\mathcal{R}.
\end{equation}
\vspace{0.1in}

\subsection{Construction of Lundgren spiral vortex tube}
Next, we derive the specific form of the radial correction $\omega_r(s,r,\theta)$ in Lundgren spiral vortex tube and give the expression for the vorticity $\boldsymbol{\omega}(s,r,\theta)$ in \eqref{eq:S_www}.
The axial component of vorticity $\omega_s$  is determined by the following equations (details can be found in Section 2.1 of the main text):
\begin{equation}\label{eq:S_Lundw}
    \begin{cases}
\begin{aligned}
&\omega_s(r,\theta,\tau(s)) = \omega_0(r) + 2 f(r) \sum_{n=1}^{N_\varepsilon} \mathcal{D}_n(r, \tau(s)) \cos \Upsilon_n (r, \theta, \tau(s)),\\
&\omega_0(r) =\frac{\Gamma}{\pi (1+h) r_0^2} \left( 1 + h (r/r_0)^2 \right) \exp \left( -(r/r_0)^2 \right),\\
&f(r)=\frac{\Gamma h r^2}{2\pi (1+h) r_0^4}  \exp \left( -(r/r_0)^2 \right).
\end{aligned}
\end{cases}
\end{equation}
For the $n$th mode, the Lundgren spiral phase is $\Upsilon_n(r,\theta,\tau)=2n\left[\theta-\mu(r)\tau\right]$ and 
the Lundgren diffusion factor is $\mathcal{D}_n(r,\tau)=\exp\left[-{4n^2\mu'(r)^2\nu_L\tau^3}/{3}\right]$.
The corresponding angular velocity function and its radial derivative are
\begin{equation}
    \mu(r) = \frac{\Gamma \left[ 1 + h - \left( 1 + h + h (r/r_0)^2 \right) \exp \left( -(r/r_0)^2 \right) \right]}{2(1 + h)\pi r^2}
\end{equation}
and
\begin{equation}
    \mu'(r)=\frac{\Gamma}{(1+h)\pi r^3} \left\{ \left[ 1+h + (1+h)(r/r_0)^2 + h(r/r_0)^4 \right] \exp(-r^2/r_0^2) - (1+h) \right\}.
\end{equation}
We also introduce a space-time mapping that continuously and cyclically connects vortex structures of different ages along the same centerline
(details can be found in Section 2.2 of the main text).
Within one length cycle $\ell_{\text{cycle}} = 2\ell_{\text{up}}  = 2 l_0 \left( \text{e}^{\gamma_L \tau_L} - 1 \right)$, the age-increasing and age-decreasing mapping could be written as
\begin{equation}\label{eq:S_taus}
\begin{aligned}
    \tau(s)
    =
    \begin{cases}
    \displaystyle
    \frac{1}{\gamma_L}
    \ln\left(
    1+\frac{s}{l_0}
    \right),
    &
    0\leq s\leq\ell_{\rm up},
    \\[2.2ex]
    \displaystyle
    \frac{1}{\gamma_{L}} \ln \left( 1 + \frac{\ell_{\text{cycle}}-s}{l_0} \right) ,  
    &
    \ell_{\text{up}} < s \le \ell_{\text{cycle}}.
    \end{cases}
\end{aligned}
\end{equation}

Since only $\Upsilon_n(r,\theta,\tau)$ depends on $\theta$ in $\omega_s$ and ${\partial\Upsilon_n}/{\partial\theta}=2n$,
the derivative with respect to $\theta$ in \eqref{eq:S_wr_c} is
\begin{equation}\label{eq:S_wr_theta}
    \frac{\partial \omega_s}{\partial\theta}= -4 f(r) \sum_{n=1}^{N_\varepsilon} n \mathcal{D}_n(r, \tau) \sin \Upsilon_n(r, \theta, \tau).
\end{equation}
In WLT, the explicit variation of the local axial profile along the centerline occurs through the age mapping $\tau=\tau(s)$, so that
${\partial \omega_s}/{\partial s}={\partial \omega_s}/{\partial\tau} \cdot{d\tau}/{ds}.$
For the $n$th spiral mode, we obtain $\partial\Upsilon_n/\partial\tau=-2n\mu(r)$ and $\partial\mathcal{D}_n/\partial\tau=-4n^2\mu'(r)^2\nu_L\tau^2\mathcal{D}_n$. Applying the chain rule yields
\begin{equation}
    \frac{\partial}{\partial \tau} [\mathcal{D}_n \cos \Upsilon_n] = \mathcal{D}_n \left[ 2n\mu(r) \sin \Upsilon_n - 4n^2 \mu'(r)^2 \nu_L \tau^2 \cos \Upsilon_n \right].
\end{equation}
Thus, the derivative of the axial component of vorticity with respect to vortex age is
\begin{equation}\label{eq:S_wr_tau}
    \frac{\partial \omega_s}{\partial \tau} = 4 f(r) \sum_{n=1}^{N_\varepsilon} \mathcal{D}_n(r, \tau) \left[ n \mu(r) \sin \Upsilon_n(r, \theta, \tau) - 2 n^2 \mu'(r)^2 \nu_L \tau^2 \cos \Upsilon_n(r, \theta, \tau) \right]
\end{equation}
The derivative of the local age could be derived from \eqref{eq:S_taus} as
\begin{equation}
    \tau'(s) = \frac{d\tau}{ds} = 
\begin{cases} 
\dfrac{1}{\gamma_L(l_0 + s)}, & 0 \le s \le \ell_{\text{up}}, \\
-\dfrac{1}{\gamma_L(l_0 - s + \ell_{\text{cycle}})}, & \ell_{\text{up}} < s \leq \ell_{cycle},
\end{cases}
\end{equation}
so the derivative with respect to $s$ in \eqref{eq:S_wr_c} is
\begin{equation}\label{eq:S_wr_s}
    \frac{\partial \omega_s}{\partial s} = 4 f(r) \tau'(s)\sum_{n=1}^{N_\varepsilon} \mathcal{D}_n(r, \tau) \left[ n \mu(r) \sin \Upsilon_n(r, \theta, \tau) - 2 n^2 \mu'(r)^2 \nu_L \tau^2 \cos \Upsilon_n(r, \theta, \tau) \right]
\end{equation}
Replacing the coordinate variable $r$ in \eqref{eq:S_wr_theta} and \eqref{eq:S_wr_s} with the integration variable $\mathcal{R}$ and substituting the resulting expressions into \eqref{eq:S_wr_c}, we could get the radial correction in Lundgren spiral vortex tube as
\begin{equation}\label{eq:S_wr_correctionL}
    \begin{aligned}
\omega_r(s, r, \theta) = & - \frac{4}{J_cr} \int_0^r \mathcal{R}f(\mathcal{R}) \sum_{n=1}^{N_\varepsilon} \mathcal{D}_n(\mathcal{R}, \tau(s)) 
 \Big\{n\left[\tau'(s) \mu(\mathcal{R})+\tau_c(s)\right]  \sin \Upsilon_n(\mathcal{R}, \theta, \tau(s)) \\&- 2 \tau'(s) n^2 \mu'(\mathcal{R})^2 \nu_L \tau(s)^2  \cos \Upsilon_n(\mathcal{R}, \theta, \tau(s)) \Big\} d \mathcal{R} .
\end{aligned}
\end{equation}

\begin{figure}  
	\centering
	\begin{overpic}
		[width=1.0\textwidth]{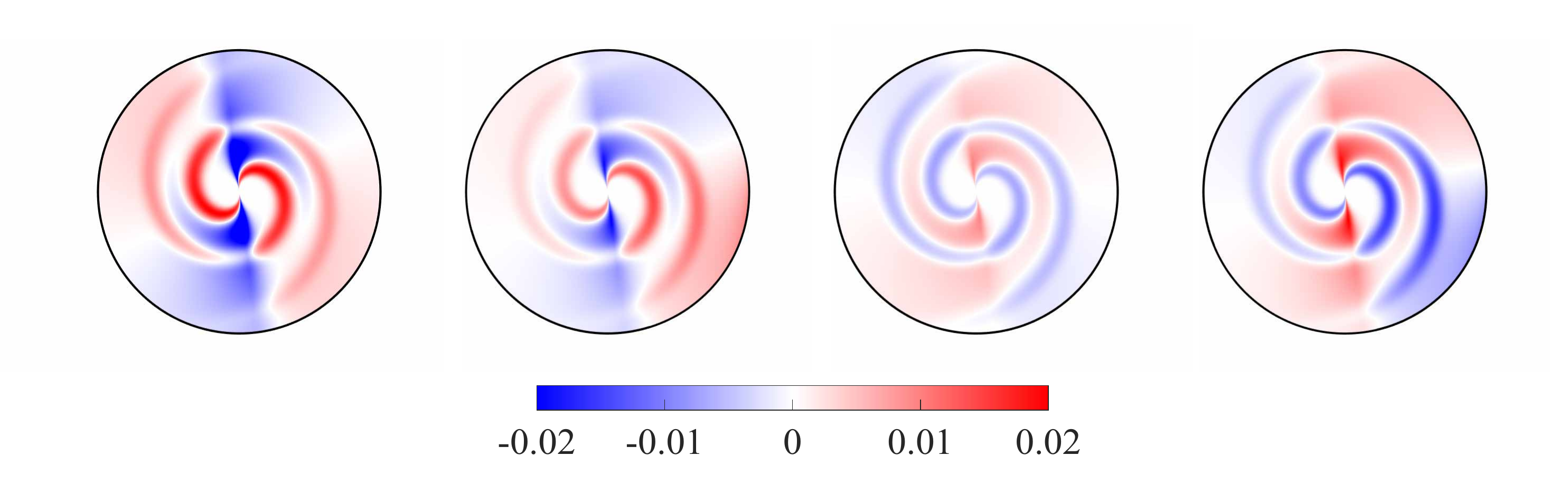}
        \put(4,29){\footnotesize{(a)}}
        \put(27,29){\footnotesize{(b)}}
        \put(50.5,29){\footnotesize{(c)}}
        \put(74,29){\footnotesize{(d)}}
    \end{overpic}
    \caption{Contours of the radial correction $\omega_r$ in \eqref{eq:S_wr_correctionL} for different centerline parameters: (a) $\kappa_c=0$ and $\tau_c=1.5$; (b) $\kappa_c=0.3$ and $\tau_c=1.5$; (c) $\kappa_c=0$ and $\tau_c=-1.5$; (d) $\kappa_c=0.3$ and $\tau_c=-1.5$.
   Here, the black circle marks the boundary of the constructed vortex tube, and outward-directed vorticity is defined as positive (inward-directed vorticity as negative).}
    \label{fig:w_rr}
\end{figure}

\begin{figure}  
	\centering
	\begin{overpic}
		[width=0.72\textwidth]{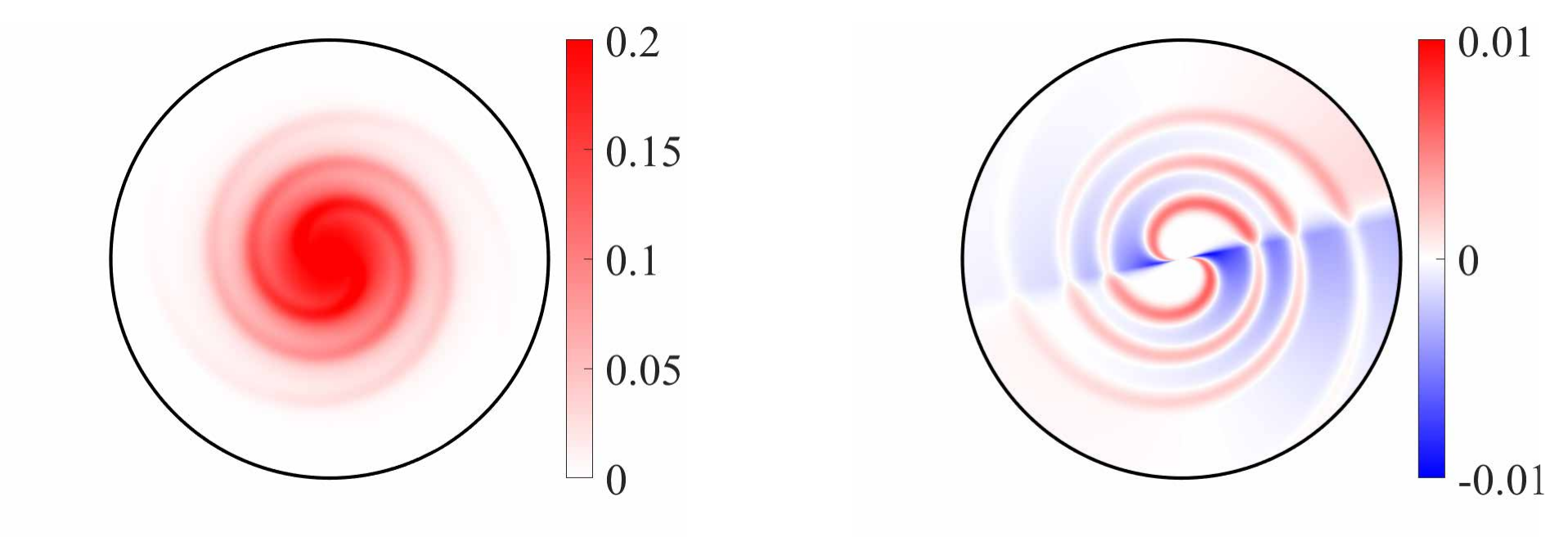}
        \put(2,32){\footnotesize{(a)}}
        \put(56,32){\footnotesize{(b)}}
    \end{overpic}
    \caption{Contours of different vorticity components in Lundgren spiral vortex tube: (a) axial vorticity $\omega_s$ in \eqref{eq:S_Lundw}; (b) radial vorticity $\omega_r$ in \eqref{eq:S_wr_correctionL}, where the outward-directed vorticity is defined as positive (inward-directed vorticity as negative).
    The black circle marks the boundary of the constructed vortex tube.}
    \label{fig:w_sr}
\end{figure}

We use figures~\ref{fig:w_rr} and \ref{fig:w_sr} to illustrate the vorticity distribution within the tubular region.
The cases shown in figure~\ref{fig:w_rr} are based on the following dimensionless parameter values: $r_0 = 1$, $\Gamma=1$, $\nu_L= 5 \times 10^{-4}$, $ h = 0.5$, $\tau=75$ and $\tau'(s)=10$.
In the geometric construction of the spiral vortex tube, the topological parameters $\kappa_c$ and $\tau_c$ of the vortex centerline affect the magnitude and direction of the radial correction vorticity.
The case shown in figure~\ref{fig:w_sr} is based on the following dimensionless parameter values: $r_0 = 1$, $\Gamma=1$, $\nu_L= 1 \times 10^{-4}$, $ h = 0.5$, $\tau=100$, $\tau'(s)=10$, $\kappa_c=0.5$ and $\tau_c=4$.
Compared with the radial component $\omega_r$, the axial component $\omega_s$ dominates the vorticity contribution within the vortex tube.

Consequently, incorporating \eqref{eq:S_Lundw} and \eqref{eq:S_wr_correctionL} into \eqref{eq:S_www}, we obtain
\begin{equation}\label{eq:S_Lundgrentubes}
    \begin{aligned}
\boldsymbol{\omega}(s,r,\theta)
    = &[\omega_0(r) + 2 f(r) \sum_{n=1}^{N_\varepsilon} \mathcal{D}_n(r, \tau(s)) \cos \Upsilon_n (r, \theta, \tau(s))]\cdot  \boldsymbol{e}_s \\&- \Big\{\frac{4}{J_cr} \int_0^r \mathcal{R}f(\mathcal{R}) \sum_{n=1}^{N_\varepsilon} \mathcal{D}_n(\mathcal{R}, \tau(s)) 
 \Big\{n\left[\tau'(s) \mu(\mathcal{R})+\tau_c(s)\right]  \sin \Upsilon_n(\mathcal{R}, \theta, \tau(s))\\& - 2 \tau'(s) n^2 \mu'(\mathcal{R})^2 \nu_L \tau(s)^2  \cos \Upsilon_n(\mathcal{R}, \theta, \tau(s)) \Big\} d \mathcal{R}\Big\}  \cdot \boldsymbol{e}_r.
\end{aligned}
\end{equation}
Equation \eqref{eq:S_Lundgrentubes} gives the vorticity within a single Lundgren spiral vortex tube in WLT.
The constructed vortex tube has a dominant axial vorticity component and a radial correction component given by \eqref{eq:S_wr_c}. Substituting \eqref{eq:S_Lundgrentubes} into the divergence expression \eqref{eq:S_w2} yields $\nabla\cdot\boldsymbol{\omega}=0$, demonstrating that its vorticity strictly satisfies the divergence-free condition.


\section{Construction of the vorticity and velocity fields in WLT}
We denote the vorticity in \eqref{eq:S_Lundgrentubes} as $\boldsymbol{\omega}_L$.
For the numerical construction of WLT, we prescribe the vorticity $\boldsymbol{\omega}^{(i)}, i=1,2,\cdot\cdot\cdot,N_L$, for each spiral vortex tube constructed along the centerline
$\boldsymbol{\mathcal{C}}^{(i)}$ as follows:
\begin{equation}
    \boldsymbol{\omega}^{(i)} = \begin{cases} \boldsymbol{\omega}_L, & r \in [0, R_L], \\ 0, & r \in (R_L, +\infty). \end{cases}
\end{equation}
Here, $N_L$ denotes the number of intertwined vortex tubes in the field, and $R_L$ denotes the truncation radius of the tubular region. The radius is sufficiently large to include almost all the vorticity in \eqref{eq:S_Lundgrentubes}.
Finally, we obtain the total vorticity field
\begin{equation}
   \boldsymbol{\omega}= \sum_{i=1}^{N_L} \boldsymbol{\omega}^{(i)}
\end{equation}
by summing the vorticity fields of the individual spiral vortex tubes. 
The resulting field satisfies the divergence-free condition as well.

The vorticity field $\boldsymbol{\omega}$ is constructed in the periodic domain, and the associated incompressible velocity field $\boldsymbol{u}$ is recovered using the spectral Biot--Savart relation~\citep{Xiong2019,Shen2023}:
\begin{equation}
   \boldsymbol{u}
   = \mathcal{F}^{-1}\!\left(
       \frac{\mathrm{i}\,\boldsymbol{k}\times\tilde{\boldsymbol{\omega}}}{|\boldsymbol{k}|^2}
   \right).
\end{equation}
Here, \(\mathcal{F}\) denotes the Fourier transform operator, \(\mathcal{F}^{-1}\) denotes the inverse Fourier transform, and 
\(\tilde{\boldsymbol{\omega}}=\mathcal{F}(\boldsymbol{\omega})\) is the Fourier transform of the vorticity field \(\boldsymbol{\omega}\).

\vspace{0.7in}


\phantomsection
\addcontentsline{toc}{section}{References}

\bibliographystyle{jfm}
\bibliography{jfm}